\documentclass[preprint]{aastex}

\newcommand{\CII}{C~{\sc ii}}
\newcommand{\CIII}{C~{\sc iii}}
\newcommand{\CIV}{C~{\sc iv}}
\newcommand{\SiII}{Si~{\sc ii}}
\newcommand{\SiIII}{Si~{\sc iii}}
\newcommand{\SiIV}{Si~{\sc iv}}
\newcommand{\NIV}{N~{\sc iv}}
\newcommand{\NV}{N~{\sc v}}
\newcommand{\HeI}{He~{\sc i}}
\newcommand{\HeII}{He~{\sc ii}}
\newcommand{\HeIII}{He~{\sc iii}}
\newcommand{\HI}{H~{\sc i}}
\newcommand{\HII}{H~{\sc ii}}
\newcommand{\AlII}{Al~{\sc ii}}
\newcommand{\AlIII}{Al~{\sc iii}}
\newcommand{\FeII}{Fe~{\sc ii}}
\newcommand{\OI}{O~{\sc i}}
\newcommand{\OVI}{O~{\sc vi}}
\newcommand{\NiII}{Ni~{\sc ii}}
\newcommand{\MgII}{Mg~{\sc ii}}

\slugcomment{Submitted to the Astrophysical Journal Supplement}

\shorttitle{QSO Metal Line Absorption Systems}
\shortauthors{Boksenberg, Sargent, \& Rauch}

\begin{document}

\title{PROPERTIES OF QSO METAL LINE ABSORPTION SYSTEMS AT HIGH REDSHIFTS: NATURE AND
EVOLUTION OF THE ABSORBERS AND THE IONIZING RADIATION BACKGROUND\altaffilmark{1}}

\author{Alec Boksenberg}
\affil{University of Cambridge, Institute of Astronomy, Madingley Road, 
Cambridge, CB3 0HA, UK}
\email{boksy@ast.cam.ac.uk}

\author{Wallace L. W. Sargent}
\affil{Palomar Observatory, 105-24 California Institute of Technology,
Pasadena, CA 91125}
\email{wws@astro.caltech.edu}

\and

\author{Michael Rauch}
\affil{The Observatories of the Carnegie Institution of Washington, 813 
Santa Barbara St., Pasadena, CA 91101}
\email{mr@ociw.edu}

\altaffiltext{1}{The data presented herein were obtained at the W.M. Keck 
Observatory, which is operated as a scientific partnership among the 
California Institute of Technology, the University of California and the 
National Aeronautics and Space Administration. The Observatory was made 
possible by the generous financial support of the W.M. Keck Foundation.}

\begin{abstract}

A sample of 908 \CIV\ absorber components clumped in 188 systems outside
the Lyman forest in the redshift range $1.6 \lesssim z \lesssim 4.4$ have
been identified in Keck HIRES spectra of nine QSOs. These and
corresponding lines of \SiIV, \CII, \SiII\ and \NV\ have been fitted with 
Voigt profiles to obtain column densities and \CIV\ Doppler parameters.
The properties of the \CIV\ absorbers are almost constant although 
their system velocity spreads tend to increase with decreasing redshift.
We find a mild increase in \CIV\ comoving mass density with decreasing redshift
with a mean 
$\langle \Omega_{\scriptsize\textrm{\CIV}} \rangle = (3.8\pm0.7) \times10^{-8}$
(1$\sigma$ uncertainty limits; spatially flat $\Lambda$CDM cosmology with 
$\Omega_{\Lambda} = 0.7$, $\Omega_{\rm M} = 0.3$ and $h = 0.71$), 
in broad agreement with earlier work. Corresponding values of 
C/H in the Lyman forest based on $\Omega_{\rm b}$ from the 
CMB and ionization fractions from our data are 
[C/H]$_{\langle z \rangle = 4.0} \geq -3.11^{+0.14}_{-0.19}$ and
[C/H]$_{\langle z \rangle = 2.1} \geq -2.64^{+0.15}_{-0.22}$, suggesting a rise 
by a factor $\sim 3$. Relating the hydrogen mass density more directly to regions 
containing the \CIV\ absorbers our values for [C/H] become $\gtrsim -2.2$ at 
$\langle z \rangle = 4.0$ and $\gtrsim -2.0$ at $\langle z \rangle = 2.1$. 
\CIV\ absorber components exhibit strong clustering out to 
$\Delta_v \lesssim 300$ km s$^{-1}$ but there is no clustering on any scale 
between \emph{systems}. We argue that for our sample the \CIV\ clustering is 
entirely due to the peculiar velocities of gas present in the outer extensions of 
galaxies. We find no change in the median column density ratio \SiIV/\CIV\ with 
redshift, particularly no large change near $z = 3$, contrary to previous 
observations; other ionic ratios vary continuously with redshift. We show that 
these are only partial indicators of ionization state and remedy this by use of 
specific pairs of ionic ratios. We demonstrate that the majority of absorbers are 
photoionized and find that at $z \lesssim 2.65$ QSOs dominate the ionization of 
the absorption systems whereas at $z \gtrsim 3.4$ an additional, dominant 
contribution from galaxies with specific spectral characteristics and high 
radiative escape fraction in the energy range 1--4 Ryd is required. These results 
also indicate that [Si/C] $=$ 0.0--0.4 fits the data well. Between $z = 2.65$ and 
$z = 3.4$ there is evident transition in the ionization properties of the 
absorbers, with large scatter. The UV spectral properties required for the 
galaxies are not reproduced by standard stellar population synthesis models. We 
conclude that the heavy element absorbers at $z \gtrsim 3.4$ are located close 
to galaxies and irradiated dominantly by them, consistent with our independent 
conclusion from clustering properties.

\end{abstract}


\section{INTRODUCTION}

The spectra of exceptional quality delivered by the Keck~I High Resolution 
Spectrograph (HIRES; Vogt et al. 1994) have revealed individual metal absorption
features related to the high redshift Lyman forest for a large fraction of the 
stronger lines \citep{cow95,tyt95,soc96,wsl96}. By redshift {\it z} $\sim 3$ such 
absorbers are found to have a median carbon abundance approximately $10^{-2}$ of solar 
(with substantial scatter) and Si/C similar to Galactic halo stars, although these 
values are based on rather uncertain ionization corrections \citep{soc96,rhs97}. There 
is evidence for some metal enrichment also at considerably lower \HI\ column densities 
\citep{ell00,sch00}. While it is still unclear how the pollution of the forest material 
has come about, the observed metal absorbers provide a powerful probe of early stages 
in the growth of structure and the formation of galaxies and give an observational 
approach to determining the spectral character of the cosmological ionizing background 
at those times.

In recent years, detailed hydrodynamical simulations of cosmological structure
formation in the presence of a photoionizing background which yield direct 
quantities for comparison with observations have been performed by several groups 
\citep{cen94,zan95,her96,dat01,vie02}. In these simulations it is straightforward to
compute the neutral hydrogen absorption that would be produced in the light of a 
background QSO along an arbitrary line of sight through the simulation volume. It 
is impressive that such results can reproduce the evolving spectral appearance and 
statistical properties of cosmologically distributed \HI\ absorbers in considerable 
detail, spanning the range from the weakest detected to those showing damped 
Lyman~$\alpha$ profiles.

An important insight gained from the simulations is that galaxies and \HI\ absorbers 
develop naturally together in the hierarchical formation of structure. High column 
density lines ({\it N}(\HI) $\gtrsim 10^{17}$ cm$^{-2}$) arise from radiatively cooled 
gas associated with forming galaxies in collapsed, high density, compact regions. 
Lower column density absorption ({\it N}(\HI) $\lesssim 10^{15}$ cm$^{-2}$) occurs in 
the shallower dark matter potential wells, containing gas in various stages of 
gravitational infall and collapse; typically these are in the form of flattened or 
filamentary structures of moderate overdensity with Doppler parameters that are often 
set by peculiar motions or Hubble flow in addition to thermal broadening. 
Gravitational, pressure and ram-pressure confinement all play significant roles. Such 
a scenario in which metal absorption arises in gas assumed to be homogeneously 
enriched has been discussed by Rauch et al. (1997a). In their simulation the large 
velocity widths of some metal absorbers arise from interactions between associated 
protogalactic clumps or from alignments of groups of such objects along chance 
filaments in the line of sight. This simple model does not include stellar ``feedback'' 
of energy and momentum from galaxies which could strongly modify the local gas 
distribution and kinematic state by producing outflows opposing the general infalling 
motion as well as contributing to the enrichment of the gas and fundamentally 
influencing the local ionizing conditions. However, increasing attention is now being 
given to accounting for stellar processes in galaxy formation simulations 
\citep{maw02,the02,cro02,sah02,nsh03}. Evidence for stellar feedback in the 
intergalactic medium has recently been reported, for example, by Rauch, Sargent, \& 
Barlow (1999; 2001), \citet{bon01}, and \citet{ade03}.
 
The strength and spectrum of the metagalactic ionizing radiation background at high
redshift and the nature of the sources which ionized the intergalactic medium are 
outstanding issues in cosmology. Although QSOs have long been accepted as the main 
contributors to the metagalactic ionizing radiation, there has been speculation on 
whether they dominate at the highest redshifts 
\citep{bec87,das87,sag87,bdo89,mam93,ham96}. Most of recent studies suggest that QSOs 
fall short of producing enough flux to satisfy measurements of the ``proximity effect''
at redshifts significantly beyond $z \sim 3$ where the space density of bright QSOs is 
sharply decreasing \citep{cec97,rau97,mhr99,sco00,bck01}. The presence of a strong 
population of high redshift Lyman-break galaxies \citep{ste96,ste99} and the detection 
of significant flux beyond the Lyman limit escaping from such galaxies \citep{spa01} 
lend support to the idea that star-forming regions dominated the ionizing background at 
early times \citep{hae01}. The spectral shape of the ultraviolet background radiation 
should be reflected in the ionization pattern of QSO metal system absorbers
\citep{cha86,bes86,sas89,var93,gas97,bsr01} and this can be used to identify the 
character of the ionizing sources if the spectral modifications due to propagation of 
the radiation through the intergalactic material are properly accounted for 
\citep{ham96}. In turn, this can lead to robust determinations of heavy element 
abundances in galaxy halos and in denser regions of the intergalactic medium.

In this paper we study the ionization state and kinematic properties of a large sample 
of metal absorbers and trace their evolution in redshift, greatly extending our earlier 
work (Boksenberg 1997; Boksenberg et al. 2001). In \S2 we describe the observations and 
initial data reduction and in \S3 outline our analysis of the absorbers, represented as 
multi-phase-ionization systems containing individual single-phase component regions. In 
\S4 we give full tables of results with supporting information, comments and displays. 
In \S5 we define the samples which we use in \S6 to derive statistical properties and 
the redshift evolution of absorber quantities, in \S7 for clustering studies from which 
we deduce the nature of the absorbers, and in \S8 for redshift distributions of ionic 
ratios and redshift-selected samples of ionic ratio combinations. In \S8 we also verify 
that collisional ionization is not an important ionizing mechanism for the observed 
species and show the influence of changes in column density and metallicity. In \S9 we 
study the properties of absorbers near the observed background QSOs. In \S10, from 
comparison of the characteristics of the observed ionic ratio combinations with results 
of photoionization modelling using the Cloudy code \citep{fer96}, we draw specific 
conclusions about the sources contributing to the ionizing radiation environment and 
how the effective balance of these changes with redshift. We summarize our results in 
\S11. In a further paper (in preparation) we determine in greater detail the parameters 
of the best observed absorbers.

\section{OBSERVATIONS AND DATA REDUCTION}

The work presented in this paper is based on our HIRES observations of a set of nine 
QSOs with redshifts $2.32 < z_{\rm em} < 4.56$, listed with associated information in 
Table 1. Most of the spectra were obtained using a slitwidth of $0\farcs86$ yielding a 
resolution $\sim 6.6$~km~s$^{-1}$ FWHM covered by roughly three pixels. The exception 
is for the gravitationally-lensed object Q1422$+$2309 \citep{pbw92} of which we take 
the data from image component C using a narrower slitwidth, $0\farcs574$, yielding 
$\sim 4.4$~km~s$^{-1}$ FWHM, and obtained in excellent seeing ($\lesssim 0\farcs6$) 
with position angle set to minimize contamination from the closely-spaced neighbouring 
components A and B (Rauch et al. 1999).

Two partially overlapping configurations for each wavelength region were used to give 
complete coverage of the free spectral range for the HIRES echelle format. The data 
were reduced as described in \citet{bas97}, with the individual exposures for each QSO
wavelength-shifted to heliocentric, vacuum values and added together with weights 
according to their signal-to-noise ratio (S/N). Continuum levels were 
delineated by means of polynomial fits to regions apparently free of absorption 
features and these were used to produce continuum-normalized spectra in preparation 
for the analysis of the absorption systems.

Due to the spectral variations in instrument efficiency and the sharply uneven 
exposures over the range resulting from the overlapped setups, in combination with 
the intrinsic spectral variation of signal intensity from the QSOs themselves, the 
spectra show complex and quite large variations in S/N along their lengths. To 
account for these individual patterns of S/N, matching statistical 1$\sigma$ error 
arrays are built up during the reduction stages and associated with each completed 
spectrum file. A rough indication of minimum signal quality is given in Table 1 by 
S/N values sampled at a few rest-wavelength positions (avoiding emission lines) in 
common to each spectrum.

\section{DETERMINATION OF ABSORPTION-LINE PARAMETERS}

\subsection{Selection Strategy}

The metal absorbers typically appear as well-defined clumps with velocity structure 
ranging up to a few hundred km s$^{-1}$ in width. In general there are wide expanses 
of apparently clear redshift space between such clumps. We classify these absorbing 
entities as {\it systems} and identify them by the presence of \CIV. Within each 
system we define a population of physical ``clouds'', termed {\it components}, each 
having a Gaussian velocity distribution of arbitrary width which collectively produce 
the velocity structure in detail.
  
Because of blending with \HI\ absorption for metal lines in the Lyman forest, which 
becomes increasingly severe to higher redshifts, our data sample is built primarily 
on measurements made outside the forest; only in some exceptionally favourable cases 
(indicated in the tables) are metal lines in the forest included. Outside the forest 
\CIV\ absorption is prevalent and is the only ion detected in the weakest systems. 
Stronger systems also contain lines of some, or occasionally most, of the species 
\SiIV, \CII, \SiII, \NV, \OI, \AlII, \AlIII, \FeII\ or \NiII, if available in the 
observed spectral range. 

For the work presented in this paper we concentrate on the lines 
\CIV~$\lambda\lambda$1548.195,1550.770, \SiIV~$\lambda\lambda$1393.755,1402.770, 
\CII~$\lambda$1334.5323, \SiII~$\lambda$1260.4221 and 
\NV~$\lambda\lambda$1238.821,1242.804.\footnote{All rest-frame vacuum wavelengths and 
related atomic data used here are from \citet{mor91} except for those which have 
revised {\it f-}values as given in the compilation of \citet{tls95}.} If a line is not 
detected at the wavelength expected from the presence of other species in the same 
system, an upper limit is determined, except for lines falling in the Lyman forest. In 
systems with \SiII~$\lambda$1260 in the forest, we substitute $\lambda$1526.7066 and 
include $\lambda$1304.3702 if available, but these lines are relatively much weaker 
than $\lambda$1260 and detected only when \SiII\ is strong; when not detected, upper 
limits obtained from these alternative lines are generally too high to be useful.

\subsection{Profile-Fitting Analysis}

For the analysis of our spectra we applied the Voigt profile-fitting package VPFIT 
developed by R. F. Carswell and collaborators \citep{car87,car02} and kindly made 
available to us. VPFIT is a $\chi^{2}$-minimization program capable of making detailed 
fits to the absorption profiles of several different transitions simultaneously. It 
estimates redshift, $z$, Doppler parameter, {\it b} and column density, $N$, with 
their associated errors, for the individual components of the systems in the defined 
fitting regions. For the instrumental resolution included in this procedure we took
${\it b}_{instr} = 2.83$ km~s$^{-1}$ for Q1422$+$239C and 
${\it b}_{instr} = 3.96$ km~s$^{-1}$ for the rest. Since the $\chi^{2}$-minimization 
technique operates on the reduced spectra there is a degree of correlation between 
neighbouring pixels arising from the rebinning of the data in the reduction process. 
To compensate for this smoothing we derived an error-correction file for each spectrum 
obtained by comparison of values at many wavelengths in the accumulated error array 
with directly measured values of the root-mean-square fluctuation in the final 
continuum. The error-correction factor is relatively small, typically in the range 
1.1--1.3, and is applied automatically within VPFIT. 

While the lower ionization species often are dominated by narrower components and the 
higher ionization ones by broader, we found, consistent with the S/N, that a range of 
component widths invariably are present in both sets. We conclude that the different 
ionic species in a system trace the same physical cloud regions; depending on the 
ionization states of these regions, each species shows its individual balance of 
component absorption strengths. We infer that the individual components identified in 
the metal systems represent \emph{single-phase-ionization absorbing regions} 
co-existing in multi-phase system complexes. We demonstrate with a simple example case 
that the absorbing regions indeed conform quite closely with this idealized model. 

Using a system at ${\it z} = 2.285$ in Q1626$+$6433, we show in Figure 1 how the 
derived components distribute themselves among the low to high ionic species. While 
different components dominate in \CII\ and \NV, both sets of components co-exist as 
blended, strong constituents of the \CIV\ overall profile well-separable in our 
analysis. Thus the components of this system straightforwardly represent individual 
regions having quite different degrees of ionization unambiguously traceable through 
the species. We deal with this point more quantitatively later in the paper. A 
considerably more complex example similarly demonstrating such separation into 
physically simple component entities is discussed in \S4.

The unifying assumptions we therefore adopted in our analysis are ({\it a}) that the 
component redshift structure seen in one ion corresponds exactly to that in any other 
ion of the same system, while allowing that in general the relative strengths of the 
components be different, and ({\it b}) that the Doppler parameters of corresponding 
components in different ionic species (in general containing thermal broadening and 
turbulent motion terms) are physically related. 

For each QSO the first step in our procedure was to identify all \CIV\ absorption 
doublets evident outside the forest. In VPFIT, doublet or other members of the same ion 
automatically are fitted with the same parameters when the wavelength regions where 
they occur are specified. We began our analysis by deriving the parameters of all 
necessary components that could be identified in the fitting of the profiles of the 
\CIV\ doublets. In the relatively few cases where components of one doublet member were 
excessively confused by blending with interloping species at other redshifts, or were 
otherwise severely contaminated, we used the remaining member.

Next, all other members of our defined set of ionic species potentially present in the 
available wavelength range, \emph{whether apparent or not}, were assigned the same 
initial set of component redshifts and linked to track together with \CIV\ in the 
subsequent fitting stages. The component {\it b-}values for all ionic species of the 
same atom (\CIV\ with \CII\ and \SiIV\ with \SiII) also were linked to track together. 
To enable VPFIT to derive mutually consistent {\it b-}values among atoms ({\it i.e.} 
each containing appropriate contributions of individual thermal broadening and common 
turbulence broadening for each cloud region) it was necessary both to assign realistic 
component temperatures (see below) and to relate the {\it b-}values of all species 
present. (Because of the relative atomic weights, changes in temperature have only a 
slight effect on the relative {\it b-}values for C and N but can produce quite marked 
changes in those for Si.) VPFIT then was allowed to attempt simultaneous fits to the 
line profiles in a first complete pass. Potential components not detected ``dropped 
out'' of the analysis and subsequently were assigned upper limits in the manner 
described below.\footnote{In some instances a minor velocity shift relative to \CIV, 
typically a small fraction of 1 km~s$^{-1}$, was made to one or other of the profiles 
of widely separated species in the same system to correct for slight departures from 
the nominal global fit to the wavelength scale \citep{bas97}.} 

Generally our procedure resulted in a set of profiles which corresponded well to the 
data for all of the components detected in each species. We found it beneficial to 
iterate the process by making two or three VPFIT runs while refining the nominal 
component temperatures between the runs. In the few cases when a component in \CIV\ was 
weak while \CII\ was strong, \CII\ was substituted as the prime species in the 
analysis. For some strong, well-separated, components, reliable {\it b-}values 
sometimes could be obtained independently for Si as well as C, although these cases 
were relatively rare. 

To obtain nominal component temperatures we used results given by \citet{rau96} 
from a formal decomposition into thermal and Gaussian non-thermal motions in a sample 
of related \CIV\ and \SiIV\ absorption components, shown in their Figure 3 as a plot 
of \CIV\ thermal against total {\it b-}values. They derive a mean temperature of $3.8 
\times 10^{4}$ K but note that their analysis is dominated by narrower components which 
have smaller measurement errors than the more uncertain broader features. They also 
suggest that a tail in the {\it b-}value distribution towards large values apparently 
indicating temperatures beyond $10^{5}$ K and increased non-thermal contributions may 
well represent blends of components. Our analysis supports this view. Since \CIV\ is 
more common and generally stronger than the other species, a blend of components 
representing a typical mix of relatively high ionization regions tends to appear 
significantly broader in \CIV\ than \SiIV. If interpretated as a single feature this 
indeed is likely to indicate an erroneously high temperature. We also show in \S8 that 
collisional ionization at temperatures near $10^{5}$ K cannot be significant for the 
absorbing regions we observe. Consequently we use the plot given by Rauch et al. as an 
aid to set temperatures only for the numerous narrower components in our sample 
({\it b} $\lesssim 10$ km~s$^{-1}$). 

For the more rarely-occurring broader components we found our fits to the observed 
profiles to be consistent with there being relatively little difference in the 
{\it b-}values among the species, consequently leading to large turbulent 
contributions. We set the temperature of these nominally at the mean 
$3.8 \times 10^{4}$ K given by Rauch et al. To explore the validity of this 
approximation we made VPFIT trials using a complex system at $z = 2.291$
in Q1626$+$6433 (described in \S4.2) to indicate the effect on the achieved column 
density values resulting from successive changes in assigned component temperatures. 
This system contains several broad components which overlap with numerous narrow ones. 
It is an appropriate example because in the profile-fitting process components are not 
treated in isolation but adjusted relative to one another to achieve the fit to the 
data; the degree to which a given component is influenced by others in this process 
then depends on their relative strength over its range of overlap. We compared the 
column density results for all components in the complex for two widely-spaced trial 
temperatures, $1\times10^{4}$ K and $1\times10^{5}$ K, assigned to the three broadest 
components, numbered 2, 4 and 14 in Figure 5 (where we have used the nominal 
temperature $3.8\times10^{4}$ K). These temperatures exceed the range expected for 
clouds of low metallicity and density photoionized by the intergalactic ultraviolet 
background radiation and so is a stringent test. The initial temperature assignments of 
the remainder of the components were treated equally in the two cases, following the 
procedure already described. In Figure 2 we show the resultant column densities for 
\CIV, \SiIV, \CII\ and \SiII\ obtained for the two cases. The two sets of values are 
not significantly different, nor from our results using the adopted nominal 
temperature. It is apparent that the derived column densities for the individual
components do not depend strongly on the thermal properties of the \emph{broad} 
components in an absorption complex and gives us confidence that our necessarily 
approximate approach is a sound procedure for obtaining reliable column densities. 

On the same basis, the more constrained temperature bounds set by the widths of the 
narrower components make the derived column densities for these less sensitive still 
to changes in assigned temperature. The relative insensitivity to profile width 
applies equally for the instrumental profile, which varies by $\sim \pm3.8\%$ in 
velocity width over the spectral range \citep{bas97}, and allowed us to use a single, 
averaged, figure for each spectrum, given above.

Most system members are well isolated from lines at other redshifts and the analysis 
using the described procedure is generally straightforward. When blending does occur, 
if some lines belong to doublets whose other members are uncontaminated, or if they 
are linked with accessible lines of related transitions, we found reliable values 
usually could be obtained for blended lines by a simultaneous analysis of all the 
systems present. Component parameter values judged to be too uncertain due to blending 
were excluded from the subsequent scientific analysis.

Component parameter errors as given by VPFIT are nominally $1\sigma$ values, but 
confusion between too-closely overlapping components with comparable parameters can 
give very large apparent errors.\footnote{Nevertheless, the combined column densities 
in such cases remains accurate and VPFIT has a procedure which can be used to give the 
correct error for the \emph{total} column density of a set of adjacent components in 
a complex.} Consequently, in making our profile fits we avoided ``overfitting'' and 
adopted the general rule to end with the minimum number of components that gave a 
reasonable fit after achieving a reduced $\chi^{2}$ close to 1 per degree of freedom 
globally for the set of spectral regions linked in the analysis. 

At the conclusion of the fitting process for each system we obtained the associated 
errors on the component column densities alone by fixing the corresponding values 
of $z$ and {\it b} in a final iteration. In this operation we also derived upper 
limits for all potential components within the different species which had been too 
weak to survive the first pass of the analysis. We did this by re-introducing them 
with the appropriate fixed values for $z$ and {\it b} and a small assigned column 
density well below the threshold of significance. The associated error values which 
are obtained become the adopted 1$\sigma$ upper limits.

The ability of the Voigt profile-fitting technique to separate different absorbing 
regions, even though it is sometimes arbitrary, gives it significant advantage over 
the apparently more direct technique of computing optical depths throughout an 
absorption complex \citep{sas91,son98} because the latter cannot account for 
overlapping blends of adjacent components, interlopers from other systems or 
differential temperature broadening between species of the same physical component. It 
is also clear that the wide range of ionization conditions generally found within a 
system means that reliable determinations of ionic \emph{ratios} cannot be obtained 
simply from ratios of \emph{total} system quantities as used by \citet{soc96} and 
\citet{son98}; to be physically meaningful they must be determined individually from 
the components of the multi-phase systems as we do later in this paper.

\section{PRESENTATION OF RESULTS}

\subsection{The Data}

Following the prescription outlined in the previous section, for each system 
we obtained excellent simultaneous fits over all species with a single pattern of 
component redshifts and appropriately linked {\it b-}values. As a full example of 
this we show in Figure 3 our VPFIT results superimposed on the observations for all 
the detected systems in Q1626+6433 having more than just \CIV\ accessible. In Tables 
2--10 we list the derived values for all available components of our target 
transitions in the spectra of the nine QSOs. Column 1 gives the absorption redshift 
$z$; column 2 gives {\it b-}values for C and column 3, imposed {\it b-}values for Si 
as described in \S3.2 shown bracketted and independent values when left as a free 
parameter, unbracketted\footnote{We show only the more reliable values for these; 
values with large uncertainties are left bracketted.} ({\it b-}values used for N are 
not given but are close to those of C when representing the same physical component); 
columns 4 to 8 give column densities; and column 9 identifies components by number 
within a system. Each system is headed in the tables with the mean redshift of its 
constituent components. Where there is severe contamination, strong saturation or too 
much confusion from blending by lines at other redshifts to give useful quantities, no 
entry is given as explained in a footnote. In the few marginal cases of weak lines when 
VPFIT yielded a column density lower than the associated error (instead of making a 
rejection) these two quantities are added in quadrature and included in the table as 
an upper limit. Redshifts are vacuum, heliocentric values and are based on the \CIV\ 
components, but apply consistently to all species, as described in \S3. The listed 
{\it b-}values for C almost always are the result of fits to \CIV\ which usually 
dominates in strength over \CII. In rare cases, when component regions have low 
ionization and \CIV\ is very weak, independent {\it b-}values are derived solely from 
\CII\ (and \SiII\ when left as a free parameter). For the well separated, stronger 
components the formal $1\sigma$ error in {\it z} is typically $\lesssim$ 0.000005 and 
in {\it b} (for C, and Si when independently derived), $\lesssim$ 0.5 km~s$^{-1}$.

Very few indeed of the absorption features in our spectra remain unidentified and of 
these most are very weak and some may be spurious. No dense forest of weak metal 
lines is seen even in the spectra having the highest S/N in our sample. To allow ready 
comparison among the QSOs, in Figure 4 we show ``spike diagrams'' displaying the 
column densities of all detected components identified in Tables 2--10; note that the 
vertical scales for \SiIV, \SiII\ and \NV\ are lower by 1 dex than the others. The 
missing indeterminate values indicated in the tables are relatively few and have only 
a minor effect on the appearance of the spike diagrams. The coverage in redshift 
outside the forest for the main species of our sample occurs between the two dotted 
vertical lines shown in each frame. For \SiII\ this applies for $\lambda$1260 only 
but we also show values where usefully obtained from strong $\lambda$1527 and these,
of course, appear at redshifts which would be in the forest for $\lambda$1260; such 
cases are clarified in the footnotes to Tables 2--10. For all species except \CIV\ 
the relatively few values reliably obtained from metal lines in the Lyman forest also 
are shown in the diagrams.

\subsection{Broad Absorption Features}

A substantial number of systems contain one or more broad (up to a few 
$\times 10$ km~s$^{-1}$), generally high ionization, components self-consistently
present in both members of the \CIV\ doublet, that overlap in velocity space with 
many of the more numerous narrower components. Often a broad feature protrudes (in 
velocity) at the edge of a system boundary and can be seen directly as having a 
structureless appearance. In some cases the presence of a significant broad feature 
immersed among the body of narrow components in a system is required to ``depress'' 
the profile of one or more of the observed species to account for differences in the 
characteristics of the corresponding profiles. While we could contrive to construct 
the broad features from a solid blend of numerous narrow components we do not believe 
that this recourse is justified by the data, and good fits using single broad 
components were obtained in general. Nevertheless, it must be remembered that the 
implicit model envisaged in the VPFIT profile constructions characterizes each assumed 
cloud in a system only by temperature and Gaussian turbulence broadening, while 
significant velocity gradients could also feature in the true overall absorption 
profile. The broad, high ionization components thus might represent spatially 
associated but physically distinct regions of low volume density dominated by bulk 
motions \citep{rau96}. Limitations to the detection of broad components are discussed
in \S6.2.

A simple example with partially exposed high ionization broad components is the 
${\it z} = 2.056$ system in Q1626$+$6433, for which the constituents of the 
\CIV~$\lambda$1548 profile are shown in the upper panels of Figure 5. Here, component 
1, with {\it b} $\sim$ 21 km~s$^{-1}$, is well enough separated from the others for 
its smooth outline to be clearly seen and component 4, with {\it b} 
$\sim$ 55 km~s$^{-1}$, although more immersed (and partially overlapping with 
component 1) reveals an extended shallow wing. The two panels on the right separately 
show combinations of the broad and narrow components. 

In the lower panels of the same figure details of the ${\it z} = 2.291$ system in 
Q1626$+$6433 are shown as a more complex as well as more comprehensive example, with 
broad features which are more immersed in the system. We go first to the panels on the 
left. The narrower components, {\it b}(\CIV) $\lesssim$ 10 km~s$^{-1}$ (see Table 2), 
here mostly have quite low ionization, with \CII\ relatively strong; components 11 and 
17 are the exceptions. Of the broader components, 2 and 4 unusually also have 
significant strength in \CII, while the remainder have relatively high ionization with 
\CIV\ strong and \CII\ very weak or undetected. 

The associated three sets of panels on the right of Figure 5 highlight different 
subsets of these components. While the contributions of the combined high ionization 
components dominate the overall profile of \CIV\ (middle set), it is particularly 
striking that the embedded combined subset of narrow, lower ionization components 
(top set) closely resembles the \SiIV\ overall profile and indeed the \CII\ profile. 
As before, this emphasises the successful segregation of differently ionized regions. 
The broadest high ionization component (14, with {\it b} $\sim$ 32 km~s$^{-1}$) is the 
strongest by far in \CIV\ and is the only significant component identified in the weak 
\NV\ profile shown in Figure 3.

\section{DEFINITION OF SAMPLES}

From the data in Tables 2--10 we define two samples {\it sa} and {\it sb}. We use 
sample {\it sa} for the statistical and clustering investigations in \S6 and \S 7 and 
{\it sb} for the ionization balance presentations in \S8. Sample {\it sa} is
statistically complete. Sample {\it sb} is used to probe the ionization state of the 
gas from individual ionic \emph{ratios} which, as will become clear in later sections, 
does not require a statistically homogeneous population of absorbers.

To avoid significant proximity ionization effects both samples are selected to have 
system velocities $\gtrsim$ 3000 km s$^{-1}$ from the nominal redshifts of the 
background QSOs \citep{par01}. We find no significant differences in our results or 
conclusions based on these samples if we extend the limit to 5000 km s$^{-1}$. 

It would be useful to determine individual \HI\ column densities for each of the
components we detect in \CIV\ in order to set thresholds for $N$(\HI) in our samples. 
Although this is possible in some cases it cannot be done reliably in general. The 
large thermal broadening experienced by \HI\ relative to the metal species studied 
here causes severe confusion among adjacent components in a great many of the systems; 
this is compounded by the strongly saturated nature of most of the Lyman~$\alpha$ 
system profiles, as can be seen from the examples in Figure 3. Higher members of the 
Lyman series are not uniformly available in the redshift range of our data. 
Consequently we did not attempt to set formal $N$(\HI) thresholds, except in the 
particular cases of components showing discernible Lyman~$\alpha$ damping wings as 
described in the sample definitions below. However, we found from many VPFIT trials 
using selected higher redshift systems having several available Lyman series members 
(for which we assume the \HI\ profiles contain the component population identified 
in the metals) that the absorbers in our chosen samples generally are optically thin 
in the Lyman continuum. Later we show through Cloudy modelling how the optical 
thickness, as well as increasing metallicity, influences the derived quantities.

The samples are further defined as follows:

sample {\it sa}---contains all \CIV, \SiIV, \CII, \SiII\ and \NV\ lines which fall 
outside the Lyman forest, while limiting \SiII\ only to the strong transition 
$\lambda$1260; includes the apparent  partial Lyman limit system at $z = 3.381$ in 
Q1422$+$2309C;\footnote{For this relatively simple system there is no evidence in the 
line ratios for the presence of a low ionization region (see footnote to Table 7).} 
excludes all components in a system showing significant Lyman~$\alpha$ damping wings 
for any of its components; \footnote{The high \HI\ column density may have been a
reason for obtaining the spectra for these systems, which then do not represent a 
statistically homogeneous contribution to our data set.}

sample {\it sb}---includes all components in sample {\it sa} with the exception of the 
few having ionic members which are saturated (shown in square brackets in the tables); 
adds lines of \SiII~$\lambda\lambda$1304,1527 if outside the forest (but upper limits 
are obtained only from $\lambda$1260 when it is accessible outside the forest); adds 
strong, unambiguous lines in relatively clear regions of the forest for any species 
except \CIV; includes some components from systems showing relatively mild damping 
wings in Lyman~$\alpha$ ($z = 2.761$ in Q1107$+$4847, $z = 2.904$ in Q0636$+$6801 and 
$z = 2.770$ in Q1425$+$6039: see the footnotes to the tables) but, in order to limit 
self-shielding effects, uses only components well-separated in velocity from those 
having high $N$(\HI).

Sample {\it sa} contains 867 \CIV\ components in 185 systems and sample {\it sb}, 
908 \CIV\ components in 188 systems. When using these samples, in most cases we also 
set appropriate column density thresholds on \CIV\ and sometimes additionally on other
species (specified in \S\S6--8) to avoid bias due to variations in S/N across the 
redshift range.

\section{EVOLVING STATISTICAL PROPERTIES} 

\subsection{Redshift Sampling of C~{\scriptsize IV}, Si~{\scriptsize IV}, 
C~{\scriptsize II}, Si~{\scriptsize II} and N~{\scriptsize V}}

The distribution in redshift of the \CIV\ component column densities, $N$(\CIV), 
for sample {\it sa} is shown in the top panel of Figure 6. For each system 
the values appear as vertical distributions of data points. To avoid confusion in the 
crowded figures, errors, listed in Tables 2--10, are not indicated in this and 
subsequent similar figures; mostly these are relatively small. The histogram at the
bottom of each panel
shows the number of sightlines covered at each bin in redshift from the nine 
sightlines of this sample. The second and third panels compare component subsets from 
the simpler and the more complex systems contained in the \CIV\ sample, arbitrarily 
taken as systems respectively having number $n \leqslant 6$ and $n \geqslant 7$ 
identified components. The column densities \emph{summed} over each system are 
displayed in the bottom panel. 

The panels in Figure 7 give displays for the \SiIV, \CII, \SiII\ ($\lambda$1260) and 
\NV\ component column densities, in this case corresponding to the top panel of 
Figure 6 only (note the vertical scale shifts for \SiIV, \SiII\ and \NV\ 
relative to \CIV). The coverage in redshift outside the Lyman forest as indicated in 
the histograms is different for each ion (see Figure 4). Within the permitted redshift 
ranges all system components detected in \CIV\ are represented in the other species 
either as determined values or upper limits; in both cases \SiIV\ and \NV\ are accepted 
even if only one member of a doublet is outside the Lyman forest. Only \SiIV\ and \CII\ 
adequately cover the complete redshift range, while \SiII\ and \NV\ are rather poorly 
sampled. By comparing with \CIV\ in Figure 6 it is evident that proportionally fewer 
of the simpler systems contain detected \SiIV\ components, and fewer still \CII. \SiII\ 
mimics \CII\ quite closely within the redshift intervals in common. Despite the meagre 
coverage, it is interesting to note that at the lower redshifts \NV\ is detected 
in most of the few narrow windows available, suggesting that \emph{in this range the 
ion is quite prevalent}.

\subsection{C~{\scriptsize IV} Component Doppler Parameter--Column Density Relationship}

The top panel in Figure 8 shows values presented in the {\it b}--log$N$ plane for all 
\CIV\ components in sample {\it sa}, extending over the range $1.6 < z < 4.4$. The 
large majority of components are well resolved. The lower bound in {\it b-}value seen 
in the diagram for these data comes both at about the minimum value resolvable and 
near the level of thermal broadening for gas at $\sim 10^{4}$ K. In the lower two 
panels the plot is given separately for $z < 3.1$ and $z > 3.1$ where the data divide 
into roughly equal numbers of components (and, it so happens, systems). We see that 
there is no marked difference between the lower and higher redshift plots except 
perhaps for a mild extension to higher column densities at the lower redshifts. 

We note that observational biases are present or potentially present in these data 
\citep{rau92}. Foremost, the unfilled triangular zone tending to the top left of the 
diagrams arises because at lower column densities the broader, and therefore shallower,
components become relatively more difficult to detect above the noise. Thus, the 
broadest components are detected only at the higher column densities. In addition,
there is a tendency for weak, broad components to be masked by more numerous, narrower 
components which, in general, overlap with them. Broad, shallow systems can also be 
confused with small residual undulations in the continuum level. Indeed, trial 
simulations indicated the magnitude of these effects is greater than expected simply 
from the limitation of S/N in the data.

A second possible bias is for the intrinsic narrowness of some lines to be hidden when 
approaching line saturation. This would affect the {\it b-}values for the narrower, 
higher column density components at the bottom right of the diagrams. However, trials 
showed this is not a significant effect within the range of column densities in our 
data set. 

A potential third effect can be expected from unseparated blends of closely 
overlapping components which would then appear as single components with larger 
{\it b-}values and higher column densities. While recognising that defining 
components as entities is an approximate process, and that we have aimed to 
introduce the minimum number of components to fit the spectral profiles, we believe 
that beyond our resolution limit such blending occurs relatively rarely in our high 
quality data.

There is some evidence for a small rise in the {\it b-}values at the lower boundary of
the plots with increasing $N$(\CIV), amounting to $\sim 1$ km s$^{-1}$ over a factor 
of more than 100 in column density, possibly reflecting the latter two bias effects at 
some level. This is somewhat less than the effect noted by \citet{rau96} using a 
smaller data set.

In following sections we keep in mind particularly the first of these bias effects. 
 
\subsection{C~{\scriptsize IV} Component Column Density and Doppler Parameter 
Distributions}

Figure 9 shows the distributions of $N$(\CIV) and {\it b}(\CIV) for all components of 
sample {\it sa}. Again we compare values for the ranges $1.6 < z < 3.1$ and 
$3.1 < z < 4.4$ (having means $\langle z \rangle = 2.51$ and 3.58) and also separately 
show plots for the {\it simple} and {\it complex} systems as  defined in \S6.1. 

For $N$(\CIV) there is a mildly significant difference between the component 
distributions of the {\it simple} and {\it complex} systems, with the latter peaking 
at a value about three times higher than the former. In contrast, the distributions of 
{\it b-}values are quite similar. It is clear, however, that the apparent shapes of 
these distributions are strongly influenced by the incomplete sampling in the 
``exclusion'' zone explained in the last section, which severely distorts the true 
shapes of the distributions. This effect is manifested in $N$(\CIV) by the sharp fall 
towards smaller column densities beyond the peak at log $N$(\CIV) $\sim 12.5$. In 
{\it b}(\CIV) the presence of a peak and the fall to small values beyond it are 
probably real, while the sampling deficiency brings about a too-rapid fall in the 
tail of the distribution extending to larger {\it b-}values.

Confining attention to the well sampled regions, here we see more quantitatively than 
in Figure 8 that there is little significant bulk change for either {\it b}(\CIV) or 
$N$(\CIV) from higher to lower redshifts. In particular, there is no discernible change 
in the behaviour of the components of simpler and more complex systems with redshift. 

\subsection{C~{\scriptsize IV} System Column Density and Velocity Spread Distributions} 

In Figure 10 we give histograms similar to those in Figure 9 but now for the total
system \CIV\ column density, $N_{syst}$(\CIV), and the total spread in velocity between 
system components, $\Delta v_{syst}$(\CIV). Because of the way they are defined both 
parameters show a marked difference between the distributions for {\it simple} and 
{\it complex} systems. 

The effect of incomplete sampling of low column density components is more complicated 
in these cases than for the \emph{component} distributions just described. For 
\emph{systems} the sampling deficiency is diluted because the weak components are not
in general concentrated near low $N_{syst}$(\CIV) but are spread among the systems, 
most of which contain several stronger components. Only the single-component systems 
directly mirror the sampling effect; the presence or absence of weak components 
embedded in systems with relatively high aggregate column densities cannot 
significantly influence the detection of these systems. It is possible that a
proportion of the fall to small values of total column density is real. Similarly, for 
$\Delta v_{syst}$(\CIV) there would be only a mild effect due to incomplete sampling, 
dependent on how the broad and narrow components are distributed within a 
multi-component system.

There is no large change with redshift in the distributions for $N_{syst}$(\CIV) but 
there is a hint that in the higher redshift data set there is a bias towards 
relatively smaller values. For $\Delta v_{syst}$(\CIV) there is a clearer change, with 
\emph{systems at the higher redshifts covering a smaller velocity range}.

\subsection{Relations between C~{\scriptsize IV} System Column Density, Velocity 
Spread and Number of Components}

Some significant relationships between $N_{syst}$(\CIV), $\Delta v_{syst}$(\CIV), and 
the number of components detected in a system, $n_{syst}$(\CIV), are shown in Figure 
11, again comparing values for $1.6 < z < 3.1$ and $3.1 < z < 4.4$. $N_{syst}$(\CIV) 
is strongly dependent both on $n_{syst}$(\CIV) and $\Delta v_{syst}$(\CIV) and, in 
turn, there is a strong proportionality between the latter two. Note that in these 
logarithmic plots, systems with only one detected component ({\it i.e.} nominally of 
zero velocity extent) are excluded. \citet{pab94} found similar relationships 
involving total \emph{equivalent width} rather than column density. We find no marked 
systematic changes with redshift in any of the relationships.

\subsection{C~{\scriptsize IV} Component and System Differential Column Density 
Distributions} 

Figure 12 gives the \CIV\ differential column density distribution function 
$f(N,X)$ for both the individual components of sample $sa$ and for these summed as 
systems, each in the same two redshift subsets as before. The data are summed over 
the bin size $10^{0.3}N$ as shown and the vertical error bars are $\pm 1\sigma$ 
values derived from the number of absorbers in each bin. The function $f(N,X)$ is 
defined as the number of absorbers per unit column density per unit absorption 
pathlength, $d^2{\cal N}/dNdX$ (this accounts for the multiple redshift coverage from 
the different sightlines), where for a given redshift interval $dz$, the $\Lambda$CDM
cosmology-corrected absorption pathlength interval $dX$ is given by

\begin{equation} 
dX = \frac{(1 + z)}{\sqrt{\Omega_{\rm M}(1 + z) + \Omega_{\Lambda}/(1 + z)^{2}}}\phd dz 
\phantom{.........} (\Omega = 1)
\end{equation}

\noindent where $\Omega = \Omega_{\rm M} + \Omega_{\Lambda}$ and we use 
$\Omega_{\rm M} = 0.3$, $\Omega_{\Lambda} = 0.7$. 

First, for the \emph{components}, there is remarkably little difference between the 
two redshift subsets. The incompleteness at low column densities seen in Figure 8 
causes the turnover below $N \sim 10^{12.5}$ but, otherwise, the observed distribution 
can be approximated as a power-law $f(N,X) \propto N^{-\beta}$ up to 
$N \sim 10^{13.5}$ when the observations again fall away because very few components 
have higher column density. There is a hint that this fall-off is greater at high 
redshift. The power law slope is closely similar for both redshift subsets; we obtain 
$\beta = 1.84$ for these combined and show this in the figure. For a lower resolution 
sample with $\langle z \rangle = 2.65$ \citet{pab94} obtain $\beta = 1.64$.

For the \emph{systems}, there is similarly little difference between the two redshift 
subsets. The drop below $N \sim 10^{12.5}$, evidently proportionally greater 
than for the components, again contains the incomplete sampling effect explained
in \S6.4. A power-law fit over the range 
$N = 10^{13.0}$--$10^{14.3}$ with $\beta = 1.6$, as shown, is a good representation. 
With sparser data \citet{ell00} obtained $\beta = 1.44$ for systems near $z= 3.2$ while 
\citet{son01} from more extensive data found $\beta = 1.8$ for her sample over the 
range $1.5 \lesssim z \lesssim 4.5$. We notice Songaila's results extend to somewhat 
higher column densities than we find in our system data set. Including the seven 
complex systems with mildly damped Lyman $\alpha$ profiles (and separated from the 
emission redshift by $\gtrsim$ 3000 km s$^{-1}$) makes very little difference to our 
results, either in the system or component cases.

\subsection{Redshift Evolution of System Ionic Number Densities and Population Total 
Column Densities}

In Figure 13 we display redshift evolution plots for all the observed species in 
the systems of sample {\it sa}. As a baseline we select systems having total 
column density $N_{syst}$(\CIV) $> 1 \times 10^{12}$~cm$^{-2}$, yielding 179 systems, 
and additionally apply the individual thresholds indicated in the figure for \SiIV, 
\CII, \SiII\ and \NV. Imposing these thresholds gives close to homogeneous sampling 
for each ion over the observed redshift range.

Looking first at \CIV, the six panels give the total number of systems per unit 
redshift interval $d{\cal N}_{syst}/dz$ and the aggregated column density per unit 
redshift interval $dN_{tot}/dz$ as a function of redshift for the sets of {\it all}, 
{\it simple} and {\it complex} systems in sample {\it sa} as defined previously. The 
data are corrected for multiple redshift coverage and summed over the arbitrarily 
adopted bins indicated by the horizontal bars. The indicated $\pm1\sigma$ uncertainties
for $dN_{tot}/dz$ are based on the number of systems present in each bin weighted by 
system total column density, and are dominated by the few systems with highest column 
densities. For $d{\cal N}_{syst}/dz$ the number alone of systems detected above the 
column density threshold defines the indicated uncertainties. The resultant 
distributions in redshift are remarkably constant, although some mild trends are 
apparent. 

To examine the evolution of the number density of systems with redshift, ${\cal N}(z)$ 
($\equiv d{\cal N}_{syst}/dz$), in the comoving volume we use \citep{mis02}

\begin{equation} 
{\cal N}(z) = {\cal N}_0\frac{(1 + z)^{2+\epsilon}}{\sqrt{\Omega_{\rm M}(1 + z)^3 + 
\Omega_{\Lambda}}} 
\phantom{.........} (\Omega = 1),
\end{equation}

\noindent where ${\cal N}_0$ is the local value of ${\cal N}(z)$ and we take 
$\Omega_{M} = 0.3$, $\Omega_{\Lambda} = 0.7$ as before. If the absorbers have constant 
comoving volume density and constant proper size then $\epsilon = 0$. The same form 
can be used for the comoving total column density of the absorber population by 
substituting $N(z)$ ($\equiv dN_{tot}/dz$) and $N_0$, and we put $\kappa$ in place of 
$\epsilon$ to identify the evolution in this case. 

In the left panels ($d{\cal N}_{syst}/dz$) we show in dotted lines fits to the three 
data sets for unevolving populations. In dashed lines we give actual fits to the data, 
with $\epsilon = 0.35$ ({\it all}), $\epsilon = 0.5$ ({\it simple}) and 
$\epsilon = -0.25$ ({\it complex}). The latter fits give only a small improvement over 
the unevolving cases and within the uncertainties the data are consistent with 
\emph{no evolution in number density}. \citet{mis02} using their combined sample EM15 
including data from \citet{sbs88} and \citet{ste90} obtain $\epsilon = -1.18$, quite 
strongly evolving in the sense of increasing number density with cosmic time as the two 
earlier studies had found (largely with the same data). However the two samples are 
very different. Ours is a very sensitive survey of relatively few QSOs and contains a 
large number of weak systems; the other surveys have much lower resolution and yield 
only strong systems, with rest-frame equivalent width $W_0 > 0.15$ \AA\ (implying 
$N_{syst}$(\CIV) $\gtrsim 5 \times 10^{13}$~cm$^{-2}$), from nearly an order of 
magnitude more QSOs yet with fewer (136) redshifts in the combined sample. Systems in 
our sample with column density greater than this high threshold show, within rather 
large errors, \CIV\ evolutionary behaviour consistent with the earlier studies, 
although we agree with Misawa et al. that the number density is about twice that found 
by \citet{ste90}.

In the right panels ($dN_{tot}/dz$) we show dotted line fits to unevolving populations 
as before. This describes the {\it simple} subset well but there appears to be a 
departure for {\it complex} systems which show an excess below $z = 3$, reflected also 
in the full sample. The dashed line actual fits to the data have $\kappa = -1.2$
({\it all}), $\kappa = -0.56$ ({\it simple}) and $\kappa = -1.4$ ({\it complex}). 

For the {\it complex} systems the derived positive evolution in total \CIV\ column 
density with cosmic time yet with constant number density indicates an increasing mean 
column density per system. In contrast, the {\it simple} systems (these make up the 
bulk of the systems for \CIV) are consistent with being a \emph{fully unevolving} 
population. A difference in evolution between weaker and stronger systems was noticed 
by \citet{ste90} in his high column density sample. 

We note that changes in the threshold column density produce little, if any, 
significant effect on the total column density values because the stronger systems 
dominate in the totals. For the subset of {\it complex} systems changes in the 
threshold also have little effect on number density because they are detected well 
above the threshold imposed for the full sample. For the {\it simple} subset the 
column densities range down to low values so the number density here is sensitive to 
the adopted threshold; nevertheless, doubling the threshold to 
$N_{syst}$(\CIV) $> 2 \times 10^{12}$~cm$^{-2}$, for example, makes relatively little 
change to the total number of systems detected in the full sample (164 from 179) and 
does not change our conclusions. 

Like \CIV, \SiIV\ shows little evidence for evolution in number density (dotted 
line) for the full sample. There is a hint that this is the result of countervailing 
trends in the {\it simple} and {\it complex} subsets (for all species these are defined 
by the parent number of \CIV\ components) but the data are also consistent with 
unevolving contributions from these subsets. In total column density \SiIV\ appears to 
evolve in similar fashion to \CIV. However, in the {\it complex} subset the unusually 
strong (for \SiIV) system at $z = 1.927$ in Q1626$+$6433 (Figure 3) has a large effect 
on the observed trend. To illustrate this the lowest redshift bin is plotted in two 
forms, one containing the $z = 1.927$ system and one beginning at $z = 2.0$ which 
excludes it. The resulting, unevolving, trend in the latter case may be more typical 
of the redshift evolution of \SiIV\ total column density. This alternative binning is 
also shown in the full data set, with the same conclusion. (In \CIV, however, we find 
the observed rise at low redshift is not significantly influenced by the $z = 1.927$
system.)

Changes with redshift appear greater in \CII, although here the data are rather 
limited and the errors are large. {\it Complex} systems now strongly dominate the 
population (note the large scale change in $dN_{tot}/dz$ for the subset of {\it simple} 
systems); consequently, these and the full sample follow closely similar behaviour. 
They indicate strong positive evolution with cosmic time both in number density and 
total column density, although not with high significance. (In \CII\ the strong 
$z = 1.927$ system is in the Lyman forest, so not included.) 

The data are even more limited for \SiII\ and accidental sampling of strong systems 
(see Figure~7) enhances the lower bin in $z$. It is interesting that \NV, sampled 
even more poorly in redshift, indicates a steep rise in total column density per mean 
system with cosmic time. This trend implies increasing levels of ionization and seems 
to conflict with \CII\ which indicates the reverse. 

The lack of evolution in \CIV\ total column density in terms of the {\it simple} 
systems and the mild evolution of the {\it complex} systems is surprising because it 
depends not only on the evolution of overall abundance of carbon, but also on the shape 
and normalization of the ionizing radiation background and the density of the absorbing 
regions. Over the extensive redshift range observed, large changes would be expected 
{\it a priori} in response to the evolving baryon density and character of the ionizing 
sources and the development of structure, even for constant metallicity. \SiIV, insofar 
as the data allow, presents a broadly similar picture to the behaviour of \CIV. 
Nevertheless, \CII\ and \NV\ do indicate evolutionary changes, although these appear
to be contradictory. However, we need to be cautious in interpreting all the 
observations in Figure 13 in this way because we are not here identifying the detailed 
ionization balance in isolated absorbing regions but taking a more global view of the 
individual properties of the ionic species. We return to this issue in our analysis of 
ionization conditions in the system \emph{components} in \S8.

\subsection{Evolution of C~{\scriptsize IV} Mass Density}

The comoving \CIV\ mass density is given by

\begin{equation} 
\Omega_{\scriptsize\textrm{\CIV}}(z) = 
\frac{H_0 \textrm{\phantom{.}}m_{\scriptsize\textrm{\CIV}}}
{c \textrm{\phantom{.}}\rho_{crit}}
\textrm{\phantom{.}} \frac{\sum N_{tot}(\textrm{\CIV},z)}{\Delta X(z)},
\end{equation}

\noindent where $\rho_{crit} = 1.89 \times 10^{-29}h^2$ g cm$^{-3}$ is the cosmological
closure density, $m_{\scriptsize\textrm{\CIV}}$ is the mass of the ion and 
$H_0 = 100h$ km s$^{-1}$ Mpc$^{-1}$. We calculate $\Delta X(z)$ from equation (1) and 
$\sum N_{tot}(\textrm{\CIV},z)$ from the data in Figure 13. The four bins we show for 
\CIV\ in the figure have mean redshifts $\langle z \rangle = 2.1$, 2.7, 3.2, 4.0. 
For the full sample (top panel) and using $h = 0.71$ \citep{spe03} we obtain respective 
values $\Omega_{\scriptsize\textrm{\CIV}} = (5.7\pm2.3$, $3.6\pm1.3$, $3.2\pm0.8$, 
$2.9\pm1.0)\times10^{-8}$ (1$\sigma$ uncertainties). These values are 
formally consistent with being invariant over the range $1.6 < z <4.4$ but also 
suggest a mild evolutionary increase in $\Omega_{\scriptsize\textrm{\CIV}}$ with 
decreasing redshift, in contrast to system number density. Songaila (2001, 2002) also 
found a more or less constant trend in the same redshift range; she assumed a 
$q_0 = 0.5$, $\Lambda = 0$ cosmology and when this is adjusted to the cosmology used 
above our values of $\Omega_{\scriptsize\textrm{\CIV}}$ agree with Songaila's within 
the errors.

The mean of our values over the redshift range is 
$\langle \Omega_{\scriptsize\textrm{\CIV}} \rangle = (3.8\pm0.7)\times10^{-8}$ at 
$\langle z \rangle = 3.1$. If we compare the lowest redshift bin with the mean of the 
other three (over which the values are very similar) we get formally 
$\Omega_{\scriptsize\textrm{\CIV}} = (5.7\pm2.3)\times10^{-8}$ at 
$\langle z \rangle = 2.1$ and
$\langle \Omega_{\scriptsize\textrm{\CIV}} \rangle = (3.2\pm0.6)\times10^{-8}$ at 
$\langle z \rangle = 3.3$.

The subset of {\it simple} systems, showing no evidence for evolution over the full 
redshift range, yields
$\langle \Omega_{\scriptsize\textrm{\CIV}_{spl}} \rangle = (1.5\pm0.4)\times10^{-8}$ 
at $\langle z \rangle = 3.1$. For the {\it complex} systems over the same redshift 
range
$\langle \Omega_{\scriptsize\textrm{\CIV}_{cplx}} \rangle = (2.5\pm0.6)\times10^{-8}$; 
however, given the positive evolution in \CIV\ mass density, we again compare the 
lowest redshift bin with the mean of the other three and get formally
$\Omega_{\scriptsize\textrm{\CIV}_{cplx}} = (4.1\pm2.2)\times10^{-8}$ 
at $\langle z \rangle = 2.1$, and
$\langle \Omega_{\scriptsize\textrm{\CIV}_{cplx}} \rangle = (1.8\pm0.5)\times10^{-8}$ 
at $\langle z \rangle = 3.3$.

To estimate the total carbon mass density, $\Omega_{\rm C}$, we need to know the 
\CIV\ ionization fraction 
$f_{\scriptsize\textrm{\CIV}} = n_{\scriptsize\textrm{\CIV}}/n_{\rm C}$. In \S10 
we find that the metagalactic ionizing radiation field at $z \lesssim 2.7$ is dominated 
by QSOs, from which we determine that $f_{\scriptsize\textrm{\CIV}}$ 
peaks at 0.25 for optically thin, low metallicity absorbers (\citet{son01} takes 0.5 
for the ionization fraction). Using our value as an upper limit gives 
$\Omega_{\rm C} \geq (2.3\pm0.9)\times10^{-7}$ for the full sample in 
the $\langle z \rangle = 2.1$ bin. At higher redshifts we deduce in \S10 that the
absorbing clouds are ionized by a galaxy-dominated radiation field and in this case 
find $f_{\scriptsize\textrm{\CIV}}$ peaks at 0.38, giving 
$\Omega_{\rm C} \geq (7.7\pm2.7)\times10^{-8}$ for the full sample in 
the $\langle z \rangle = 4.0$ bin.

We now estimate the average metallicity of the gas in all \CIV\ systems, making the
assumption that our \CIV\ absorbers represent the cool phase of intergalactic matter 
which also produces strong \HI\ absorption (as distinct from the ``warm-hot'' phase, 
characterized by the strong presence of \OVI). We take the baryon density in the Lyman 
forest, $\Omega_{\rm Ly}$, relative to the total baryon density obtained from 
measurements of the cosmic microwave background, $\Omega_{\rm b} = 0.0224h^{-2}$ with 
$h = 0.71$ \citep{spe03}, as $\Omega_{\rm Ly}/\Omega_{\rm b} \geq 0.9$
\citep{rau97,wei97} and obtain (using the He mass fraction $Y = 0.238$ \citep{osw00})
$\Omega_{\rm C}/\Omega_{\rm H} \geq (7.5\pm3.0)\times10^{-6}$ at 
$\langle z \rangle = 2.1$ and $\geq (2.6\pm0.9)\times10^{-6}$ at 
$\langle z \rangle = 4.0$. Taking the solar value (C/H)$_{\odot} = 2.72 \times 10^{-4}$ 
by number from \citet{hol01} and \citet{ala02}, we obtain\footnote{In the usual 
fashion we express the logarithmic abundance of element X relative to element Y 
compared with the solar values as [X/Y] $=$ log(X/Y) $-$ log(X/Y)$_{\odot}$.} 
[C/H]$_{\langle z \rangle = 2.1} \geq -2.64^{+0.15}_{-0.22}$ and 
[C/H]$_{\langle z \rangle = 4.0} \geq -3.11^{+0.14}_{-0.19}$.

Thus our data suggest a distinct, although not highly significant, rise by a factor 
$\sim 3$ in the average metallicity of carbon with cosmic time over our observed 
redshift range $z$ = 4.4--1.6. As we show in the next section, most if not all of the 
metal systems we detect are closely associated with galaxies so we do not in general 
probe the metal content of the more widespread intergalactic medium. Consequently, our 
values derived by comparison with the total intergalactic hydrogen mass density are 
underestimates of the metal/hydrogen ratios of the average absorber in our \CIV\ 
sample. 

Pursuing this, most of our \CIV\ systems relate to Lyman~$\alpha$ absorbers with 
relatively high column density, as can be inferred from Figure 3. \citet{cas97} find 
from hydrodynamical simulations that the fraction of mass in Lyman~$\alpha$ clouds with 
$N$(HI) $\geq 3 \times 10^{14}$ cm$^{-2}$ is approximately 0.25 at $z = 2$ and 0.13 at 
$z = 4$. Using these estimates our values for [C/H] become $\gtrsim -2.0$ at 
$\langle z \rangle = 2.1$ and $\gtrsim -2.2$ at $\langle z \rangle = 4.0$, now 
suggesting a constant metallicity of carbon. The comparison with simulations is quite 
uncertain, however, because the physical state of the Lyman~$\alpha$ clouds in the 
vicinity of galaxies is not typical of the general intergalactic medium \citep{ade03}. 
Nevertheless, these results are not very different from the value [C/H] $\sim -2.5$ at 
$z \sim 3$ (with scatter of about 1 dex) obtained from more direct estimates (Songaila 
\& Cowie 1996; Rauch et al. 1997a).

\section{C~{\scriptsize IV} COMPONENT AND SYSTEM CLUSTERING}

Previous studies of the two-point correlation function have shown that \CIV\ components 
cluster strongly on velocity scales $\lesssim 200$ km s$^{-1}$, with significant 
clustering out to a few 100 km s$^{-1}$ more, and in some cases extending to larger 
scales of order 1000--10000 km s$^{-1}$ (Sargent et al. 1980; Young, Sargent, \& 
Boksenberg 1982; Sargent et al. 1988a; Steidel 1990; Petitjean \& Bergeron 1994; 
Rauch et al. 1996; Womble at al. 1996). The extension to larger velocities is not a 
general property and can be traced to a few unusually complex groups of systems 
\citep{sas87,hhw89,dai96,qvy96}. Clustering similar to that observed in \CIV\ has been 
measured in \MgII\ \citep{ssb88,pab90,sas92,cvc03}. It has been suggested that the 
clustering signal reflects galaxy clustering, clustering of clouds within the same 
galactic halo, or a combination of the two. The issue is complicated by the wide 
disparity in velocity resolution, data quality, and sample size among the different 
observations. The clustering seen in metals seems to contrast with that observed in 
\HI\ which in general shows no significant clustering signal \citep{sar80}, or 
comparatively weak clustering \citep{mcd00} at larger column densities \citep{cri97}. 
However, in \HI\ much of the small-scale structure is obscured by unresolved blending 
of overlapping velocity components due to the greater thermal broadening \citep{fso96} 
and the often saturated absorption profiles (see Figure 3; the \HI\ profiles shown 
there are selected by detection of associated \CIV\ absorption and generally are 
stronger than the majority of the forest lines). We shall return to this point in the 
discussion below.

\subsection{Two-Point Velocity Correlation Function}

In Figure 14 we give velocity two-point correlation functions (TPCF) for the \CIV\ 
absorbers. In the standard manner these are normalized to the expected number of pairs 
per bin computed for a set of random distributions in redshift space matching the 
individual wavelength ranges and number of \CIV\ absorbers observed for each of the
nine QSOs in the sample. The simulated randomly distributed absorbers were collated 
by pairwise velocity separation in 1000 realizations and compared with the 
distribution of velocity separations in the data to derive the correlation 
$\xi(\Delta v) = \{N_{\rm data}(\Delta v)/N_{\rm random}(\Delta v)\} - 1$, where 
$\Delta v$ is the velocity in km s$^{-1}$ of one cloud as measured by an observer 
in the rest frame of the other.

In the top panel of the figure we show the result for the 867 individual 
\emph{components} of sample {\it sa} (without a column density threshold limit) 
spanning the total range $1.6 < {\it z} < 4.4$ with velocity resolution 15 km s$^{-1}$ 
for $\Delta v \leq 370$ km s$^{-1}$ and 20 km s$^{-1}$ for 
$\Delta v \geq 370$. The data points indicate the middle of the bins, with the 
first bin excluded because its width is comparable to the width of the \CIV\ 
components. The $\pm1\sigma$ errors in the simulated random distribution are smaller 
than the data points shown. The mean redshift is $\langle z \rangle = 3.04$. Following 
the usual pattern, the clustering signal in our data is strong at small velocity 
separations while declining steeply with increasing separation. The velocity 
correlation length, defined as the pair separation for which $\xi(\Delta v_0) = 1$, is 
$v_0 = 230$~km~s$^{-1}$, with significant signal extending only to 
$\sim 300$~km~s$^{-1}$.

The second panel includes the same data and also adds individual results for the two 
subsets with $z < 3.1$ and $z > 3.1$ and respective means $\langle z \rangle = 2.51$ 
and 3.58. There is a clear difference between these subsets, with stronger correlation 
at the lower redshifts. We need to treat this result with some caution because the 
data sets belonging to the individual QSOs show substantial variance in the overall 
\emph{shape} of the TPCF. However, on recomputing the TPCF after removing the data 
from one or more of the QSOs in several different trial combinations we found that a 
substantial redshift variation in the sense shown is always present. 

The third panel gives the result when all components of the systems exhibiting 
significant Lyman~$\alpha$ absorption damping wings (see the footnotes to Tables 2--10
for approximate \HI\ column densities) are added to sample {\it sa}, which we then 
call sample {\it ds$+$sa}. As before, we include only systems separated from the 
emission redshift by $\gtrsim$ 3000 km s$^{-1}$. This larger sample of 1020 components 
may be statistically less homogeneous than the original (see below). The profile is 
somewhat lumpier than for sample {\it sa}, now with $v_0 = 330$ km s$^{-1}$ and 
significant signal extending to $\sim 400$ km s$^{-1}$. The increase in $\Delta v$ 
relative to sample {\it sa} is consistent with the generally large velocity widths and 
richness of the added highly complex systems (see Tables 2--10). 

Structure similar to that for sample {\it ds$+$sa} has been found for \CIV\ mixed 
samples at intermediate and high redshifts in earlier studies (Petitjean \& 
Bergeron 1994; Songaila \& Cowie 1996; Womble at al. 1996; Rauch et al. 1996). 
Petitjean \& Bergeron fitted the shape of the TPCF by using two Gaussians, obtaining a 
best fit with velocity dispersions $\sigma = 109$ and 525 km s$^{-1}$ for their full 
sample and $\sigma = 95$ and 450 km s$^{-1}$ for a selected subset. With a higher 
resolution sample Rauch et al. found need for a three-component Gaussian fit with 
$\sigma = 22$, 136 and 300 km s$^{-1}$. Following the same procedure for our samples 
{\it sa} and {\it ds$+$sa}, we parameterized the TPCF as a multi-component Gaussian

\begin{equation} 
\xi(\Delta v) = A_{1}\phd exp \left(- \frac{\Delta v^{2}}{2\sigma_{1}^{2}}\right)
               + A_{2}\phd exp\left(- \frac{\Delta v^{2}}{2\sigma_{2}^{2}}\right)
                 \ldots ,
\end{equation}

\noindent where $A_{n}$ is the amplitude of the $n^{th}$ component; the results are 
shown in Figure 14. First, for {\it sa}, we achieve a very good two-component fit 
with $\sigma_{1} = 47.5$ km s$^{-1}$, $A_{1} = 8.2$, $\sigma_{2} = 112$ km s$^{-1}$, 
$A_{2} = 10.2$. A one-component fit is ruled out and more than two components is 
unnecessary. For the separate contributions in the two redshift ranges $z < 3.1$
and $z > 3.1$ we obtain fits each with a narrow component identical to that in the 
full sample and one broader one with $\sigma_{2} = 100$ km s$^{-1}$, $A_{2} = 14.5$ 
and $\sigma_{2} = 124$ km s$^{-1}$, $A_{2} = 6.7$, respectively. For {\it ds$+$sa} we 
obtain a good two-component fit with $\sigma_{1} = 47.5$ km s$^{-1}$, $A_{1} = 10.3$, 
$\sigma_{2} = 150$ km s$^{-1}$, $A_{2} = 10.5$; here the narrow component has the same 
width as before but a slightly larger amplitude. A three-component fit, shown in 
the figure, does slightly better overall, with a narrow component again identical to 
that in {\it sa} and others with $\sigma_{2} = 85$ km s$^{-1}$, $A_{2} = 5.0$, 
$\sigma_{3} = 170$ km s$^{-1}$, $A_{3} = 7.5$, although there is a slight excess in 
the tail of the function which is better matched by the two-component fit.

The bottom set of panels in the figure gives results for the data treated as 
{\it systems}; here the $\pm1\sigma$ errors in the random distributions (significantly
larger than for the components because of the smaller numbers) are shown by bounding 
thin lines. In the first panel, the result obtained using the 192 system redshifts for 
sample {\it ds$+$sa} with velocity resolution 500 km s$^{-1}$ and extending to 
$\Delta v = 12000$ km s$^{-1}$ is shown on an expanded vertical scale. The TPCF is 
notably flat over the whole range of separations, demonstrating clearly that the 
\emph{systems} of \CIV\ absorption lines are randomly distributed. Using the purer 
sample {\it sa} gives an almost identical result because the seven complex, mildly 
damped systems in sample {\it ds$+$sa} contribute to the total with equal weight to 
the rest of the systems (we have given the {\it ds$+$sa} case to demonstrate that this 
result is not specific to sample {\it sa}). We find the same pattern for velocity 
resolutions below 500 km s$^{-1}$ as well as up to several 1000 km s$^{-1}$ and 
$\Delta v$ beyond 50000 km s$^{-1}$; a large-scale example is in the second panel of 
the set. 

Typically a system spans a few 100 km s$^{-1}$. Almost all the systems extend over less 
than 300 km s$^{-1}$ and most extend less than 150 km s$^{-1}$ (see Figure 10). 
Searching more finely on the smaller scale, in the third panel of the bottom set we 
show the TPCF for a subset of the sample in which we include only the 146 systems 
having velocity spread $\Delta v_{syst}$(\CIV) $< 150$ km s$^{-1}$, and use velocity
resolution 150 km s$^{-1}$. Here again the result is indistinguishable from a random 
distribution. 

Summarising the above results, for \emph{components}, we observe, as others have 
already noted, that the detailed shape of the TPCF depends on the specific contents of 
the samples used, although broadly the results are similar. Our larger sample both 
gives better statistical definition than before and excludes QSO spectra containing 
known, highly untypical, complex groups of strong systems, such as for PKS 0237$-$233, 
which can dominate the correlation signal particularly at larger $\Delta v$ (Heisler 
et al. 1989). Furthermore, by also excluding the few unusual systems containing 
significant Lyman~$\alpha$ damping wings, we regard our sample {\it sa} as representing 
the ``normal'' great majority of \CIV\ absorption systems. With this sample of 
components we obtain a more compact TPCF than in most other work and find no evidence 
of clustering at large $\Delta v$. Specifically, the shape of our observed TPCF does 
not require Gaussian components as broad as those found by \citet {pab94} or 
\citet{rau96}. Churchill et al. (2003) came to a similar conclusion for \MgII\ 
absorbing clouds. Most importantly, we do not see evidence of \emph{system} clustering 
in our sample \emph{on any scale} from $\Delta v$ as small as 150 km s$^{-1}$, where 
the general \emph{components} clustering signal is still strong, out to very large 
separation values. We now use these results to investigate the origin of the 
correlation signal.

\subsection{Comparison with Galaxy Clustering}

It seems that for all our component samples in the top three panels of Figure 14 the 
TPCF has a common narrow core component, with broader components whose characteristics 
depend on the precise definition of the sample. While this gives a clue to the 
dynamical makeup of the absorber population we should bear in mind that this is a 
contrived description of the TPCF (we found an exponential fit is almost as good in all 
cases). 

For galaxies, the TPCF is a simple, fundamental statistic of the galaxy distribution. 
Estimations of {\it real-space} galaxy clustering yield a TPCF very close to power-law 
form over a broad range of scales \citep{lov95,zeh02,haw03}, although on detailed 
examination this is seen to be largely fortuitous \citep{zeh03}. The parameters of the 
power-law depend on the characteristics of the sample galaxies, dominantly on their 
luminosity \citep{nor02}. Galaxy clustering in {\it redshift-space}, on the other hand, 
is strongly affected by gravitationally induced distortions. On small scales, random 
peculiar velocities will cause clustering to be underestimated, while on large scales 
coherent infalling bulk flows will lead to an overestimate of the clustering amplitude 
\citep{kai87}. Consequently, the form of the redshift-space TPCF departs considerably 
from a simple power-law \citep{zeh02,haw03}. Large scale $\Lambda$CDM simulations show 
good correspondence with the observed galaxy clustering \citep{ben01,wei03}. In 
particular, these simulations predict remarkably little comoving clustering evolution 
from high redshifts to the present epoch (although the underlying dark matter 
clustering evolves strongly), agreeing well with a recent estimation of the spatial 
TPCF from a sample of Lyman-break galaxies at $z \sim 3$ \citep{ade03}.

Figure 15 reproduces the sample {\it sa} component TPCF in logarithmic form; we include 
the two-component form of Gaussian fit as in Figure 14. The stepped horizontal lines 
give the system TPCF $+1\sigma$ bin errors from the bottom set of panels of Figure 14 
taken as upper limits, in to 500 km s$^{-1}$ from the first panel then to 
150 km s$^{-1}$ from the third. We convert the velocity scale for our sample at 
$\langle z \rangle = 3.04$ to $h^{-1}$ comoving Mpc using $\Omega_{\Lambda} = 0.7$, 
$\Omega_{\rm M} = 0.3$ and Hubble constant $h$ in units of 100 km s$^{-1}$ Mpc$^{-1}$ 
and show this at the top of the figure for comparison with the galaxy estimations. The 
two differently-scaled short-long-dashed lines are results from the large 2dF Galaxy 
Redshift Survey \citep{haw03} for the galaxy real-space TPCF as fitted by the 
power-law $(r/r_0)^{-\gamma_r}$ with $r_0 = 5.05 h^{-1}$ Mpc, $\gamma_r = 1.67$ and 
the corresponding redshift-space TPCF as fitted peacemeal by the power-law 
$(s/s_0)^{-\gamma_s}$ with $s_0 = 13 h^{-1}$ Mpc, $\gamma_s = 0.75$ at small scales, 
and $s_0 = 6.82 h^{-1}$ Mpc, $\gamma_s = 1.57$ around $s_0$. The dotted line is the 
spatial TPCF for Lyman-break galaxies from \citet{ade03}; in redshift-space this 
should show distortion of similar character to the nearby samples. 

Figure 15 shows clearly that above $\Delta v \sim 150$ km s$^{-1}$ the 
(one-dimensional) TPCF for the \CIV\ components dips significantly below the TPCF 
found for galaxy clustering. This effect is more pronounced for the galaxy  
redshift-space TPCF which is the more appropriate for comparison. In addition, the lack 
of significant absorber \emph{system} clustering observed above 
$\Delta v = 150$ km s$^{-1}$ gives a very substantial deficit in clustering amplitude 
relative to galaxies out to at least $\Delta v \sim 1000$~km~s$^{-1}$. 

\subsection{Absorbers and Galaxies}

To complete the picture we now consider observations associating absorbers with 
galaxies. Attempts to establish directly how metal absorption systems and  galaxies 
are connected have focussed on searches for galaxies near the line of sight to QSOs 
with the same redshifts as the absorbers \citep{bas78,bab91,ste94,leb97,ste02}. Most 
work of this kind has been done at $z \lesssim 1$ for relatively strong systems 
selected by the presence of \MgII\ and associated with gas having 
$N$(\HI) $\gtrsim 10^{17}$ cm$^{-2}$ ({\it i.e.} Lyman limit systems). \citet{ste02} 
investigated the kinematical properties of several such cases with projected impact 
parameter $20 \lesssim d \lesssim 100$ kpc. The identified galaxies appear to be 
relatively normal spirals, with circular velocities 
$100 \leq v_c \leq 260$ km s$^{-1}$. While the absorber characteristics are consistent 
with rotation being dominant also for the absorbing gas, the total range of velocities, 
typically 200--300 km s$^{-1}$, and their placing to one side of the galaxy systemic 
redshift, is not consistent with simple disc rotation viewed along a single sightline. 
(The expectation that a sightline through a galaxy at large radius would show only a 
small differential rotational velocity, contrary to the observed velocity spread of
the absorption systems, earlier had counted against the idea that the velocity 
structure in the systems is due to motions associated with single galaxies (Sargent et 
al. 1988a).) Steidel et al. suggest that models to explain their observations 
require either extremely thick rotating gas layers, rotation velocities that vary with 
height above the extrapolated galactic plane, or a combination of both, with rotational 
motion dominating over radial infall or outflow even for gas well out of the galactic 
plane. At higher redshifts, the velocity structure of some of the strongest \MgII\ 
absorbers is suggestive of superwinds arising in actively star-forming galaxies 
\citep{bon01} and more generally there is mounting evidence for the importance of 
large-scale galactic winds \citep{pet01,ade03}. \CIV\ absorption is kinematically 
strongly correlated with \MgII\ and usually extends more widely in velocity 
\citep{chu01,cas02}.\footnote{The kinematics of the absorbing gas observed in 
strongly-damped Lyman~$\alpha$ systems (rare, metal systems for which 
$N$(\HI) $\geq 2 \times 10^{20}$ cm$^{-2}$) similarly appears consistent with rotating 
thick disc geometries \citep{paw98}, although not uniquely \citep{hsm98,mam99}, but 
there is difficulty reconciling the high ionization species with the low in the same 
model \citep{wap00}.} More substantially, from a large observational sample 
\citet{ade03} demonstrate a close association between \CIV\ absorption systems and 
Lyman-break galaxies. 
 
\subsection{Conclusions on the Identity of Metal Absorbers}

In the light of observations linking \MgII\ and \CIV\ absorption systems with specific 
galaxies close to the line of sight to the background QSO, and the fact that these
galaxies have extended kinematic structure of a few 100 km$^{-1}$ similar to the \CIV\ 
systems in our samples, it seems inescapable that our velocity correlation results for 
the absorption \emph{components} in sample {\it sa}, contrasting with the lack of 
\emph{system} clustering, are entirely due to the peculiar velocities of the gas 
present in the outer extensions of galaxies, \emph{not to general galaxy-galaxy 
clustering}. This conclusion is not changed when we substitute the component sample 
{\it ds+sa} with its bias to highly complex systems. The different broader components 
we find in the Gaussian fits to the shape of the TPCF in the various cases we described 
earlier may reflect the distribution of the more disturbed cases of outflow into the 
extended regions probed, while the ubiquitous narrow component may indicate underlying, 
more quiescent, disc-like motion.

The lack of clustering for the \CIV\ systems means \emph{there is no obvious 
observational distinction between the metal systems and strong \HI\ systems}. Because 
any complex component structure in \HI\ is largely hidden, the \HI\ absorption lines 
effectively are counted as systems like the \CIV\ systems. However, the \HI\ lines are 
detected in far greater number and probe to much lower densities in the intergalactic 
medium, so in general do not represent the same population as the observed metal 
systems which are more directly associated with galaxies.

The explanation for the lack of system clustering in our \CIV\ sample, while the 
systems are known to be correlated with galaxies, must simply be geometrical: the 
single sightline available to each background QSO, although highly extended in 
redshift, 
samples the smaller-scale galaxy population so sparsely that it is a rare occurrence 
for the gaseous extent of more than one galaxy to be probed in a given cluster or 
group. The situation for measurements of galaxy redshifts in clustering studies is 
quite different: here, all galaxies in the plane of the sky are included, with defined 
sample specifications, and the spatial variance is completely sampled 
three-dimensionally on all scales.  

It is intriguing that in the range $\Delta v \sim$ 30--150 km s$^{-1}$ the \CIV\ 
absorption component correlation function closely coincides with the galaxy 
redshift-space correlation function where this is heavily distorted by peculiar 
velocities (Figure 15). On our interpretation this correspondence must be fortuitous. 
Indeed, we find no clustering between \emph{systems} at $\Delta v = 150$ km s$^{-1}$
in the same sample.

We do not, of course, rule out the possibility of more extended correlated structures
appearing when a sightline fortuitously passes along a large-scale filamentary or 
sheetlike structure of galaxies in a supercluster, bringing about a rich complex of 
absorber systems well extended in redshift, but the incidence of such occurrences 
must be low (Sargent \& Steidel 1987; Dinshaw \& Impey 1996; Qhashnock et al. 1996). 
\citet{rau96} considered a less extreme version of such a model to explain the 
tail of the TPCF found in their smaller \CIV\ sample by Hubble flow velocity 
extension. On the other hand, \citet{ssr02} find that for the components of a limited 
sample (12) of \OVI-selected systems tracing the warm-hot intergalactic medium the 
velocity TPCF is similar to the spatial power-law form seen for galaxies, with a 
comoving correlation length $\sim 7h^{-1}$ Mpc. They conclude that for this population 
the signal is dominated by large-scale structure. \CIV\ is also present in most of the 
sample of \OVI\ absorption systems. We stress again, however, that in the present work 
we seek to characterize the ``normal'' situation obtaining for the large bulk of weak
metal absorption systems in the cooler phase identified by \CIV. In this respect our
work is very different from large-scale studies using sparse samples of strong \CIV\ 
systems. Quashnock et al. (1996) used 373 QSO sightlines having a total of 360 strong 
\CIV\ absorption systems covering the redshift range 1.2--4.5 (in contrast, our sample 
has an average of about 20 systems per sightline). With a velocity resolution of about 
600 km s$^{-1}$ they found weak clustering, $\xi \sim 0.4$, on scales of superclusters, 
with significant signal contributed by groups of absorbers in only 7 of the sightlines. 
\citet{qav98} extended this study to smaller scales using velocity resolution 
180 km s$^{-1}$ and a restricted sample of 260 strong \CIV\ systems drawn from 202 
sightlines, finding significant clustering of power-law form on scales of clusters and
superclusters. Our sample of numerous relatively weak \CIV\ systems in few sightlines 
clearly does not probe such large-scale structure. Nor, as we have seen, do our systems 
strongly sample galaxies within clusters. Our high velocity resolution data perform a 
complimentary role of finely probing the environment in the vicinity of \emph{single} 
galaxies, through the properties of the components of the systems. The smooth 
distribution in the TPCF (Figure 15) shown by the components of the full population of 
systems ranging from the weakest (very simple) to the strongest (very complex), and 
the lack of distinction in clustering between such systems, suggest that spatially all 
systems are similar objects.

Finally, we consider the apparent evolution in redshift shown in the second panel in 
Figure 14. Although, as we have found, the comoving number density of \CIV\ systems 
does not change with redshift, we shall demonstrate later in this paper that the 
radiation environment to which the absorbers are exposed changes markedly with redshift 
and is very different for the two redshift subsets used. At the lower redshifts the 
ionizing radiation from the background QSOs dominates, while at higher redshifts 
galaxies close to the absorbing clouds dominate the ionizing flux. The more diffuse 
metagalactic illumination of the absorbers on the one hand and the more locally 
confined exposure on the other, could influence the observable geometrical spread of 
the systems. We see in Figure 10, for example, that the distribution in system velocity 
spread changes with redshift. Although velocities do not simply translate to spatial 
distribution, the apparent evolutionary change in component clustering presumably 
reflects such differences.

\section{EVOLVING IONIZATION BALANCE} 

\subsection{Redshift Evolution of Ionic Ratios}

In Figure 16 we show the distribution in redshift of the column density ratio 
\SiIV/\CIV\ for absorbers in which \SiIV\ is accessible. As before, the obvious 
vertical associations are members of the same system. In all cases we take 
$N$(\CIV) $> 1 \times 10^{12}$~cm$^{-2}$ as an acceptance threshold.

The top left panel displays the component values for the full sample {\it sb}. The 
related panel on the right gives median values obtained over the redshift spans defined 
by the horizontal bars; the indicated uncertainties are $1\sigma$ bootstrap values. 
Unlike the single-species redshift distributions presented in \S6, here we do not need 
to correct for multiple redshift coverage because each ionic ratio value represents an 
independent measure of absorber ionization \emph{balance} (although incomplete, as we 
explain later). The upper limits shown in the left panel are included in the median 
values; the occasional high upper limit coming above the derived median does not 
significantly influence the accuracy of the results. It is striking that within the 
small uncertainties the median value of \SiIV/\CIV\ is consistent with being constant 
over the whole observed range $1.9 < z < 4.4$. 

In the second and third sets of panels we show the component data for the subsets of 
{\it simple} and {\it complex} systems defined as before. For the {\it simple} subset 
there is a significant continuous fall in the median with cosmic time, by about 0.4
dex over the observed range, while the {\it complex} subset, again, is consistent with 
being constant over the range. 

In the bottom panels we use sample {\it sb} to show total \emph{system} ionic ratios,
obtained after first separately summing the individual \SiIV\ and \CIV\ component 
values. Once again, there is no detectable change in the median over $1.9 < z < 4.4$. 
Although we have earlier explained the shortcomings of using total system values for 
ionic ratio determinations we give these in the figure for comparison with earlier
work by \citet{soc96} and \citet{son98}. These authors have argued that their observed 
evolution of \SiIV/\CIV, which shows a jump at $z \sim 3$ by at least 0.5 dex in the 
median, requires a sudden hardening of the ionizing background that is consistent 
with an abrupt reduction at that redshift in the opacity of the evolving intergalactic 
medium to \HeII\ ionizing photons as \HeII\ completely ionizes to \HeIII. We have 
earlier reported that we have not found such a jump (Boksenberg 1997; Boksenberg et
al. 2001), a result also found by \citet{kcd02}, and we confirm this with the more 
extensive investigations here. The reason for the difference between our findings and 
the previous work is not clear. 

Nevertheless, use simply of the ratio \SiIV/\CIV\ to measure the influence of \HeII\ 
ionizing radiation on the absorbers has significant limitations and we do not make any 
claims concerning the evolution of the ambient ionizing spectrum based on Figure 16. 
While the components in a given system might be expected to receive the same exposure 
to the metagalactic radiation, their extensive vertical distributions within systems 
show directly that \SiIV/\CIV\ cannot be a unique indicator of the ionizing flux. Most 
likely this is because of differences in density among the individual absorbers within 
a system. We also note that the totalled values plotted in the bottom panel have a 
scatter of more than an order of magnitude.

Several workers have found support more directly for high \HeII\ opacity in the
intergalactic medium at $z \gtrsim 3$ \citep{rei97,hea00,kri01,sme02}. \citet{str00} 
and \citet{the02} showed that \HeII\ reionization at $z \sim 3.5$ results in an
increase by a factor 2 in the temperature of the intergalactic medium at the mean 
density and found evidence for such a jump from observations of Lyman~$\alpha$ forest 
lines. \citet{ber03} have claimed a sudden depression in the Lyman forest effective 
optical depth at $z \sim 3.2$ by $\sim 10\%$ which is consistent with \HeII\ 
reionization. However, analyses by \citet{mcd01} and \citet{zht01} do not show a 
significant temperature change at these redshifts, although the temperatures they found 
are higher than expected for photoionized gas in ionization equilibrium with a cosmic 
background, as can be explained by gradual additional heating due to ongoing \HeII\ 
reionization.

Displays corresponding to the top panels in Figure 16 are given in Figure 17 for the 
column density ratios \CII/\CIV, \SiII/\SiIV, \SiII/\CII\ and \NV/\CIV, although for 
all but the first there is more uncertainty arising from upper limits. The median 
\CII/\CIV\ values show rather more variation with redshift than \SiIV/\CIV, with a 
progressive rise by $\sim 0.4$ dex over $z \sim$ 3.7--1.9. \SiII/\SiIV\ values show a 
yet steeper trend, rising by $\sim 0.7$ dex over this range, while the trend of 
\SiII/\CII\ appears quite flat, although in both cases the data are sparse. \NV/\CIV\ 
values indicate a rising trend from $z \sim 3.5$ to lower redshifts but there are too
few data within the few narrow windows for which \NV\ can be studied to indicate the 
behaviour at the higher redshifts.

Although some collective inferences might be made about the radiation environment from 
the data in Figures 16 and 17, the fact remains that individual ionic ratios give only 
partial evidence of ionization state. In following sections we use combinations of 
these ratios as more complete indicators with which to derive information on the 
evolution of the ionizing spectral energy distribution and the physical properties of 
the absorbers.  

\subsection{Redshift Evolution of Ionic Ratio Combinations}

In Figure 18 we give two-dimensional column density ratio displays of 
\SiIV/\CIV~:~\CII/\CIV\ derived for the individual components in sample {\it sb} for 
which we can either measure or obtain an upper limit for \SiIV\ and \CII, again taking 
$N$(\CIV) $> 1 \times 10^{12}$~cm$^{-2}$ as an acceptance threshold. In the left panels 
we compare our full data set for these ratios in three redshift ranges with model 
predictions of the Cloudy code (Ferland 1996; version 90.04) computed for one-sided 
illumination of a plane-parallel slab of low metallicity gas ($Z = 0.003$ $\times$ 
solar) optically thin in the \HI\ Lyman continuum 
($N$(\HI) $= 10^{15}$~cm$^{-2}$).\footnote{We use these as nominal values convenient 
for illustration and do not imply that they are specifically demanded by the 
observations. The effects of varying the properties of the absorbers are considered in 
\S8.4.} We use a recent version of the \citet{ham96} photoionizing background kindly 
provided to us (Haardt 1998; see \S10.1) to represent the evolving contribution 
of QSOs alone at redshifts $z_{HM} = 2.3$, 3.0, 3.9, appropriate for the redshift 
ranges of the data, and give the mean intensity at the \HI\ Lyman limit, $J_{\nu_0}$, 
for these cases in the caption to the figure. We introduce the cosmic microwave 
background at these redshifts to include Compton cooling of the 
absorbers.\footnote{Compton cooling becomes significant at low densities and high 
redshifts but the effect is relatively small in the range of our data (see \S10); the 
cooling timescale may be longer than the age of the Universe at the redshift of 
interest \citep{mei94}. The cosmic microwave background is not included in the Haardt 
\& Madau radiative transfer computations, which assume a fixed temperature in the 
ionization modelling \citep{ham96}.} In this modelling we include solar relative metal 
abundances as listed in Cloudy \citep{gaa89,gan93}, but for Si additionally use a 
higher value to conform with observations relating to early, low-metallicity 
environments. Galactic metal-poor stars show [Si/Fe] extending to $\sim$ 0.3--0.4 
fairly uniformly over the range $-3.5 \lesssim$ [Fe/H] $\lesssim -1$, in a common 
pattern with other $\alpha$-elements \citep{mcw95,rnb96} and is thought to reflect the 
yield from Type II supernovae. Depending on the interpretation of corrections for the 
presence of dust there is a similar excess in damped Lyman~$\alpha$ absorption systems 
\citep{lsb96,vla98,pet00}. On the other hand, C/Fe measurements in metal-poor stars 
generally give approximately solar values although with considerable scatter 
\citep{mcw95,mcw97,ilr00}. Based on these data we take [Si/C] $= 0.4$ as a 
representative upper level. The recent revisions of photospheric abundance values by 
\citet{hol01} and Allende Prieto et al. (2002) combine to increase the Si/C solar 
value by 0.19 dex while uncertainties in dielectronic recombination rate coefficients 
for \SiIV\ and \CIV\ \citep{sav00,sch03} could translate to a significant systematic 
lowering or raising of the inferred ratio as derived with Cloudy for the conditions in 
our photoionized absorbers. In view of the existing uncertainties we retain the nominal 
values in the version of the Cloudy code we have used.

Unlike the presentation in Figure 16, it is particularly striking that the data now
take on a coherent appearance. Values from components in a given system now string out 
along a track with relatively little scatter, accomodating the different ionization 
states present. For the lowest range of redshifts ($z =$ 1.9--2.65) the model 
achieves a good match to the data within the adopted Si/C bounds (interestingly, in 
view of the uncertainties mentioned in the previous paragraph), while towards higher 
redshifts there is first an upward scatter in the data ($z =$ 2.65--3.4) then a 
substantial overall rise, by a factor $\sim 3$ ($z =$ 3.4--4.4). We note that at the 
highest redshifts there are relatively more components having strong \CIV\ 
($N$(\CIV) $> 1 \times 10^{13}$~cm$^{-2}$) and detected \SiIV\ but only upper limits 
in \CII\ (the open symbols with left-pointing arrows out to the low \CII/\CIV, or
higher ionization, part of the diagram\footnote{The significance of \CII/\CIV\ as an 
indicator of the ionization state (or in a more restricted sense, the density) of the 
absorbers in the context of our observations is explained in \S8.4.}), as is already 
indicated from the data in Figures 16 and 17. Most of the components having both 
\SiIV\ and \CII\ with only upper limits have $N$(\CIV) $< 5 \times 10^{12}$~cm$^{-2}$. 
We illustrate separately the corresponding behaviour of the {\it simple} and 
{\it complex} subsets of the sample. These figures demonstrate that the components in 
the {\it simple} systems tend to populate the higher ionization regions of the diagrams 
while the {\it complex} systems have a wider range of ionization. Both subsets follow 
the same paths in the diagrams and the same general evolutionary behaviour with
redshift, again suggesting they represent similar objects.

For the undifferentiated sample {\it sb} we show in Figure 19 corresponding displays 
of two other combinations of the ionic ratios. For \SiII/\SiIV~:~\CII/\CIV\ a single 
model fit is applicable because there is no relative abundance dependence. Indeed, 
the data points tend to cluster more closely together in the lowest and highest 
redshift panels than they do in Figure 18. At the lowest redshifts the model fits the 
data well. In the highest redshift interval the data generally fall below the model, 
while in the middle range there is increased scatter with a trend in the same sense. 
Again there is distinct progressive evolution in redshift in the displayed quantities. 
For \SiII/\CII~:~\CII/\CIV\ the model prediction gives good fits to the data points, 
with no obvious evolution in redshift.

Finally, in Figure 20 we show displays of \NV/\CIV~:~\CII/\CIV. For the model curves 
we again use both the solar relative abundance of N/C (upper curve) and an upper 
bounding value appropriate for low metallicity absorbers, reflecting the early 
production of nitrogen. \citet {hek00} give data for C/O and N/O in \HII\ regions and 
stars from which we obtain [N/C] $\sim -0.2$ at low metallicities, used as the second 
value in the figure. However, it should be borne in mind that for both values there is 
considerable uncertainty in such data. Around the solar value (where metal-sensitive 
secondary production of N contributes significantly) there is scatter of several tenths 
dex in the observations, while at low metallicity (where primary N production dominates 
over secondary) additional departure to considerably lower values of N/C is expected 
as a consequence of the relatively long time delay for the release of primary N because 
of its origin in lower mass stars. Indications of the latter, with downward dispersion 
of N relative abundance by up to 1 dex, are seen in damped Lyman~$\alpha$ absorption 
systems \citep{plh95,lsb98,cen98,pil99,pet02,cen03}. We represent this downward 
dispersion by the shaded region in the figure. At the lowest redshifts, the match 
between the data and the QSO ionization model is not as good as in Figures 18 and 19. 
Nevertheless, while our data for \NV\ are more limited than for the other species and 
the relationship between the data and the model is not as clear, an evolutionary trend 
is again evident over the the whole observed redshift range. At the highest redshifts 
\NV/\CIV\ extends down to relatively low values indicated by upper limits only, as we 
have shown earlier.

It is clear from Figures 18--20 that two-dimensional analyses like those presented 
are necessary to make useful interpretations of the ionization properties of the
absorbers. We deal with this in more detail in \S10 and explore the effect of varying 
the spectrum of the photoionizing radiation on the various ionic ratios we have 
measured. 

\subsection{Photoionization Equilibrium or Collisional Ionization?}

From the data in Figures 18--20 we see that our assumption that the absorbers are 
\emph{photoionized} is well borne out by the general concordance of the observations 
with the shapes of the model curves, albeit with the additional evolutionary effects. 
It is particularly important to note that there is \emph{no significant distinction 
between the broader and narrower components} in the sample, which fit equally well. 
This result confirms that the broader components represent highly turbulent or bulk 
velocity-extended structures rather than regions at high temperature and dominantly 
thermally broadened.

We demonstrate this by modelling the additional effect of significant collisional 
heating on the ionization equilibrium. We use the lowest redshift case, $z_{HM} = 2.3$, 
for which the pure QSO form of metagalactic ionizing radiation background is dominant. 
In photoionization equilibrium, Cloudy modelling at this redshift gives mean 
temperatures $\lesssim 5 \times 10^4$ K in the \CII/\CIV\ range of our data; some 
specific examples are given later in Figure 23. In pure collisional ionization 
equilibrium the \CIV\ ionization fraction strongly peaks at $10^5$~K \citep{sad93} and 
there is a peak at the same temperature in a more complicated case of a nonequilibrium 
radiatively cooling gas \citep{sam76} as discussed by \citet{tri02}. Such a temperature 
corresponds to {\it b}~$\simeq 10$~km~s$^{-1}$ and is already ruled out for our 
narrower components (see the {\it b-}value distribution given in Figure 9). In the same 
circumstances \SiIV\ peaks at a slightly lower temperature: $0.8 \times 10^5$~K.

To identify the observable effects of collisional ionization, in Figure 21 we show in 
plots of \SiIV/\CIV~:~\CII/\CIV\ and \SiII/\SiIV~:~\CII/\CIV\ (corresponding to the 
top panels in Figures 18 and 19 but with extended axes) idealized models run in Cloudy 
at three fixed temperatures 0.8, 1.0 and $1.2 \times 10^{5}$~K in the presence of the 
same metagalactic ionizing radiation background at $z_{HM} = 2.3$ and for the same 
absorber parameters as defined in Figure 18. The pure photoionization equilibrium model 
as shown before is included for comparison. The assumed Si/C relative abundance values 
are adopted as in the previous figures. The collisional ionization curves terminate in 
the diagrams where the ionic ratios become independent of total hydrogen volume density
in the model (near $n$(H) $= 3 \times 10^{-2}$ cm$^{-3}$). In the absence of ionizing 
radiation the fixed temperature collisional models at all densities give single values 
at these termination points. 

At temperatures where collisional ionization would be dominant these models bear no 
overall resemblance to our observations. We can conclude that the majority of the 
absorbers in our samples are in photoionization equilibrium. 

\subsection{Effects of Changes in Absorber Parameters}

In Figure 22, again using \SiIV/\CIV~:~\CII/\CIV\ and \SiII/\SiIV~:~\CII/\CIV, we 
demonstrate with our $z_{HM} = 2.3$ photoionization equilibrium model as applied in 
the top panels of Figures 18 and 19, how these ratios are affected by changes in 
metallicity, [$Z$], and \HI\ column density, $N$(\HI). For clarity we restrict the 
displays here to the single value of solar Si/C relative abundance. The nominal case
with $Z = 0.003$ $\times$ solar ({\it i.e.} [$Z$] $= -2.5$) and 
$N$(\HI) $= 10^{15}$ cm$^{-2}$ is shown in thin continuous lines.

For \SiIV/\CIV~:~\CII/\CIV\ (upper panels), changes in metallicity and \HI\ column
density have little effect for log $N$(\CII)/$N$(\CIV) $\gtrsim -1$: significant
departures from the nominal case occur only at lower \CII/\CIV\ values. Comparison with 
the data in the top panel of Figure 18 suggests that most of the components in our 
sample have [$Z$] $\lesssim -1.5$ and $N$(\HI) $\lesssim 10^{16.5}$ cm$^{-2}$. In 
contrast, \SiII/\SiIV~:~\CII/\CIV\ (lower panels of Figure 22) is almost invariant 
to changes in metallicity and \HI\ column density.

Figure 23 shows corresponding changes in total hydrogen volume density, $n$(H), and 
mean temperature within the gaseous column, $\langle T_e \rangle$, over the observed 
\CII/\CIV\ range of our data plots at $z_{HM} = 2.3$. Increasing the metallicity leads 
to a reduction in temperature due to increased cooling; increasing the column density 
to about $N$(\HI) $= 10^{16.5}$ raises the temperature, while above that column density
$\langle T_e \rangle$ decreases. At the same time there is comparatively little change 
in the near-linear relationship between $n$(H) and \CII/\CIV. For a given radiation 
field \CII/\CIV\ is thus a good indicator of the gas density (or ionization parameter). 
Using this, we see from the top panel in Figure 18 that the absorbers vary in density 
over the range $n$(H) $= 10^{-3.5}$--$10^{-1.7}$ cm$^{-3}$.

For the components with Si and C {\it b-}values listed independently in Tables 
2--10 ({\it i.e.} with {\it b}(Si) shown {\it unbracketted}: see \S\S3 and 4) we 
obtain (following \citet{rau96}) an average absorber temperature of 
$(2.0\pm0.2)\times10^{4}$ K, broadly consistent with Figure 23 for intermediate to
low \CII/\CIV\ over which \SiIV\ or \SiII\ are (necessarily) relatively strong. At 
$z < 3.1$ and $z > 3.1$ the values are $(1.9\pm0.3)\times10^{4}$ K and 
$(2.1\pm0.4)\times10^{4}$ K respectively, showing no significant change with redshift. 
These values are lower than Rauch et al.'s mean of $3.8\times10^{4}$ K, but as we have 
explained in \S3, on the one hand we strictly include only the narrower components 
({\it b} $\lesssim 10$ km~s$^{-1}$), and on the other, in Rauch et al's analysis 
differential blending effects in broader components may give a tendency to estimate 
apparently higher temperatures. We explore the temperature properties and other 
physical parameters of the absorbers more fully in a further paper (in preparation).

An important conclusion from the results in Figure 22 is that the systematic 
evolutionary effects exhibited in the ionic ratios are not caused by \emph{changes in 
absorber properties}. For example, the observed trends in the ionic ratios are the
opposite of those to be expected if the mean metallicity and \HI\ column density of 
the absorbers increase with cosmic time. We must therefore look to evolution in the 
\emph{ionizing radiation environment} to explain the observations.

\section{ABSORBERS CLOSE IN REDSHIFT TO THE QSOS} 

For comparison with the results given in \S\S6 and 7, all of which use data selected 
to avoid the local influence of the observed background QSOs, we show in Figure 24, 
in similar displays, the properties of those absorbers in our sample which are in the 
near-vicinity of the QSOs. The data consistently conform with expectations for gas in 
photoionization equilibrium illuminated by a hard radiation source: \CIV\ is 
substantially stronger than \SiIV; \CII\ and \SiII\ are relatively weak or undetected; 
and \NV\ is strong and observed over the full available redshift range. For the 
comparisons with model ionic ratio combinations shown in the bottom set of panels, we 
replace the metagalactic ionizing flux with a simple power-law spectral energy 
distribution of the form $f_{\nu} \varpropto \nu^{-1.8}$ \citep{zhe97,lao97,tel02} with 
10 times the mean background intensity at $z_{HM} = 2.3$ to represent the dominating 
radiation of a local QSO, not modified by transfer processes in the intergalactic
medium. Very slight differences between the two redshift intervals, barely apparent 
even at low \CII/\CIV\ ({\it i.e.} at low density), is the consequence of different 
Compton cooling by the microwave background.

\section{EVOLVING SPECTRAL CHARACTERISTICS OF THE IONIZING RADIATION ENVIRONMENT FROM 
IONIC RATIO COMBINATIONS} 

In \S8.2 we compared our observations with predictions of photoionization models of 
low metallicity absorbers optically thin in the \HI\ Lyman continuum exposed to an 
evolving metagalactic ionizing radiation background obtained from radiative transfer 
computations. There we used a simple fiducial case including only QSOs as the source 
of the ionizing background, as in \citet{ham96} but updated \citep{haa98}. While we 
concluded in \S8.3 that there is clear indication from these comparisons that our 
observations are consistent with expectations of photoionization equilibrium, there is 
evidently also a strong evolutionary effect not reproduced by the assumed ionizing 
radiation model. Below, we explore the implications of our observations for the 
spectral characteristics of the ionizing radiation and move to a more general form of 
metagalactic radiation field containing contributions from both QSOs and galaxies, each 
with their own evolutionary behaviour. In addition to using a standard galaxy 
synthesis model we also investigate the effects of a contrived galaxy spectral energy 
distribution.\footnote{We acknowledge the generosity of Haardt \& Madau in setting up 
and processing several specific cosmological radiative transfer models for us 
\citep{haa98}.} 

We stress that in computing the mean intensity of the metagalactic radiation all 
cosmically distributed sources must be included from the outset. The results for 
different sources cannot be added since the background radiation intensity determines 
the ionization balance of the clumpy intergalactic medium which in turn determines the 
background radiation intensity. In specific examples we also include the effects of 
stars in the local environments of the absorbers. To first order such contributions 
\emph{can} be added incrementally if the flux in aggregate is insufficient to modify 
the ionization state of the general intergalactic medium significantly. However, in 
our application we do not imply any \emph{addition} of sources: in effect we are 
simply defining \emph{the location of the absorbers} relative to existing distributed 
galaxies. Thus, an isolated absorber will be exposed only to the general metagalactic 
radiation while an absorber located close to a galaxy will also experience a direct, 
possibly dominant, ``proximity effect''.

\subsection{New Haardt \& Madau Models for the Cosmic Ionizing Radiation Background
Including Template Galaxies}

A brief account of their revised radiative transfer computations is given in 
\citet{ham01}. We include here a short description of the assumed QSO and galaxy 
source contributions. Haardt \& Madau now adopt a spatially flat $\Lambda$CDM  
cosmology with $\Omega_{\Lambda} = 0.7$, $\Omega_{\rm M} = 0.3$ and use 
$H_{0} = 65$ km s$^{-1}$ Mpc$^{-1}$. 

The QSO contribution to the radiation background is based on Boyle et al.'s (2000)  
two-power-law blue luminosity function, with the redshift evolution of the break 
luminosity $L_{\rm B}^{\star}$ following the analytical fit described in Madau et al.
(1999). The assumed optical--ultraviolet spectral energy distribution is similar to 
that given in \citet{ham96} except in the region shortward of 1050 \AA\ where the form 
becomes $f_{\nu} \varpropto \nu^{-1.8}$, steeper than the original exponent. The 
extension into the X-ray region has also been revised, consistent with the X-ray data 
from recent spacebourne missions.

The galaxy contribution is based on the rest-frame luminosity at 1500 \AA, and is 
assumed to arise from a young stellar population described by a star-forming galaxy 
template spectral energy distribution computed from Bruzual \& Charlot's (1993)
isochrone synthesis evolutionary code libraries with metallicity 0.2 $\times$ solar, 
Salpeter IMF with $M_{\star}/M_{\odot} < 125$, constant star formation rate and age 0.5 
Gyrs. Evolution of the star formation rate as computed in \citet{mad96} but with the 
addition of recent high redshift observations ({\it e.g.} Steidel et al. 1999) is used 
to normalize the 1500 \AA\ rest-frame flux of stellar radiation. The emergent ionizing 
flux at 912 \AA\ is estimated from the 1500 \AA\ flux through the escape fraction, 
$f_{\rm esc}$, for Lyman limit photons, defined as the fraction of emitted 912 \AA\ 
photons that escapes the galaxy without absorption by interstellar material divided by 
the fraction of 1500 \AA\ photons that escapes (Steidel et al. 2001).

In Figure 25 we compare the rest spectral energy distributions of the new models for 
$f_{\rm esc} = 0.05$ and 0.5 at the two redshifts $z_{HM} = 2.3$ and 3.9 with the 
simple case of QSOs alone as used in Figures 18--20. We identify these models by 
QG$_{0.05}$, QG$_{0.5}$ and Q, respectively.

\subsection{Predictions of Ionic Ratio Combinations using the New Haardt \& Madau 
Models}

We now test the revised model radiation fields against the observed evolution in our 
ionic ratio combinations. As before, in our Cloudy modelling we include the cosmic 
microwave background.\footnote{However, the change in ionization balance due to Compton 
cooling losses from scattering off the cosmic microwave background, which increases to 
higher redshifts, turns out not to be significant in the range of our data.}

In Figure 26 we show our Cloudy-generated results for \SiIV/\CIV~:~\CII/\CIV\ at 
redshifts $z_{HM} = 2.3$ and 3.9 with our low and high redshift data sets respectively 
covering $ 1.9 < z < 2.65$ and $ 3.4 < z < 4.4$, given in Figure 18. For comparison, 
at each redshift we plot the results computed with both radiation models, QG$_{0.05}$ 
and QG$_{0.5}$. The small value of $f_{\rm esc}$ in the former case may be typical at 
our lower redshifts as indicated by observations of local and intermediate redshift 
star-forming galaxies \citep{lei95,hjd97,deh01,hec01} and we take the latter as 
representative at higher redshifts following the large value of $f_{\rm esc}$ obtained 
for Lyman-break galaxies at $z \sim 3.4$ (Steidel et al 2001). In the corresponding 
right panels we show the forms of these metagalactic background spectra in the 
frequency range effective for photoionization of the absorbers relevant to our species 
of interest, and identify the positions of significant ionization thresholds. Compared 
with model Q, the intensities are raised at energies below the \HeII\ ionization edge, 
over which the relatively soft galactic radiation makes its contribution, and have 
depressions at higher energies due to increased \HeII\ continuum opacity from resultant 
changes in the ionization balance of the intergalactic medium. At the higher redshifts 
the magnitude of these effects is proportionally greater because galaxies make up a 
much larger fraction of the model ionizing source population. 

For our low redshift case in Figure 26 we note that for model QG$_{0.05}$ the curve 
traced is quite close to that for model Q and still fits the data well, but for model 
QG$_{0.5}$ there is a substantial rise in the curve as it progresses into the higher 
ionization (lower \CII/\CIV) region resulting in a somewhat poorer fit to the data. 
Thus, our data are broadly consistent with a low value for $f_{\rm esc}$ and a 
dominant QSO population. Conversely, for our high redshift data set we see that a high 
value for $f_{\rm esc}$ gives a better fit in the higher ionization region of Figure 
26. Nevertheless, the overall fit remains extremely poor because the model is unable 
to match the data in the intermediate-to-low ionization part of the diagram.

In Figure 27 we show our corresponding model QG$_{0.05}$ and QG$_{0.5}$ results at 
$z_{\it HM} = 3.9$ for the three combinations \SiII/\SiIV~:~\CII/\CIV,  
\SiII/\CII~:~\CII/\CIV\ and \NV/\CIV~:~\CII/\CIV\ with the high redshift data sets 
given in Figures 19 and 20. The \SiII/\SiIV\ case shows very little change from 
model Q for either value of $f_{\rm esc}$ and again the data are poorly matched by 
the models. For \SiII/\CII\ both models give broadly consistent fits although, unlike 
the other ratios, the fit for QG$_{0.05}$ remains closer to the data, similar to model 
Q. The data in the \NV/\CIV\ case, being all upper limits, can be marginally 
accomodated by varying N/C within the permitted range for both models QG$_{0.05}$ and 
QG$_{0.5}$. However, QG$_{0.5}$ is favoured due to the greater reduction of the \HeII\ 
continuum, wherein lie the ionization edges of \NIV, \NV\ and \CIV, with resultant 
greater suppression of \NV\ and reduced loss of \CIV.

In summary, we have found that we cannot achieve good fits to the observed ionic 
ratios as a function of redshift with ``standard'' metagalactic spectral energy 
distributions. It is necessary to modify these. In following sections we investigate 
specific spectral variations in the ionizing flux.

\subsection{Effect of a Large Reduction in the He~{\scriptsize II} Continuum}

First, we ask whether substantial suppression of the metagalactic radiation intensity 
in the \HeII\ continuum as proposed by \citet{soc96} and \citet{son98} can improve the 
fit to our high redshift data for \SiIV/\CIV~:~\CII/\CIV\ (the relevant ionization 
potentials of the Si and C species related to the appearance and loss of \SiIV\ and 
\CIV\ straddle the \HeII\ ionization edge). As we have already shown in \S8.1 this is 
not indicated by the behaviour of \SiIV/\CIV\ median values as a function of redshift. 
We now test this hypothesis by arbitrarily modifying the metagalactic radiation output 
spectrum, and in the left panel of Figure 28 show two shapes, a horizontal cut and a 
deep depression, both beginning with a drop of 4 dex at the \HeII\ edge. Although 
these are not self-consistently modelled constructions they serve to give a gross 
indication of any resultant effects. We discuss model Q but models QG$_{0.05}$ or 
QG$_{0.5}$ give similar results. 

The outcomes in the right panel of Figure 28 for the two shapes differ little and show 
a large rise in \SiIV/\CIV\ in the high ionization region similar in form to that 
found in Figure 26 but more extreme, and evidently go no closer to achieving a fit in 
the intermediate-low ionization range of the data. It is clear that this 
two-dimensional display of ionic ratios has only limited value as an indicator of 
\HeII\ continuum opacity. Moreover, from the disposition of the data and upper limits 
in Figure 28 it can be seen why simply using the \SiIV/\CIV\ ratio, as in Figure 16, 
is not a useful tool for this purpose.

In Figure 29 we show corresponding results for the other high redshift ionic ratio 
combinations as before. \SiII/\SiIV~:~\CII/\CIV\ shows very little change from the 
unmodified model Q for either of the cases and clearly is not sensitive to \HeII\ 
continuum suppression. \SiII/\CII~:~\CII/\CIV\ does change markedly relative to the
model Q result and also to those for the QG models in Figure 27, rising higher to 
give an almost horizontal curve, inconsistent with the trend of the data. For 
\NV/\CIV~:~\CII/\CIV, because of the great reduction of the \HeII\ continuum, the 
model results in Figure 29 are significantly lowered and consequently are consistent 
with the data upper limits.

The results, particularly for \SiIV/\CIV~:~\CII/\CIV\ and \SiII/\SiIV~:~\CII/\CIV, 
show that modifications in the \HeII\ opacity alone do not explain our high redshift 
data, although the rise in \SiIV/\CIV\ in the higher ionization region of the diagram 
is helpful in accounting for the points with \CII\ upper limits, noted in \S8.2. 
As we have shown in \S8.4, increasing the absorber metallicity or \HI\ column density 
can have a similar effect.

\subsection{Template Galaxies Local to the Absorbers}

Next, in view of our conclusion in \S7 that the absorbers in our sample are regions
located in the outer extensions of galaxies, we attempted to account for our high 
redshift observations by augmenting the metagalactic background radiation with
radiation from local stellar sources. As we have explained, in effect this is a 
redefinition of the location of the absorber, not addition of more sources; thus, a 
large flux at the absorbers from such localized sources will not violate limits on the 
metagalactic radiation intensity defined by the \HI\ proximity effect \citep{sco00}. 
For a local component we applied the intrinsic galaxy template spectral energy 
distribution used for the dispersed galaxies in the new Haardt \& Madau QG models, but 
now not modified by passage through the intergalactic medium. For ease of reference we 
use the mean intensity at the Lyman limit for model Q at $z_{HM} = 3.9$
($J_{\nu_0} = 1.6 \times 10^{-22}$ erg s$^{-1}$ cm$^{-2}$ Hz$^{-1}$ sr$^{-1}$) as 
the ``unit'' of mean intensity at the \HI\ ionization edge and scale by factor 
$f_{\rm loc}$ in defining the relative contribution at the absorber of the radiation 
that has \emph{escaped} from the assumed local galaxy. For the actual metagalactic 
contribution we used the model Q and QG cases in different trials with 
$f_{\rm loc} = 1q$--$100q$ (we add ``$q$'' to indicate that this is a multiplier of 
the pure QSO background case). For \SiIV/\CIV~:~\CII/\CIV\ we again obtained no better 
than different degrees of progressive rise into the higher ionization region similar 
in form to those shown in Figure 26. 

In passing, we note that simply changing the \emph{intensity} of the ionizing spectral 
energy distribution at the boundary of an optically thin absorber in photoionization 
equilibrium makes no substantial difference to the character of our ionic ratio 
combination diagrams: it requires suitable alterations in \emph{spectral shape}, by
means of a contribution that spectrally overwhelms the QSO contribution to the 
background over 1--4 Ryd, to bring about the relative changes in the ratios of the 
form that we are seeking.

We conclude that in both the cosmologically distributed and the locally confined 
situations, the assumed form of the synthesized galaxy spectrum adopted in the new 
models is not able to account for the observed behaviour of our metal line ratios at 
high redshifts. 

\subsection{Contrived Stellar Sources Local to the Absorbers}

The general lack of concordance of the models with much of our data at high redshift 
is not surprising. The synthesized galaxy model that is assumed in the above 
discussion must be a gross approximation in its description of the emergent ionizing 
flux below 912 \AA, our region of interest. There are indeed large uncertainties in 
modelling the properties of the hot star population, both in the definition of 
evolutionary tracks and in treating the atmospheres 
\citep{sas97,cro00,kew01,pau01,snc02}. The Bruzual \& Charlot libraries use the Kurucz 
atmospheres given in the compilation of \citet{lcb97} (to $T_{\rm eff} = 50,000$ K) 
which ignore non-LTE effects such as metallicity-dependent wind outflows or departures 
from plane-parallel geometry. While other synthesis models are available which give 
somewhat more emphasis to the treatment of the hot star population (Fioc \&
Rocca-Volmerange 1997; Leitherer et al. 1999; Smith et al. 2002) the scarcity of 
corroborating observations at short wavelengths and in metal-poor hot, massive stars 
compounds the uncertainty. The escape fraction, itself an uncertain quantity, gives an 
overly-simplistic representation of the form of self-absorption (grey, in a region 
where it is likely to be strongly dependent on wavelength). Moreover, the star 
formation rate is likely to be varying \citep{brc93,lei99,kol99}. Because of the short 
lives of massive stars such variation is greatly amplified in the far ultraviolet and, 
to maintain a high average luminosity, the duty cycle must be high; accordingly, while 
a constant star formation rate may be a good approximation at longer wavelengths this 
may be too simplistic at the short wavelengths of interest here. 

To investigate what combination of stellar spectral features can achieve a closer 
match to the observations we sought examples from available hot stellar atmosphere 
calculations which could serve as local sources. Good fits to the data were obtained 
with selected models from the range of \citet{kur79} line-blanketed, LTE, 
plane-parallel, static atmospheres of solar metallicity readily available within the 
Cloudy code. In Figure 30 we show for \SiIV/\CIV~:~\CII/\CIV\ the predictions of a 
model star of effective temperature 45,000 K and log $g = 4.5$ (we identify this 
contrived ``galaxy'' by the letter A) taken in combination with the model Q background 
(and the cosmic microwave background), for which we achieve a good match to the 
observed data points with $f_{\rm loc} = 25q$. This combination model is identified by 
Q[A$_{25}$], with the local component indicated bracketted and the metagalactic,
unbracketted. As can be seen in the figure, relative to the model Q background, the 
locally-enhanced spectrum has a hump just beyond the \SiIII\ ionization 
edge\footnote{\citet{lev03} arbitrarily enhanced the intergalactic \HeII~$\lambda$304 
recombination emission feature present in the \citet{ham96} metagalactic background 
spectrum by a factor $\sim 4$ as a device to fit their observed ionic column 
densities in an absorption system near $z = 3$. To explain such enhancement they 
suggest the existence of strong intrinsic $\lambda$304 line emission in the 
distributed QSOs. This universal component would be incompatible with our data at 
$z =$ 1.9--3.4.} and a substantial dip at the \CIII\ edge, which together serve to 
increase the \SiIV/\CIV\ ratio.\footnote{Quantitatively this is not as straightforward 
as it appears because the photoionization cross sections are effective over quite 
extensive energy ranges \citep{ver96}. The spectral changes also influence the 
\CII/\CIV\ ratio.}

Figure 31 contains corresponding results for the other ionic ratio cases. For 
\SiII/\SiIV~:~\CII/\CIV\ use of the model Q[A$_{25}$] radiation environment now begins 
to approach a match to the data. On the other hand, for \SiII/\CII~:~\CII/\CIV\ there 
is a slight departure relative to the well-fitting model Q but the result remains  
consistent with most of the data. The upper limit values for \NV/\CIV~:~\CII/\CIV\ 
again can be formally accomodated within the wide permissable range in N/C. 

Corresponding stellar models of 40,000 K and 50,000 K are less effective. We do not 
claim that the specific 45,000 K stellar model is physically correct or uniquely valid, 
as our remarks above imply, but merely that the character of its ionizing spectrum has 
features of the required form. A variety of models from other libraries, including 
stars of lower metallicity, also approach a match to the data. While we cannot 
precisely define stellar populations, it is evident that at high redshifts the 
observed rise in \SiIV/\CIV\ in the range of intermediate-low ionization and the 
lowering of \SiII/\SiIV\ can be explained by a radiative contribution with 
\emph{stellar} characteristics which dominate in the ionizing range between the \HI\ 
and \HeII\ ionization edges. 

Although it is usefully indicative, Q[A$_{25}$] strictly is not a self-consistent model 
because the totality of ``local galaxies'' should equate to distributed galaxies as 
sources contributing to the metagalactic radiation background, which is not the case 
in this approximation. At the least, there should be a depression in the \HeII\ 
continuum relative to model Q similar to those shown by the QG models used in Figure 
26. We improve on this in the next section. 

\subsection{Contrived Galaxy Model for Metagalactic Sources}

We now investigate a model in which the distributed population of Bruzual \& Charlot 
template galaxies contributing to the metagalactic ionizing radiation background 
contained in the Haardt \& Madau QG models shown in Figure 25 is replaced by a 
population of ``galaxies'' contrived from the single Kurucz 45,000 K stellar model 
that was used in the previous section, with a \emph{source} flux scaling at the \HI\ 
ionization edge relative to the QSOs by factor $f_{\rm met}$ where as before this 
quantifies the total ionizing radiation that has \emph{escaped} from the galaxies 
\citep{haa98}. A model with $f_{\rm met} = 10Q$ (``$Q$'' indicates this is a
multiplier of the QSO \emph{source} flux), QA$_{10}$, which provides a reasonable 
match to the data in the case of \SiIV/\CIV~:~\CII/\CIV\, is shown in the top panels 
of Figure 32. However, for this model 
$J_{\nu_0} = 1.8 \times 10^{-21}$ erg s$^{-1}$ cm$^{-2}$ Hz$^{-1}$ sr$^{-1}$ which
probably exceeds the mean intensity of the metagalactic background averaged over the
range $3.4 < z < 4.4$ of our data \citep{sco00}. 

Accordingly, in the middle panels we show a match for an arbitrary parallel case with 
$f_{\rm met} = 3Q$, having 
$J_{\nu_0} = 6.6 \times 10^{-22}$ erg s$^{-1}$ cm$^{-2}$ Hz$^{-1}$ sr$^{-1}$, 
consistent with \HI\ proximity effect measurements \citep{sco00} (the spectral energy 
distribution for this metagalactic component is shown by the faint line in the 
figure), now augmented by local radiation again represented by the same hot star 
``galaxy'' model, with $f_{\rm loc} = 15q$. We call this model QA$_{3}$[A$_{15}$]. 
The combined mean intensity at the absorbers is 
$J_{\nu_0} = 3.0 \times 10^{-21}$ erg s$^{-1}$ cm$^{-2}$ Hz$^{-1}$ sr$^{-1}$. This 
model gives a somewhat better fit to the observed data points than does QA$_{10}$ 
but tends not to accomodate the points with upper limits in \CII\ quite so well 
because, as we have seen before, the lower suppression of the metagalactic radiation 
in the \HeII\ continuum produces less rise in the theoretical curve. 

However, it is expected that there would be significant variation in the local 
ionizing environments from absorber to absorber which in aggregate will broaden the 
range of model parameters and generally lead to better overall fits to the data. In 
recognition of this likely cosmic variance, in the bottom panels we extend this model 
by showing bounds for the range of local source contributions of mean intensity 
$f_{\rm loc} = 3q$--$25q$, identified as models QA$_{3}$[A$_{3}$] and 
QA$_{3}$[A$_{25}$], while respectively using [Si/C] $=0$ and 0.4 for the
absorbers, which better encompass the data. We recall that the relatively high 
ultraviolet opacity of the intergalactic medium at high redshifts means that the 
metagalactic background becomes more dominated by the closer sources of the distributed 
population \citep{ham96} and this gives further cause for variance in the radiation 
environment of the absorbers.\footnote{This horizon effect can be seen in Figure 21 
from the width of the redshift-smeared intergalactic \HeII~$\lambda$304 emission spike 
which becomes narrower with increasing redshift.}

Figure 33 shows results for the other ionic ratios corresponding to the bottom panel 
of Figure 32. \SiII/\SiIV~:~\CII/\CIV\ follows a similar trend to that noted for model 
Q[A$_{25}$] in Figure 31; further improvement still is needed. \SiII/\CII~:~\CII/\CIV\ 
is marginally consistent with the data. We note as before that the depression in the 
\HeII\ continuum tends to reduce \NV/\CIV.

\subsection{Addition of a Stellar He~{\scriptsize I} Ionization Edge}

Here we aim to improve the fits for \SiII/\SiIV~:~\CII/\CIV\ and 
\SiII/\CII~:~\CII/\CIV\ by appropriately modifying  the spectral shape of the 
radiation incident on the absorbers to enhance the ionization of \SiII\ relative to 
\CII, \CIV\ and \SiIV, thus preferentially reducing \SiII/\SiIV\ and \SiII/\CII. We 
note that the ionization potentials of \CII\ and \HeI, 24.376 eV and 24.581 eV 
respectively, nearly coincide. Main sequence stars cooler than $\sim 35,000$ K show 
an intrinsic large \HeI\ edge and rapidly falling energy to shorter wavelengths 
relative to hotter stars; these are appropriate characteristics with which to produce 
the desired ionization effects. In the upper panels of Figure 34 we use as a baseline 
the same model containing distributed and local contrived hot star galaxies with 
$f_{\rm met} = 3Q$, $f_{\rm loc} = 15q$ as shown in the lower two panels of Figure 32 
and add a cooler local stellar component, B, in the form of a synthetic galaxy spectrum 
obtained with the Starburst99 facility \citep{lei99} using a Salpeter IMF, metallicity 
0.2 $\times$ solar, ${\it M}_{\star}/{\it M}_{\odot} < 20$, and constant star 
formation rate, with scaling $f_{\rm loc} = 50q$. We identify this model by
QA$_{3}$[A$_{15}$B$_{50}$]. The local (bracketted) A and B components together
represent the combined spectral characteristics of a \emph{single} synthetic galaxy. 
The total mean intensity at the absorbers is 
$J_{\nu_0} = 1.1 \times 10^{-20}$ erg s$^{-1}$ cm$^{-2}$ Hz$^{-1}$ sr$^{-1}$.
Comparison with Figure 32 shows that the addition of component B leaves the 
\SiIV/\CIV~:~\CII/\CIV\ fit almost unaffected. In the lower panels of Figure 34 is the 
same case, but as before representing bounds in cosmic variance, with the two local 
stellar source components A and B scaled respectively by $f_{\rm loc} = 5q$, $20q$ and 
$f_{\rm loc} = 25q$, $200q$, giving models QA$_{3}$[A$_{5}$B$_{20}$] and 
QA$_{3}$[A$_{25}$B$_{200}$]. The respective total mean intensities at the absorbers are 
$J_{\nu_0} = 5.3 \times 10^{-21}$ and $3.5 \times 10^{-20}$ 
erg s$^{-1}$ cm$^{-2}$ Hz$^{-1}$ sr$^{-1}$. Again, for the absorbers in these bounding 
cases we have separately used [Si/C] $= 0.0$ with QA$_{3}$[A$_{5}$B$_{20}$] and 
[Si/C] $=0.4$ with QA$_{3}$[A$_{25}$B$_{200}$]. The points with \CII\ upper limits are 
better accomodated here than in the equivalent case shown in Figure 32. The results 
using this case for the other ionic ratios are given in Figure 35. The fits for 
\SiII/\SiIV~:~\CII/\CIV\ and \SiII/\CII~:~\CII/\CIV\ show considerable improvement over 
those in Figure 33 and \NV/\CIV~:~\CII/\CIV\ remains a good fit to the data.

The heavy dotted line in the lower right panel of Figure 34 is the result of using 
model QA$_{3}$[A$_{25}$B$_{200}$] with absorbers of metallicity [$Z$] $= -1.5$ coupled 
with the solar value [Si/C]~$=$~0.0. This does not increase the ionic model boundaries
for \SiIV/\CIV~:~\CII/\CIV\ but indicates considerable further improvement for 
\SiII/\SiIV~:~\CII/\CIV\ and \SiII/\CII~:~\CII/\CIV, as shown in Figure 35, which now 
give excellent fits to the data for this radiative model. 

Once again, we do not claim that the combination radiative model used in Figures 34 and 
35 is physically correct, only that the ionizing spectra can give a good collective 
fit to all our ionic ratio diagrams. Similar synthesis results producing a significant
\HeI\ edge can be obtained for episodic starburst galaxies with more conventional upper 
${\it M}_{\star}/{\it M}_{\odot}$. Like model Q[A$_{25}$] (Figure 30), the model used 
here is not fully self-consistent because of the incomplete match between the 
metagalactic and local radiative contributions. However, the approximation is far 
closer here and the local contribution, with the key spectral characteristics, is more 
dominant.

An important conclusion can be drawn from the broad similarity between the model 
QA$_{3}$ metagalactic background spectrum with \HeII\ continuum depression, as shown 
in Figures 32 and 34, and the corresponding model QG$_{0.5}$ in Figure 26. This 
indicates that the contrived metagalactic background we consider in our models requires 
a \emph{high escape fraction} for the ionizing radiation from the ``galaxies'' which is 
comparable with the value $f_{\rm esc} = 0.5$ used for the template galaxies. This is 
true because the specific spectral details required to explain the observed 
\emph{ionic ratios} we aim to match are not significant for deducing the gross 
ionization state of the intergalactic medium. 

We go no further in attempting to model the radiative environment or absorber 
properties. We can conclude that the observed high redshift ionic ratios collectively 
can be well explained by a metagalactic ionizing radiation background from distributed 
QSOs and galaxies of specific ionizing spectral energy distribution, with the absorbers 
placed in the close vicinity of these galaxies such that the local galactic radiation 
received strongly dominates over the metagalactic radiation.

Finally, in Figure 36 we demonstrate that our low redshift data \emph{cannot} be fitted 
with the dominant stellar contributions deduced at high redshift. In this figure we 
substitute the QSO source flux at $z_{HM} = 2.3$ but otherwise use the same stellar 
source quantities as in the lower panel of Figure 34. While galactic objects must be 
present at low redshift, the indication is that the radiative escape fraction in the 
Lyman continuum is small, less than a tenth of that we demand at high redshift, and 
possibly that star formation is considerably less active.

Our demonstration of the need for a significant efflux of ionizing radiation from 
galaxies at high redshifts supports the detection of emergent Lyman continuum 
radiation in Lyman-break galaxies at $z \sim 3.4$ reported by Steidel et al. 
(2001).\footnote{\citet{gia02} using a much smaller data set and \citet{fso03} 
employing a method based on imaging photometry of galaxies find average limits several 
times lower than Steidel et al.'s detection (although the large uncertainty 
at $z > 2.85$  in Fern\'andez-Soto et al's estimate permits a large escape 
fraction).} \citet{cao02} and \citet{fuj02} discuss specific effects of repeated 
supernova explosions giving enhanced escape of ionizing radiation from galaxies, 
effective at high redshifts. This may well be the case for the galaxies related to 
the high redshift \CIV\ absorbers. \citet{ade03} stress the role of supernovae-driven 
superwinds in their extensive study of the local gaseous environment of Lyman-break 
galaxies.

\subsection{Distance of the Absorbers from the Galaxies}

We now enquire whether the required mean intensities give realistic separations of the 
absorbers from plausible local sources. We do this by assuming an $L^{\star}$ galaxy
with $f_{\rm esc} = 0.5$ and move it to proper distances $d$ from an absorber such 
that it radiatively matches the $J_{\nu_0}$ values deduced for the \emph{local} 
component of the ionizing radiation field. These local values range from 
$J_{\nu_0} = 4.7 \times 10^{-21}$ to 
$3.4 \times 10^{-20}$ erg s$^{-1}$ cm$^{-2}$ Hz$^{-1}$ sr$^{-1}$ in the model shown
in the lower panels of Figure 34. We take $0.5 \times 10^{29}$ erg s$^{-1}$ Hz$^{-1}$ 
at the \HI\ Lyman limit as a working value for the flux of escaped galactic radiation 
\citep{ste99} and find $d = 31$--85 kpc as the implied distance range for the 
absorbers. It is encouraging that the result is consistent with the observed 
separations of such absorbers and nearby galaxies \citep{chu01,ste02,cas02,ade03} and 
with our conclusions in \S7. For galaxies fainter than $L^{\star}$ these distances 
accordingly reduce. In consequence, there is little room also to reduce significantly 
the adopted escape fraction.

\section{SUMMARY}

From high resolution spectral observations of nine QSOs we have compiled a large sample 
of metal-line systems identified as \CIV\ absorbers outside the Lyman forest in the 
redshift range $1.6\lesssim z \lesssim 4.4$ and include \SiIV, \CII, \SiII\ and \NV\ 
in these where available. By Voigt profile-fitting procedures we find we can closely 
represent these multi-phase systems as complexes of co-existing 
``single-phase-ionization'' component regions. We obtain column densities or upper 
limits for individual components of each species, with Doppler parameters for C. With 
these we study number distributions, number densities, total ion column densities, 
kinematic properties and the ionization state of the absorbers and trace their 
evolution in redshift. We arrive at the following principal conclusions:

1. With decreasing redshift, \CIV\ \emph{component} column density and Doppler 
parameter number distributions, \emph{system} column density number distribution, and 
differential column density distributions of \emph{components and systems}, remain 
almost constant while \emph{system} velocity spreads tend to increase. 

2. The \CIV\ system number density shows no cosmological evolution but there is mild 
evolution in the total population column density (which, however, obtains only in the 
more complex systems), indicating increasing column density per average system with 
cosmic time. We find a mean \CIV\ comoving mass density 
$\langle \Omega_{\scriptsize\textrm{\CIV}} \rangle = (3.8\pm0.7)
\times10^{-8}$ (1$\sigma$ uncertainty limits; spatially flat $\Lambda$CDM cosmology 
with $\Omega_{\Lambda} = 0.7$, $\Omega_{\rm M} = 0.3$ and $h = 0.71$), in broad 
agreement with Songaila (2001; 2002). \SiIV\ presents a somewhat similar picture, while
the other ions change more substantially with redshift, heralding changes in ionization 
state. 

3. Estimations at high and low redshift of the carbon cosmological mass density 
using ionization fractions from our data, relative to the hydrogen mass density in 
the Lyman forest based on $\Omega_{\textrm{\footnotesize b}}$ from the CMB, yield 
[C/H]$_{\langle z \rangle = 4.0} \geq -3.11^{+0.14}_{-0.19}$ and
[C/H]$_{\langle z \rangle = 2.1} \geq -2.64^{+0.15}_{-0.22}$, suggesting a rise by 
a factor $\sim 3$. Relating the hydrogen mass density more directly to regions 
containing the \CIV\ absorbers our values for [C/H] become 
$\gtrsim -2.2$ at $\langle z \rangle = 4.0$ and $\gtrsim -2.0$ at 
$\langle z \rangle = 2.1$, now suggesting a constant metallicity of carbon.

4. The \CIV\ components exhibit strong clustering out to velocity separations 
$\lesssim 300$ km s$^{-1}$ for our prime statistical sample but there is no clustering 
signal for \emph{systems} on any scale from 150 km s$^{-1}$ out to 50000 km s$^{-1}$. 
Neither of these (one-dimensional) distributions shows similarities with 
(three-dimensional) galaxy clustering. Contrary to some previous claims we argue that 
the results are entirely due to the peculiar velocities of gas present in the outer 
extensions of galaxies for which we adduce other evidence.

5. We find no change in the component or system median column density ratio \SiIV/\CIV\ 
with redshift and particularly no large change near $z = 3$, contrary to previous 
observations coupled with claims that this can indicate the onset of complete 
reionization of \HeII. Other ionic ratios do vary (continuously) with redshift but we 
show that these all are only partial indicators of ionization state.

6. Using combinations of ionic ratios we demonstrate that the vast majority of 
absorbers are in photoionization equilibrium, not collisionally ionized. 

7. Our data support the presence in the absorbers of a range in relative abundance 
[Si/C] $\sim$ 0.0--0.4, consistent with $\alpha$-element enhancement in galactic 
metal-poor stars.

8. We observe substantial evolution in redshift in specific combinations of ionic 
ratios, as follows: 

9. At $z \lesssim 2.65$ we find that QSOs dominate the metagalactic ionizing radiation 
background and that contributions from galaxies have minimal effect. This requires a 
low escape fraction for ionizing radiation from galaxies, $f_{\rm esc} \lesssim 0.05$, 
consistent with other observations. 

10. At $z \gtrsim 3.4$ we find that neither QSOs as dominant contributors to the 
metagalactic background, nor a high opacity in the \HeII\ continuum, can explain the 
observed ionic ratios. Between $z = 2.65$ and $z = 3.4$ there is evident transition in 
the ionization properties of the absorbers, with large scatter.

11. In the presence of the QSO population, we can match our highest redshift 
observations well by the addition of a dominant contribution from galaxies with 
specific spectral characteristics in the energy range 1--4 Ryd, although not by 
standard stellar population synthesis models. 

12. With the specific spectral features required to explain the observations the mean 
intensity at the absorbers substantially exceeds the limit imposed by the proximity 
effect \citep{sco00} if all the flux were contributed by distributed galaxies and QSOs. 
Accordingly, we conclude that the absorbers must be in the locality of the galaxies 
such that the local galactic radiation received strongly dominates over the 
metagalactic radiation, consistent with our independent conclusion from clustering 
properties.

13. At these high redshifts such sources require a much higher escape fraction than 
at our lowest redshift interval, supporting the detection of emergent Lyman continuum 
radiation in Lyman-break galaxies at $z \sim 3.4$ reported by Steidel at al. (2001).

14. Although the ionic ratio combinations basically are sensitive only to the 
\emph{shape} of the spectral energy distribution we find from comparison with the
adopted QSO contribution to the ionizing background that the resultant mean intensity 
of the radiation received from the dominant local galaxy contribution is consistent 
with observed separations from galaxies if $f_{\rm esc} \gtrsim 0.5$.

\acknowledgements

We are greatly indebted to Tom Barlow for the initial data reduction, Bob Carswell 
for providing VPFIT and being so unstinting with his time for schooling, advice and 
assistance, Francesco Haardt with Piero Madau for their generosity in providing many 
new computations of radiative transfer models for the metagalactic background 
radiation and for help and discussions, Gary Ferland for providing Cloudy and related
advice, David Valls-Gabaud and Roderick Johnstone for their considerable help with 
the use of Cloudy, Robert Lupton for advice and provision of a new facility within SM, 
Bob Abraham and Rob Simcoe for kindly providing analysis software, and Max Pettini, 
Samantha Rix, Paul Crowther, Martin Haehnelt, Stephen Smartt, Christopher Tout, Michael
Irwin, George Efstathiou and Michael Murphy for assistance and helpful discussions. 
We also thank the Keck Observatory staff for their assistance with the observations.
Finally, we extend special thanks to those of Hawaiian ancestry on whose sacred 
mountain we are privileged to be guests. Without their generous hospitality, the 
observations presented herein would not have been possible. A.B. gratefully 
acknowledges support from The Leverhulme Trust and from PPARC. W.L.W.S has been 
supported by NSF grants AST-9900733 and AST-0206067. M.R. is grateful for support from 
the NSF through grant AST-0098492 and from NASA through grant AR-90213.01-A.

%%%%%%%%%%%%%%%%%%%%%%%%%%%%%%%%%%%%%%%%%%%%%%%%%%%%%%%%%%%%%%%%%%%%%%%%% 

%%
%% Beginning of tables
%%

\clearpage
%

%%%%%%%%%%%%%%%%%%%%%%%%%%%%%%%%%%%%%%%%%%%%%%%%%%%%%%%%%%%%%%%%%%%%%%%%%

%%
%% Beginning of figures
%%

\clearpage
\begin{figure}
\figurenum{\scriptsize 1}
\epsscale{1.0}
%%\plotone{f1.eps}
%%\plotone{f1.ps2ps.eps}
\plotone{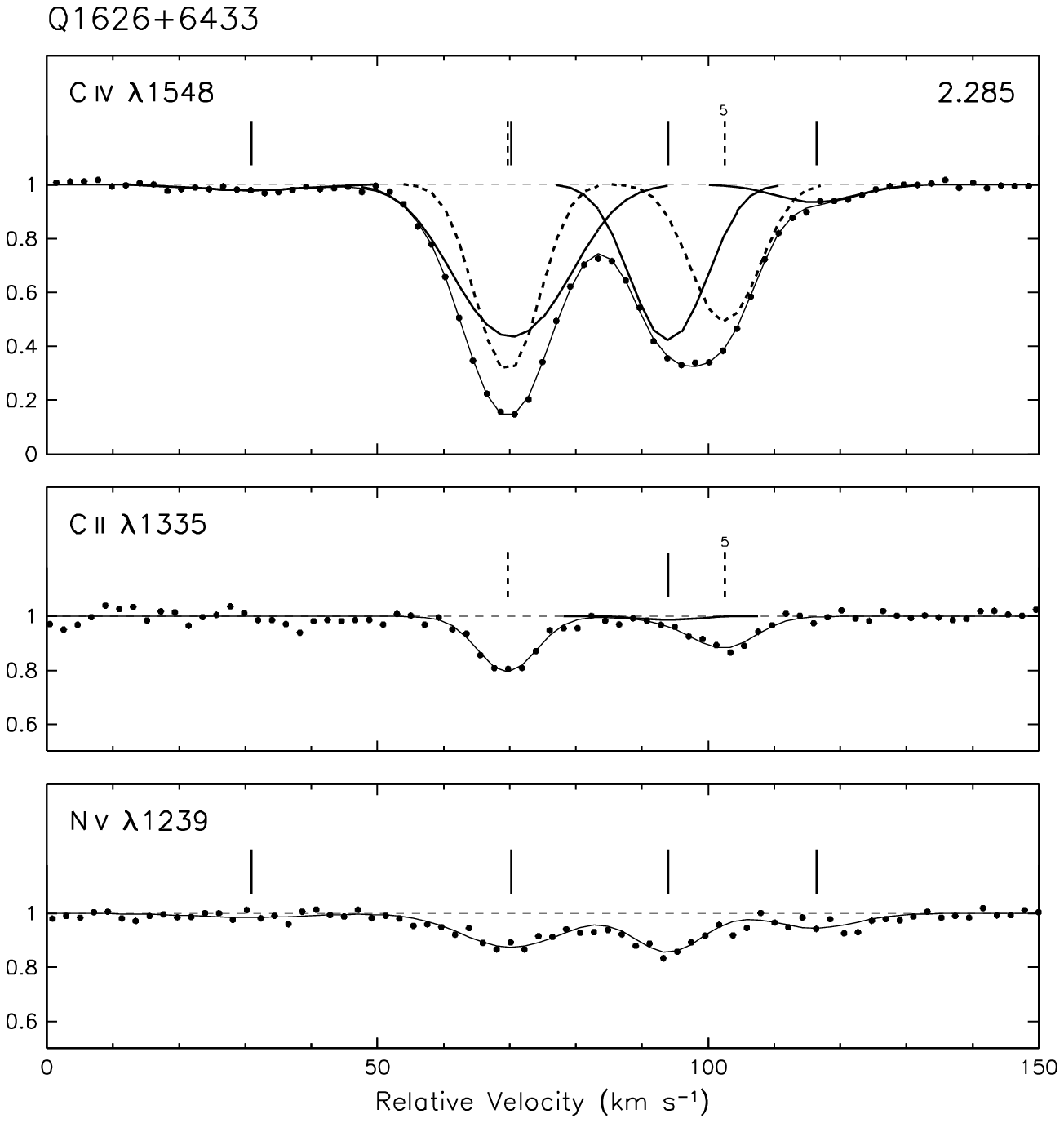}
\caption[f1.eps]{\scriptsize Continuum-normalized spectra demonstrating the 
separation of an absorption system ($z = 2.285$ in Q1626$+$6433) into 
single-phase-ionization components by means of the VPFIT analysis described in the 
text. For each profile the composite fit through the data points is shown by a 
{\it continuous thin line}. Component parameters and numbering (the fifth component
is indicated) are given in Table 2. \CIV\ is shown also with separate fits to the six 
constituent components. The different widths of the two nearly coincident components 
2 and 3 and the positions of the more spaced components 4 and 5 are mirrored 
separately in the other two species, with the \CIV\ components dominant in \CII\ (low 
ionization) identified by {\it dashed lines and vertical ticks} and those dominant in 
\NV\ (high ionization) by {\it continuous thick lines and ticks}. Ticks are shown only 
at the positions of detected components, not upper limits. Component 4 which is 
strong in \NV\ is weakly evident in the \CII\ profile and is also shown there 
separated. Note the truncation of the vertical axes in the lower two panels.}
\end{figure}

\clearpage
\begin{figure}
\figurenum{\scriptsize 2}
\epsscale{1.0}
%%\plotone{f2.eps}
\plotone{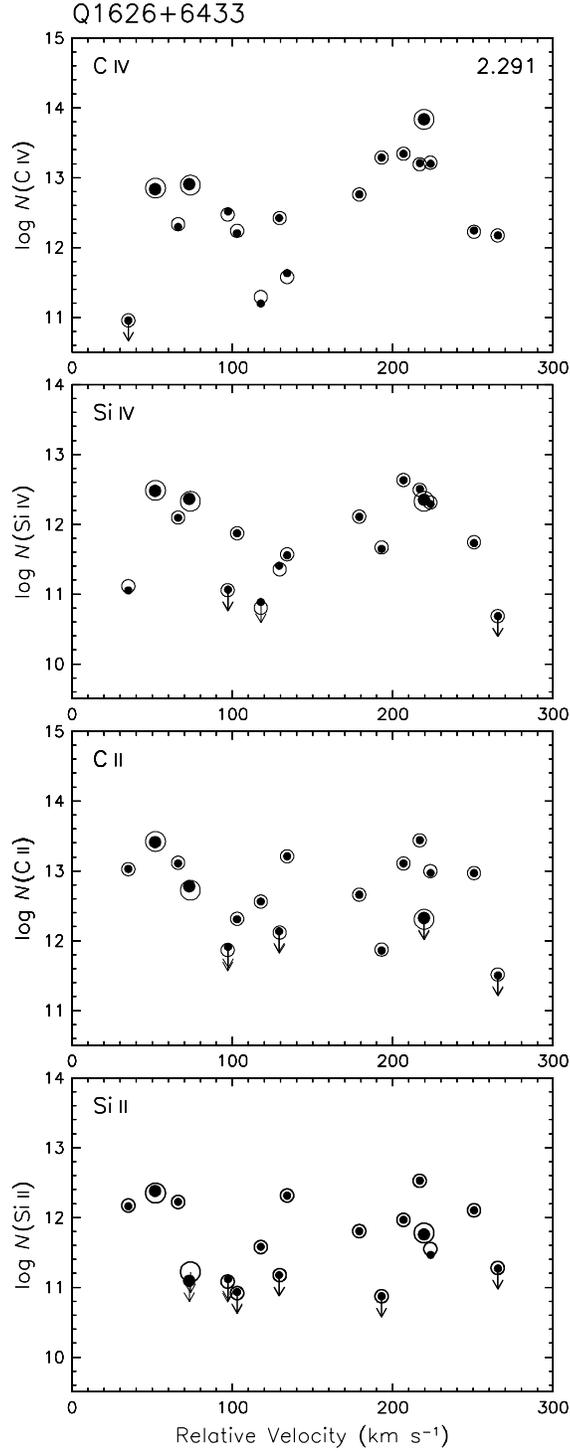}
\caption[f2.eps]{\scriptsize \CIV, \SiIV, \CII\ and \SiII\ component column 
densities $N$ (cm$^{-2}$) for the complex absorption system at $z = 2.291$ in 
Q1626+6433. Two individual runs in the VPFIT analysis show the differential effect of 
fixing the set of three broadest components (indicated by {\it enlarged symbols}---see 
Figure 5 and Table 2 for identification) at the two widely different temperatures 
$1 \times 10^{4}$ K and $1 \times 10^{5}$ K, but otherwise having the same mix of 
specific starting values for the remainder of the components (see text). All component 
values yielded in the first case are shown by {\it filled circles} and in the second 
by {\it open circles}.}
\end{figure}

\clearpage
\begin{figure}
\figurenum{\scriptsize 3}
\epsscale{1.0}
%%\plotone{f3i.eps}
\plotone{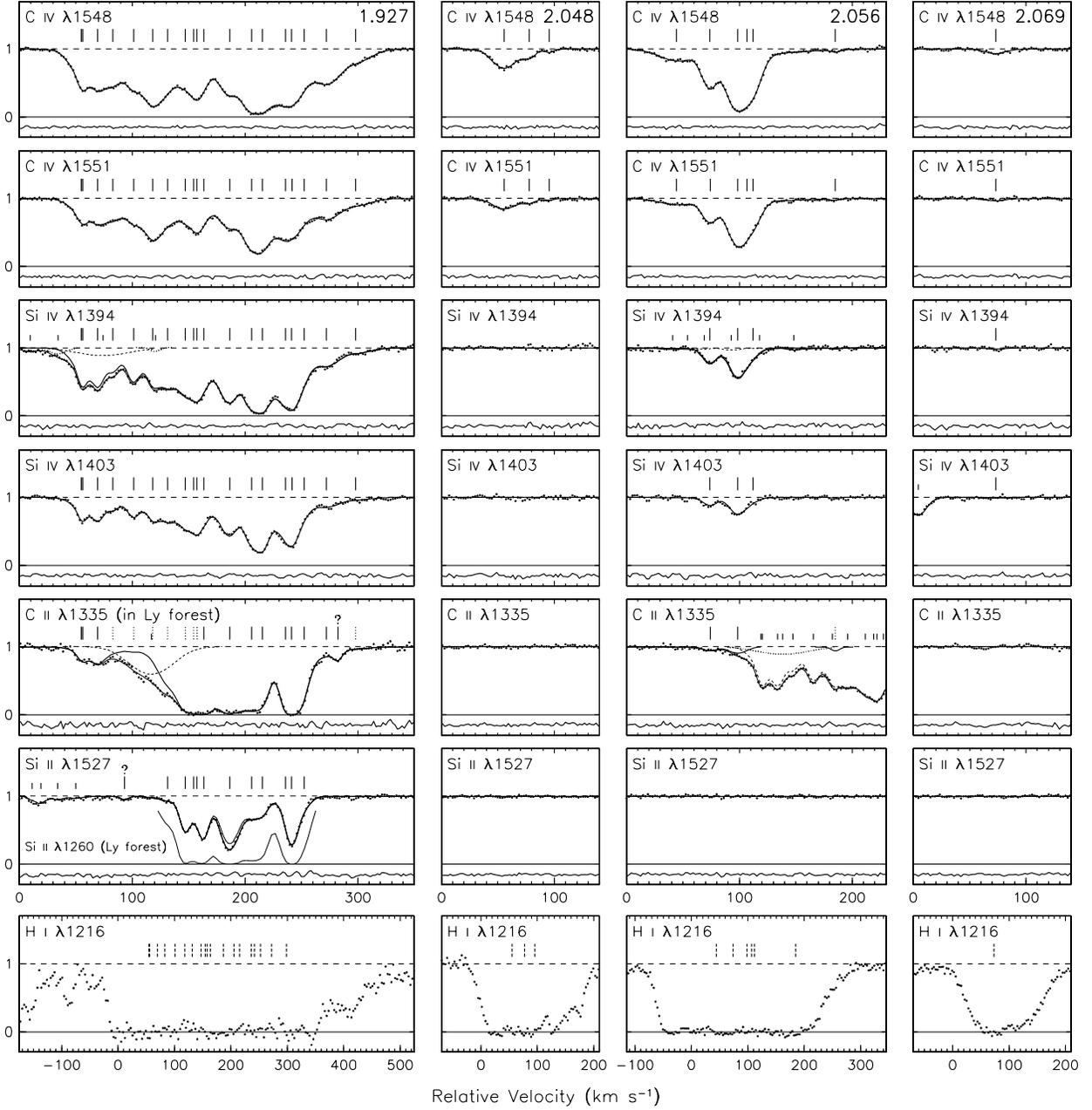}
\caption[f3i.eps]{\scriptsize Continuum-normalized spectra for all systems 
detected in Q1626+6433 ($z_{em} = 2.320$) having at least \CIV\ and \SiIV\ outside the 
Lyman forest. For each system the available metal species of interest in this paper 
are compared on a common velocity scale. The system components identified using VPFIT 
are marked with {\it long vertical ticks}; a few questionable features are so 
indicated. Components yielding only upper limits are unmarked. Component parameter 
values and upper limits are listed in Table 2; in the few cases where components 
have values which are too uncertain to be included in the table they are marked with 
{\it dotted ticks}. Blended or nearby interloper species from other systems are 
indicated by {\it short ticks} and are identified in the notes to Table 2. The data 
values are given as points and the fits obtained to these (including the convolved 
instrumental profile) are shown as {\it continuous bold lines} when there is no 
blending and {\it continuous thin lines} when blending is present; residuals 
({\it i.e.} [data] $-$ [fit]) are shown on the same scale beneath the profiles. In 
the blended cases the deconvolved fits to the appropriate species are shown by 
{\it continuous bold lines} and fits to the interlopers by {\it short-dashed or 
dotted lines}. \SiII~$\lambda$1527 at $z = 1.927$ is shown with a correction ({\it 
thin line}) accounting for mild unknown contamination deduced from related transitions; 
this and $\lambda$1527 at $z = 2.110$ have superimposed fits to $\lambda1260$ (for the 
same column densities) which match data in relatively clear regions of the forest. 
Lyman $\alpha$ observations, covering twice the velocity range, are shown unfitted, but 
with the positions of all components detected in \CIV\ indicated by {\it broken 
ticks}.}
\end{figure}

\clearpage
\begin{figure}
\figurenum{\scriptsize 3}
\epsscale{1.0}
%%\plotone{f3ii.eps}
\plotone{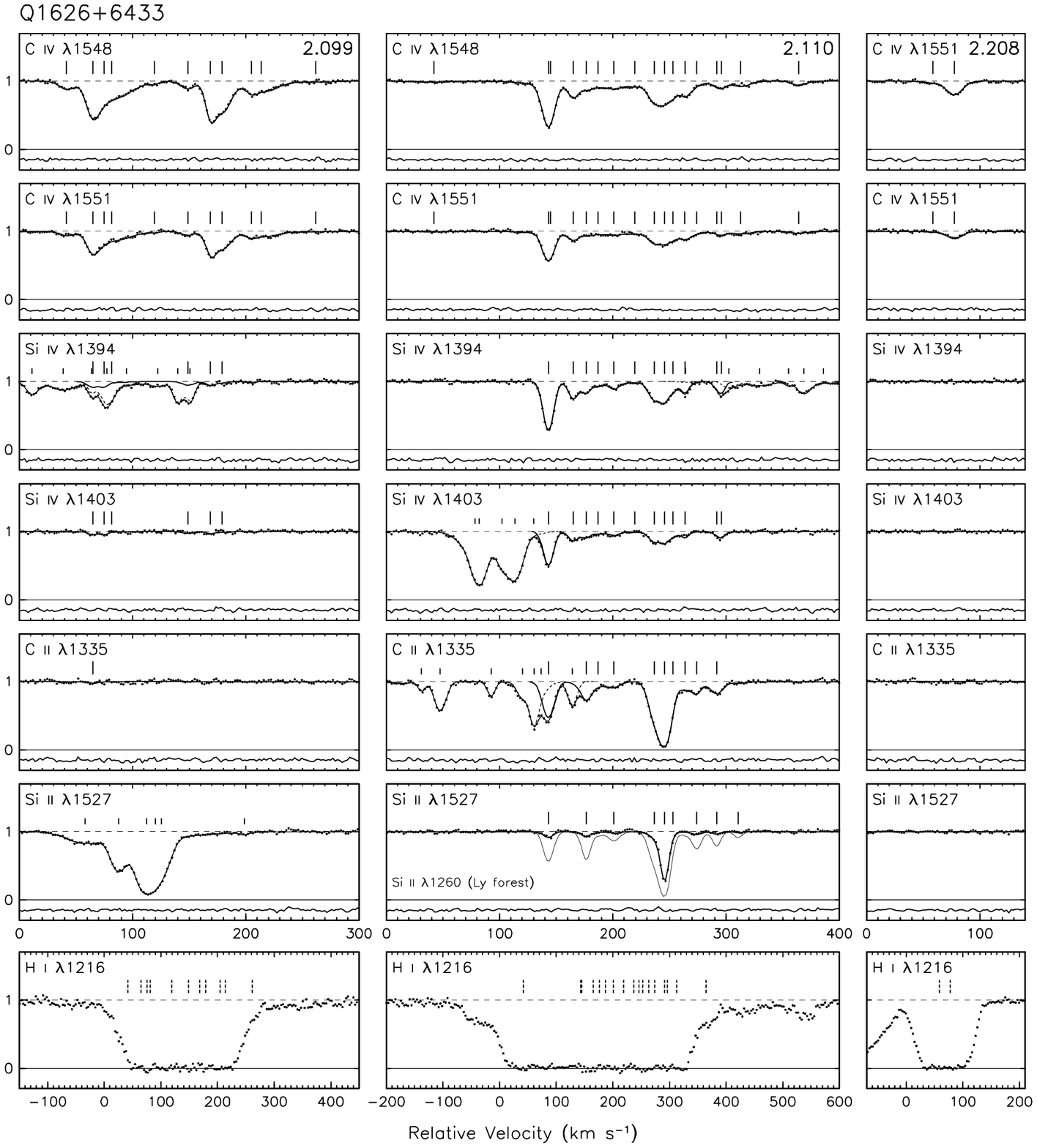}
\caption[f3ii.eps]{\scriptsize Continued.}
\end{figure}

\clearpage
\begin{figure}
\figurenum{\scriptsize 3}
\epsscale{1.0}
%%\plotone{f3iii.eps}
\plotone{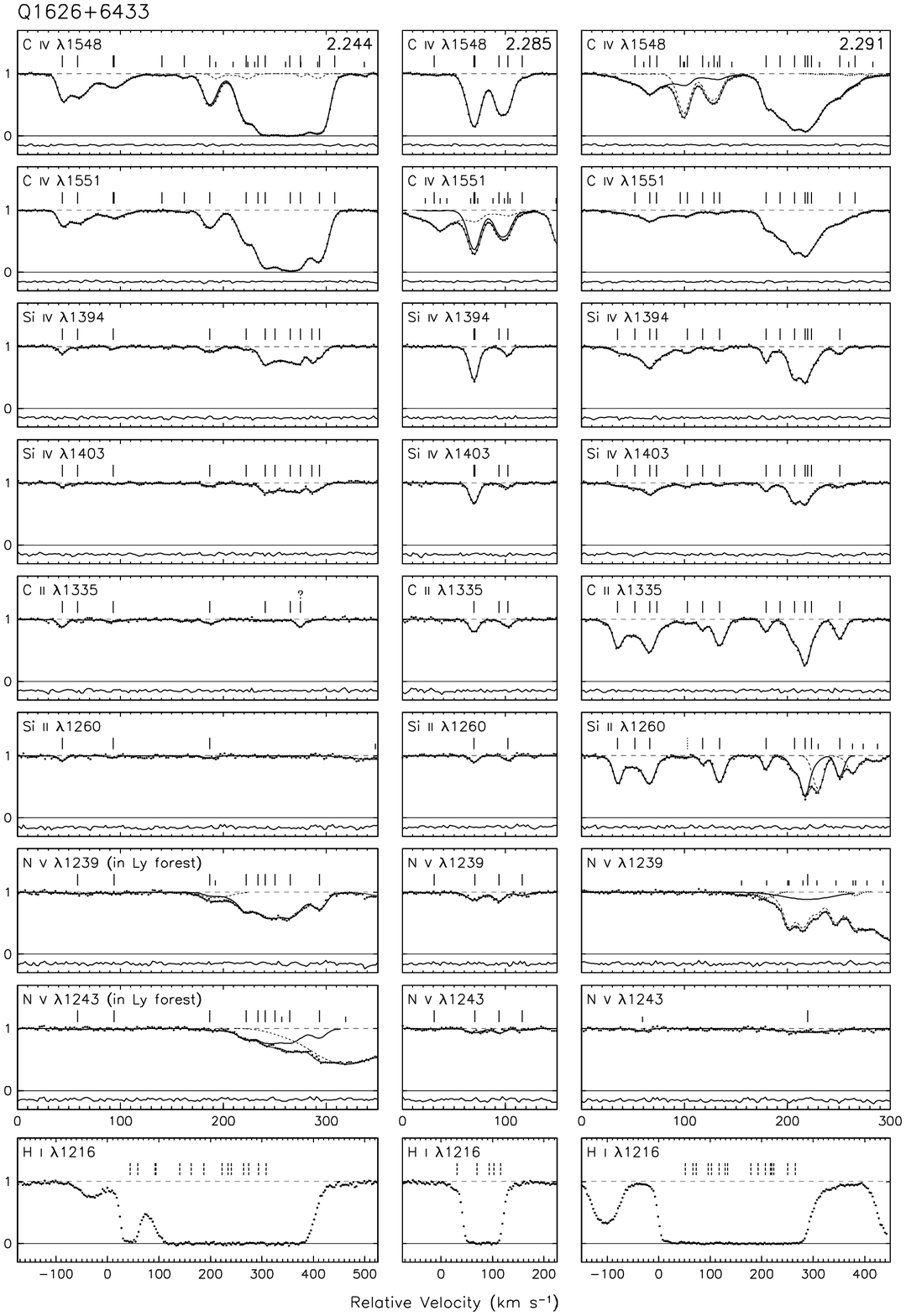}
\caption[f3iii.eps]{\scriptsize Continued.}
\end{figure}

\clearpage
\begin{figure}
\figurenum{\scriptsize 3}
\epsscale{1.0}
%%\plotone{f3iv.eps}
\plotone{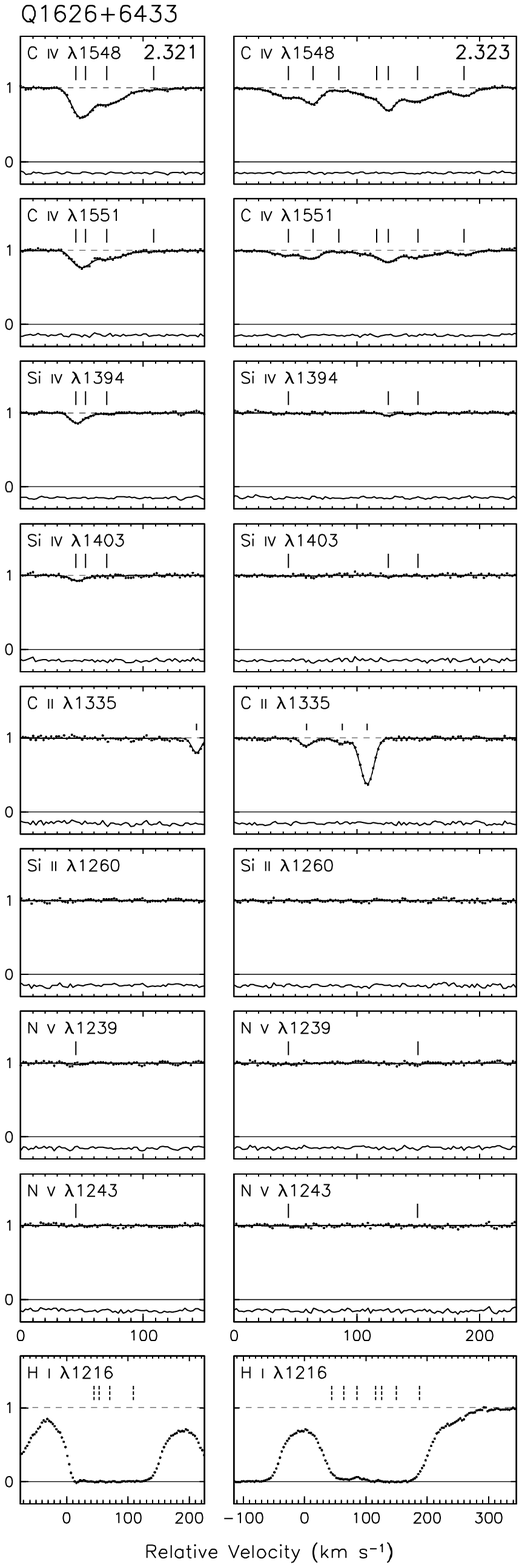}
\caption[f3iv.eps]{\scriptsize Continued.}
\end{figure}

\clearpage
\begin{figure}
\figurenum{\scriptsize 4}
\epsscale{1.0}
\plotone{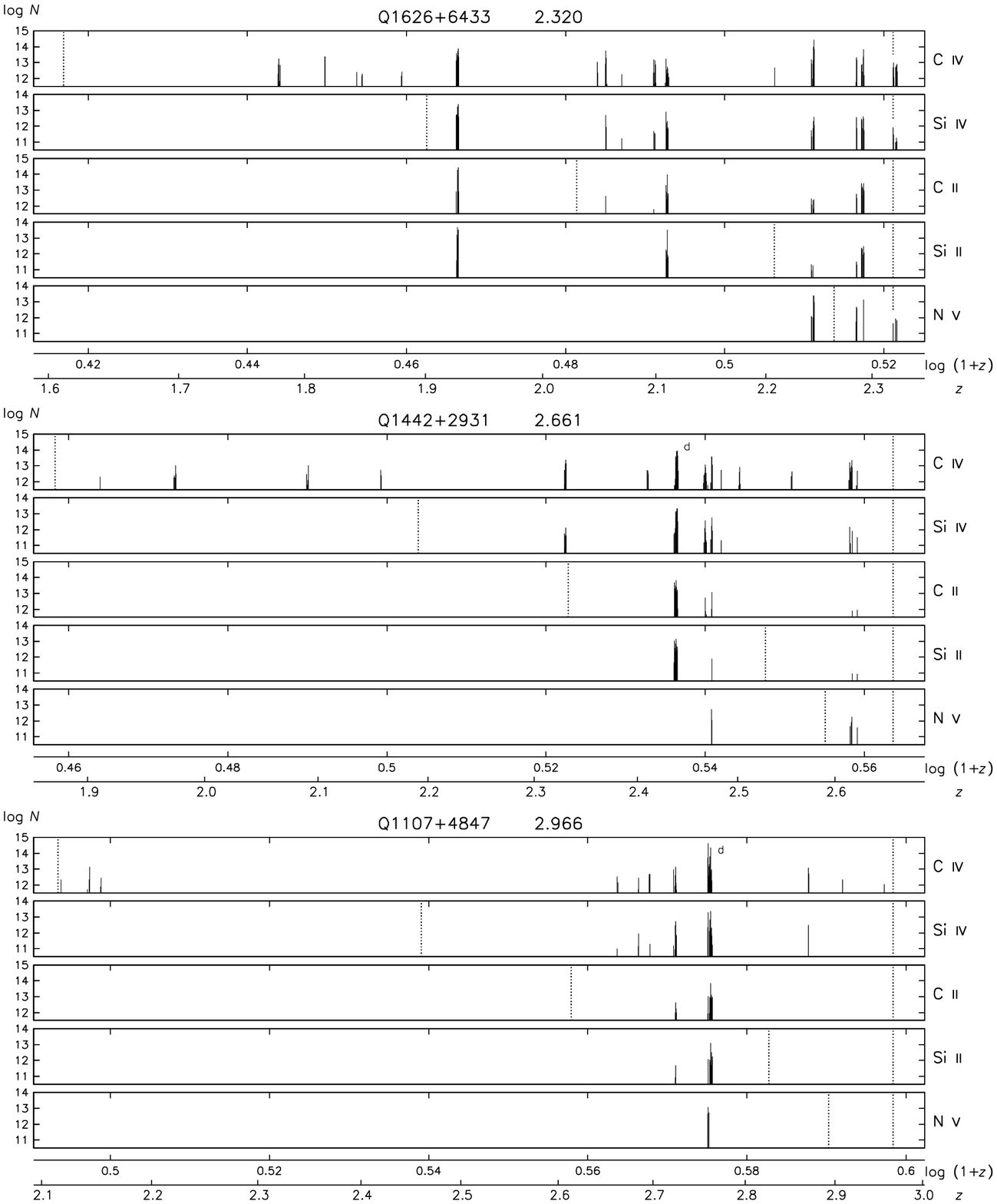}
\caption[f4i.eps]{\scriptsize \CIV, \SiIV, \CII, \SiII\ and \NV\ column 
densities $N$ (cm$^{-2}$) plotted versus redshift $z$ and log ($1+z$) for the absorber 
components \emph{detected} in the nine QSOs (see Tables 2--10). Systems with 
significant Lyman~$\alpha$ damping wings are marked {\it d}. Note the vertical scales 
for \SiIV, \SiII\ and \NV\ are shifted lower by 1 dex than the others. All frames 
cover the same extent in log ($1+z$) which for \CIV\ encompasses the region between the 
Lyman~$\alpha$ and \CIV\ emission lines with some margin. The {\it dotted vertical 
line} at the right of each frame is at the emission redshift; the similar line to the 
left marks the specific limit where a given ion falls in the Lyman forest (in \SiII\ 
this is shown for $\lambda$1260 only). Reliable values appear at redshifts in the 
forest, but for \SiII\ most of these apparent cases actually indicate values from 
strong $\lambda$1527, not from $\lambda$1260, as clarified in the footnotes to 
Tables 2--10.} 
\end{figure}

\clearpage
\begin{figure}
\figurenum{\scriptsize 4}
\epsscale{1.0}
\plotone{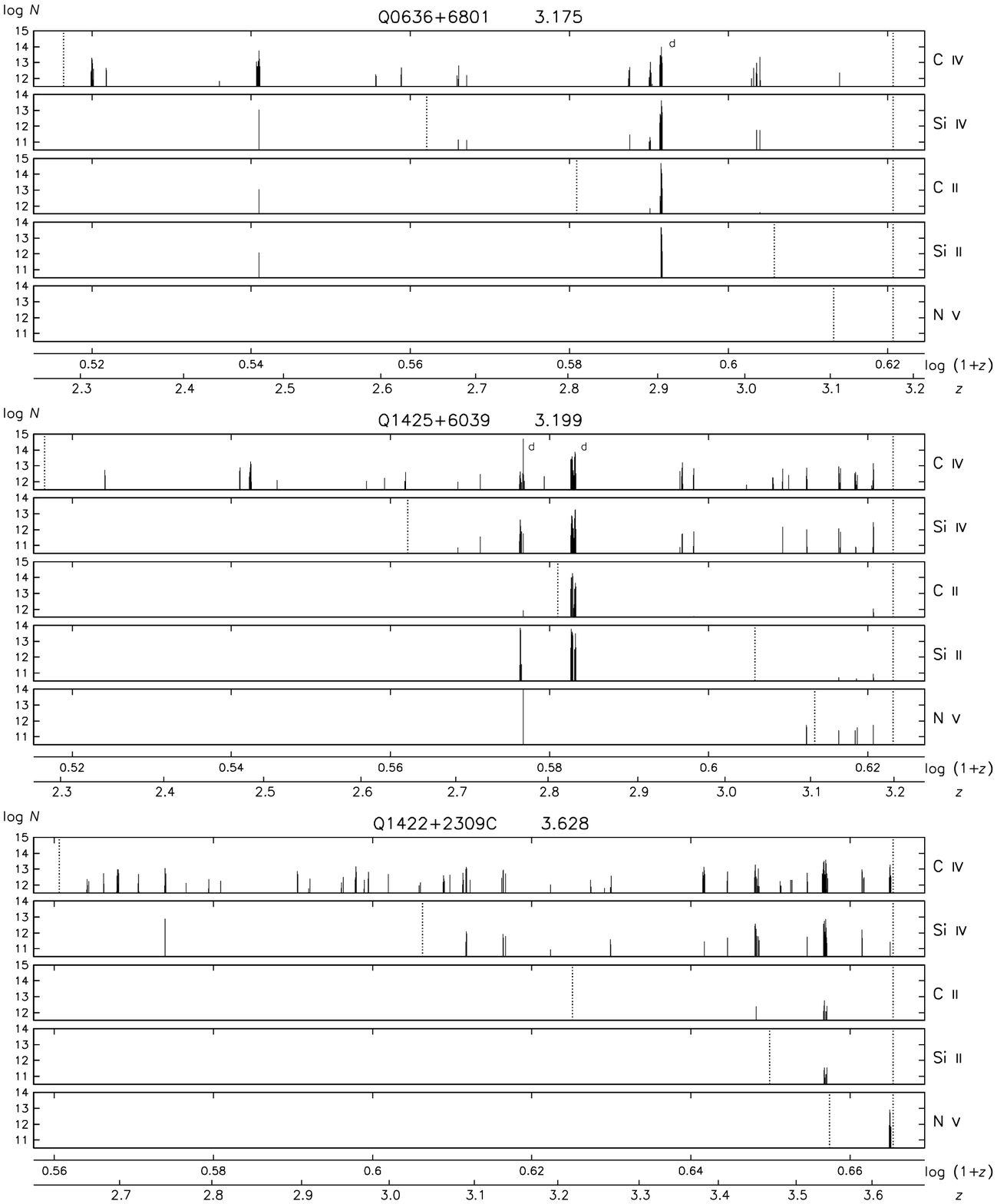}
\caption[f4ii.eps]{\scriptsize Continued.}
\end{figure}

\clearpage
\begin{figure}
\figurenum{\scriptsize 4}
\epsscale{1.0}
\plotone{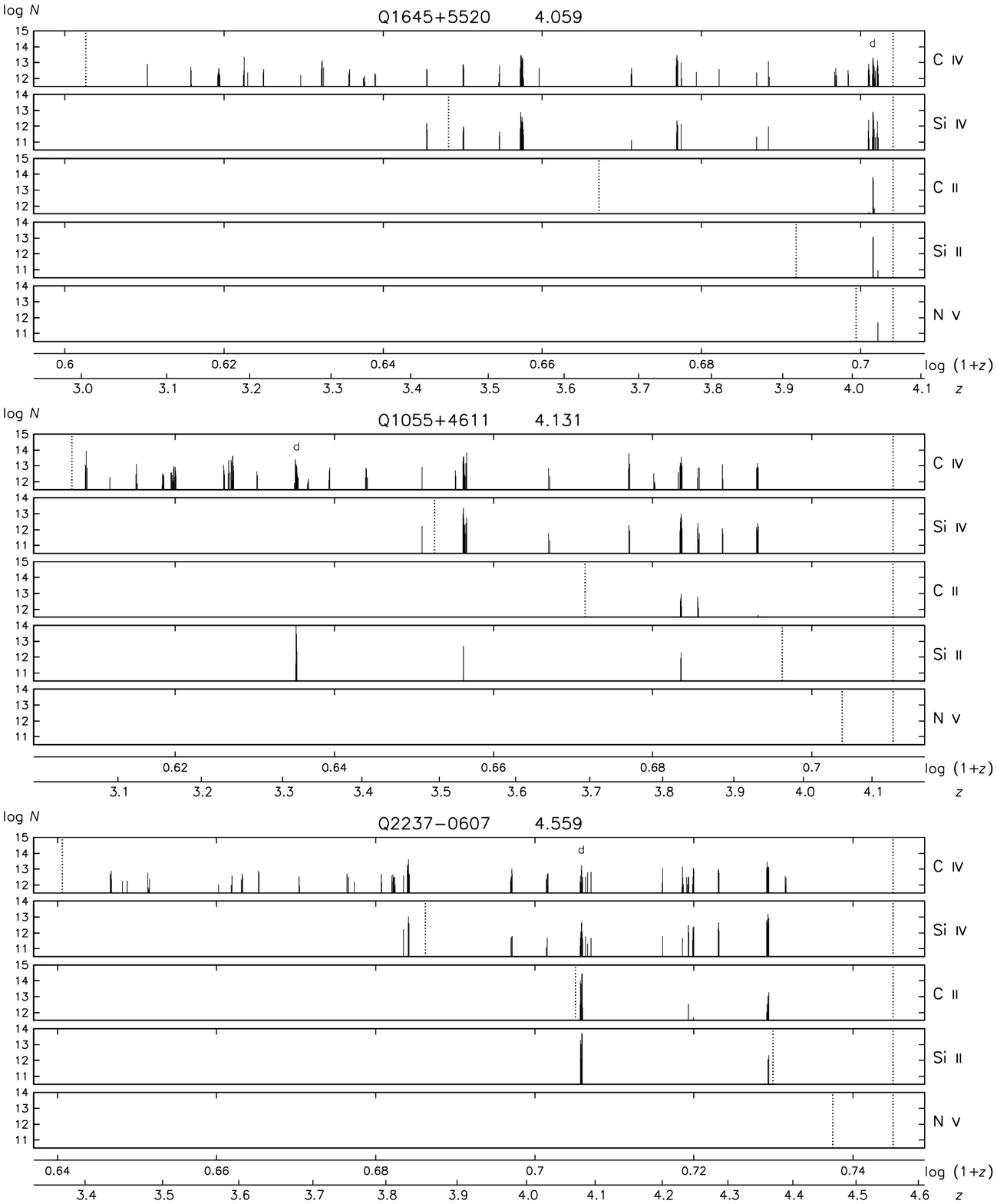}
\caption[f4iii.eps]{\scriptsize Continued.}
\end{figure}

\clearpage
\begin{figure}
\figurenum{\scriptsize 5}
\epsscale{1.0}
\plotone{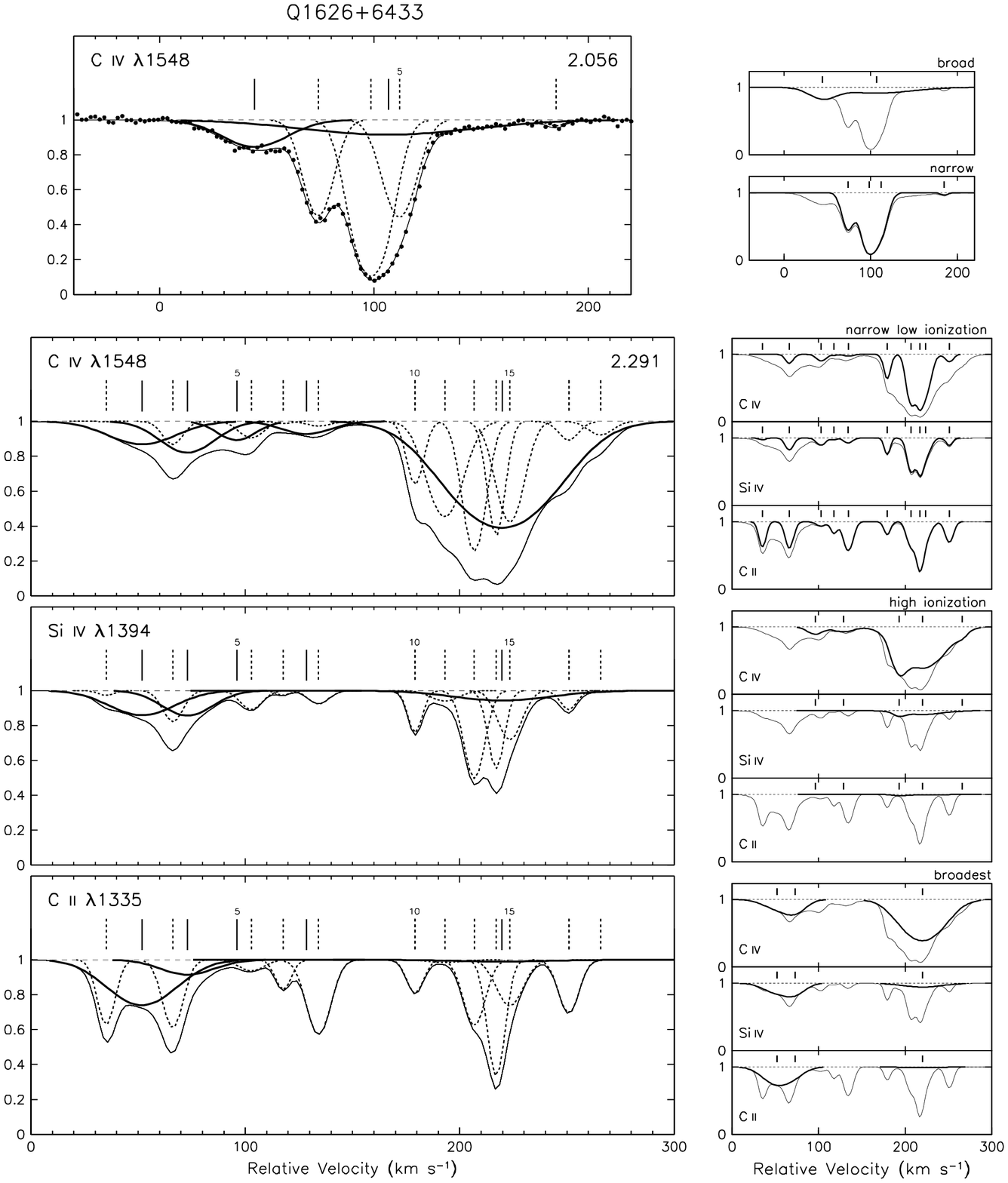}
\caption[f5.eps]{\scriptsize Two examples from Figure 3 showing component 
details of systems containing broad components ({\it continuous thick lines and ticks}) 
among the more numerous narrow components, {\it b}(\CIV) $\lesssim$ 10 km~s$^{-1}$ 
({\it short-dashed lines and ticks}), with numbering as in Table 2. The overall 
composite fits to the pure system profiles ({\it i.e.} excluding any interloper 
species) are shown in {\it continuous thin lines}. {\it Upper left panel:} \CIV\ 
profile in a simple system with exposed broad components, also showing the data points. 
{\it Upper right panels:} Separately highlighting the combined profiles of the broad 
and narrow components. {\it Lower left panels:} \CIV, \SiIV\ and \CII\ profiles in a 
complex system with immersed broad components. The ticks here include positions of 
upper limits. {\it Lower right panels:} Highlighting the combined profiles of 
{\it top}, narrow components which are strong in \CII, {\it middle}, all high 
ionization components and {\it bottom}, the broadest components.}
\end{figure}

\clearpage
\begin{figure}
\figurenum{\scriptsize 6}
\epsscale{1.0}
%%\plotone{f6.eps}
\plotone{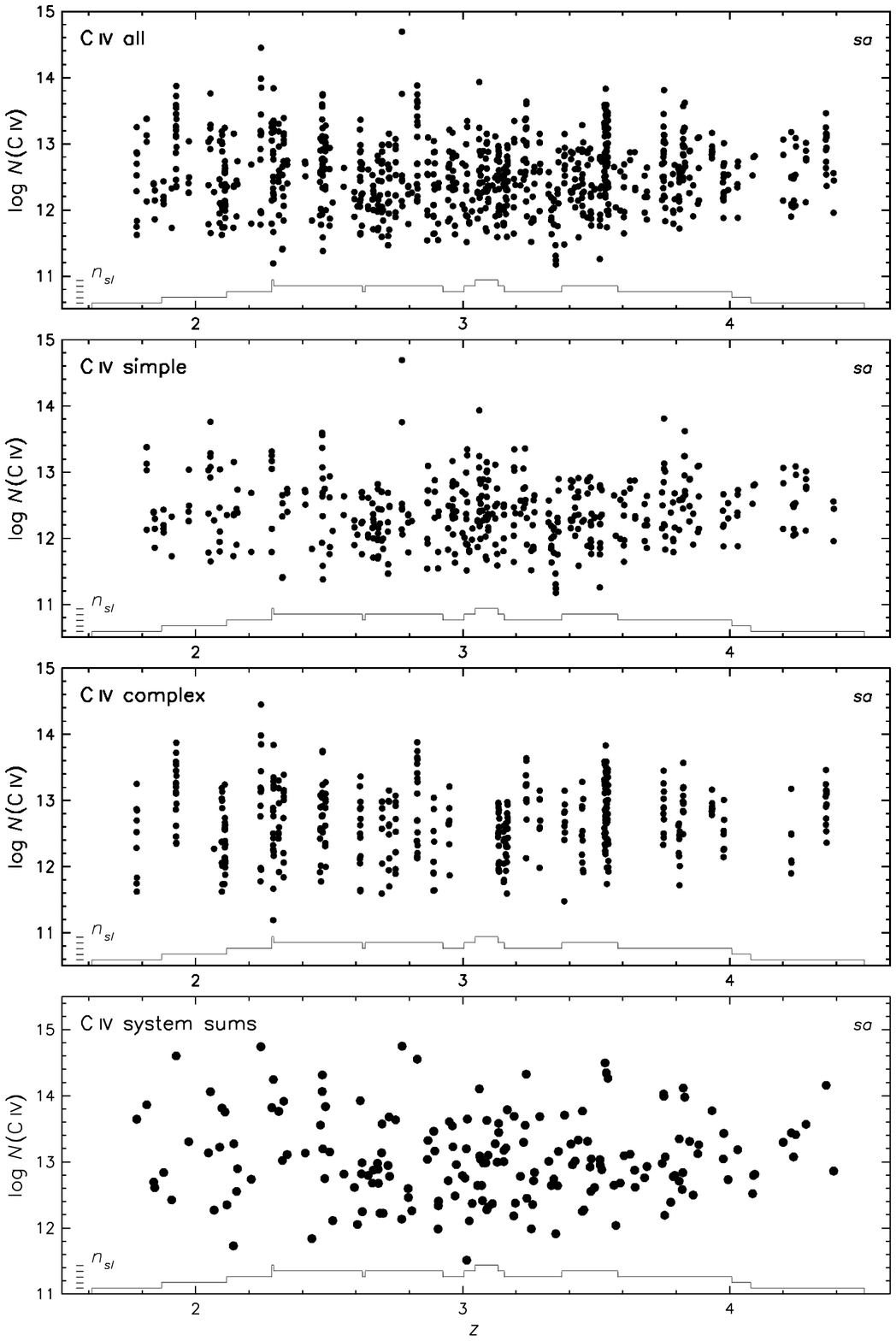}
\caption[f6.eps]{\scriptsize \CIV\ column densities $N$ (cm$^{-2}$) 
in sample {\it sa} plotted versus redshift $z$. To avoid clutter, errors 
(see Tables 2--10) are not shown. The thin continuous histograms display, in unit 
steps, the distribution of number of sightlines ($n_{sl}$) from the nine 
QSOs of the sample within the applied redshift constraints. {\it Upper three panels:} 
component values in {\it all} systems, {\it simple} systems ($\leqslant 6$ identified 
components) and {\it complex} systems ($\geqslant 7$ identified components). 
{\it Bottom panel:} {\it system} summed values for the same sample.} 
\end{figure}

\clearpage
\begin{figure}
\figurenum{\scriptsize 7}
\epsscale{1.0}
%%\plotone{f7.eps}
\plotone{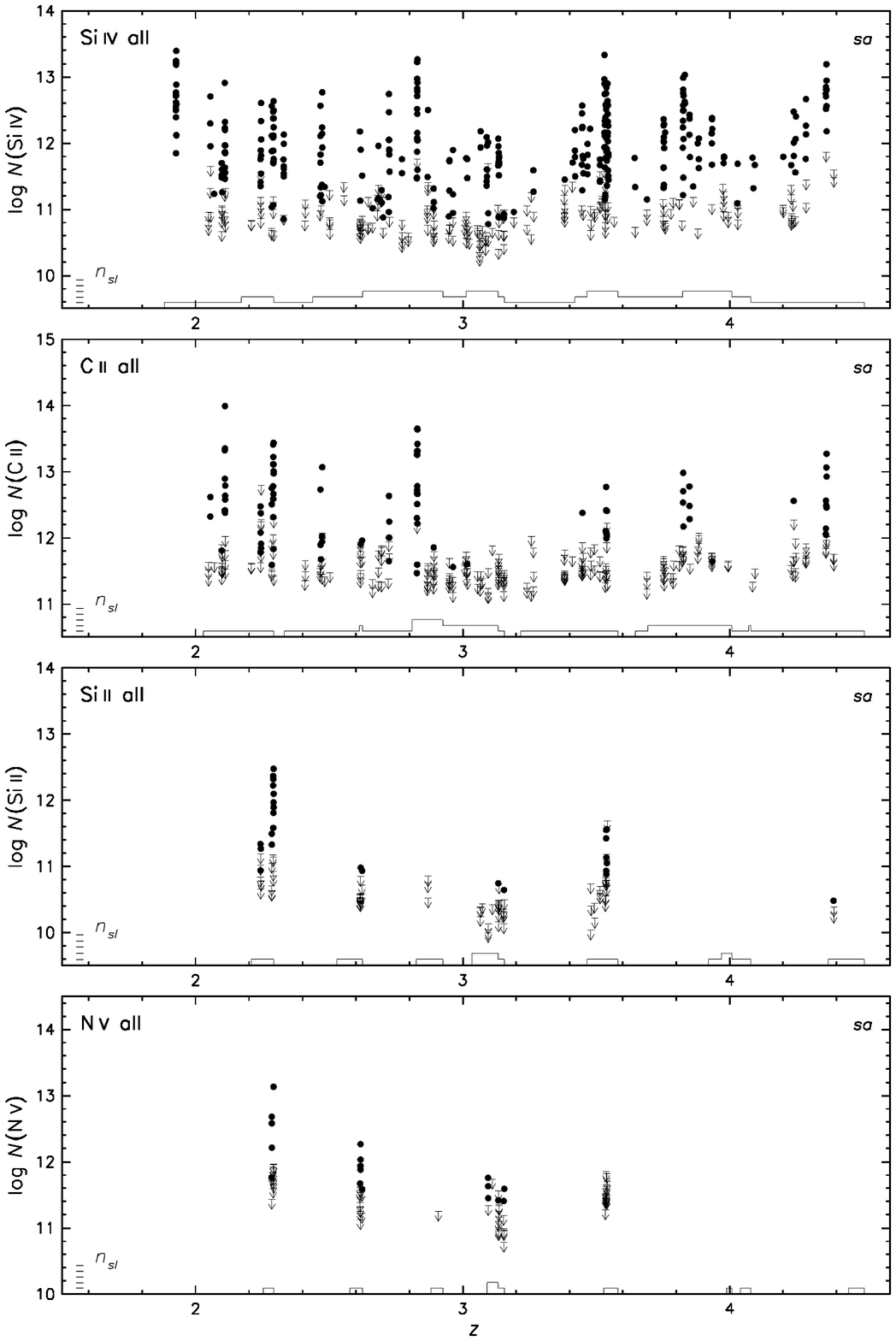}
\caption[f7.eps]{\scriptsize \SiIV, \CII, \SiII\ and \NV\ component column 
densities $N$ (cm$^{-2}$) in sample {\it sa}, following the style of Figure 6. All 
components detected in \CIV\ in the nine QSOs within the redshift intervals available 
to each species are represented. Upper limits are $1\sigma$ values.}
\end{figure}

\clearpage
\begin{figure}
\figurenum{\scriptsize 8}
\epsscale{1.0}
%%\plotone{f8.eps}
\plotone{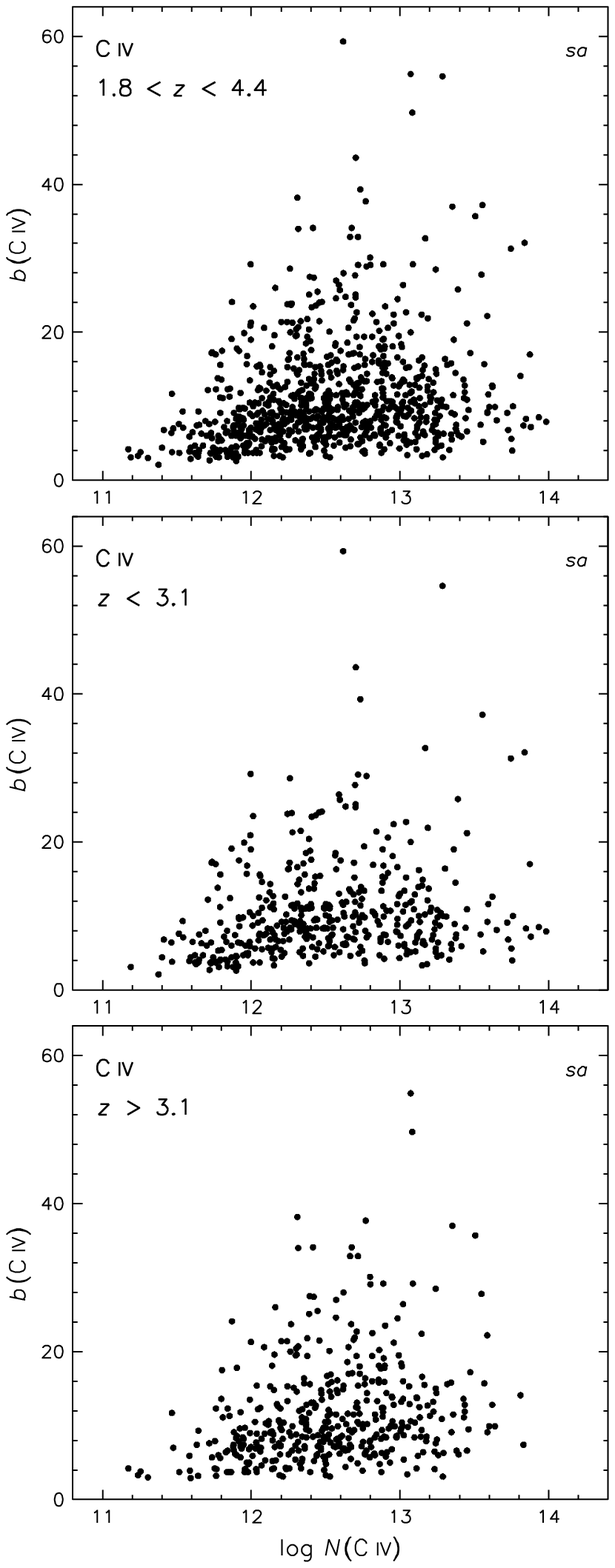}
\caption[f8.eps]{\scriptsize \CIV\ Doppler parameter {\it b} (km s$^{-1}$) 
versus column density $N$ (cm$^{-2}$) for all observed components of sample {\it sa}, 
also separately showing those with $z < 3.1$ and $z > 3.1$.}
\end{figure}

\clearpage
\begin{figure}
\figurenum{\scriptsize 9}
\epsscale{1.0}
\plotone{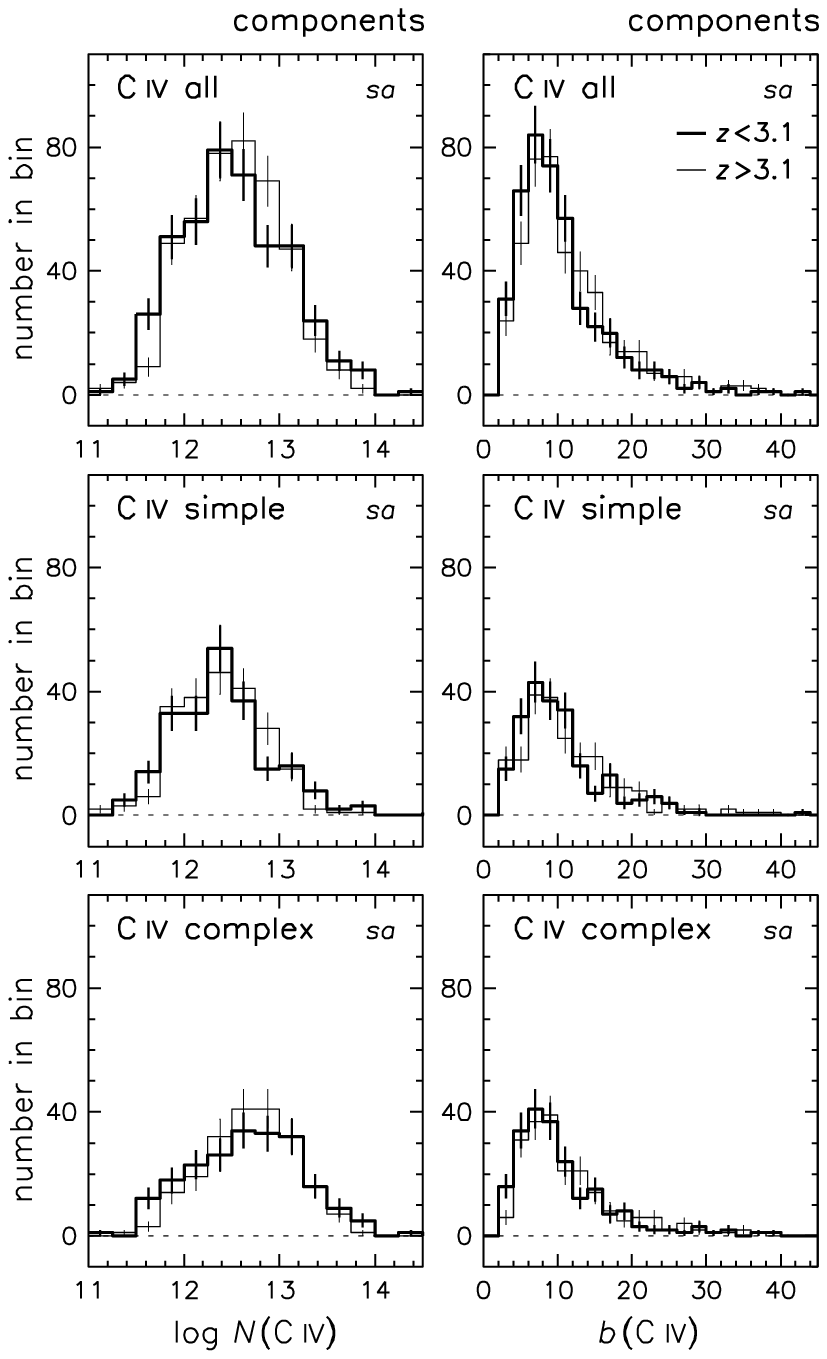}
\caption[f9.eps]{\scriptsize Histograms of \CIV\ column density 
$N$ (cm$^{-2}$) and Doppler parameter {\it b} (km s$^{-1}$) for all observed components 
of sample {\it sa}, comparing values for $z < 3.1$ ({\it thick lines}) and 
$z > 3.1$ ({\it thin lines}) and showing the data for {\it all} systems, {\it simple} 
systems ($\leqslant 6$ identified components) and {\it complex} systems ($\geqslant 7$ 
identified components).} 
\end{figure}

\clearpage
\begin{figure}
\figurenum{\scriptsize 10}
\epsscale{1.0}
\plotone{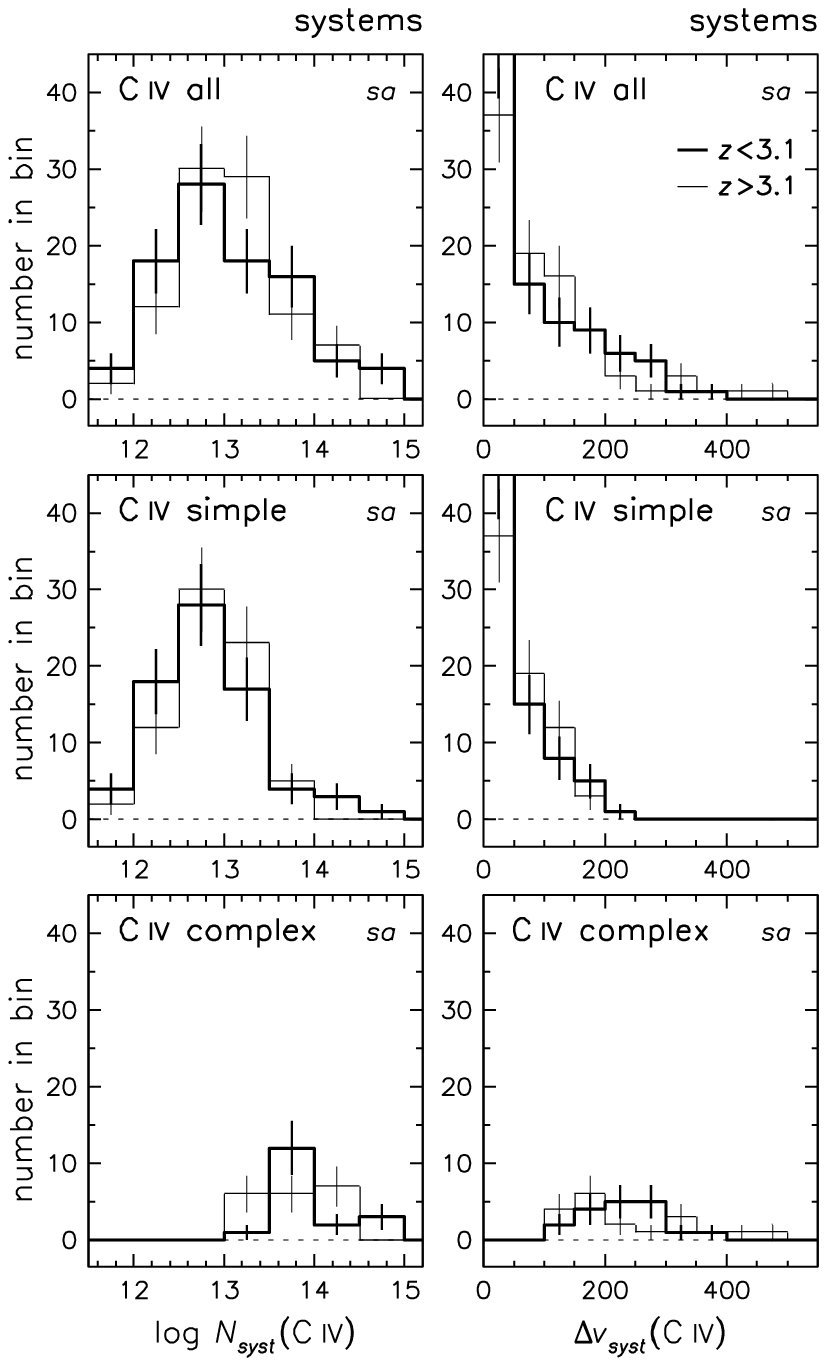}
\caption[f10.eps]{\scriptsize Histograms as in Figure 9, of system 
summed \CIV\ column density $N_{syst}$ (cm$^{-2}$) and overall velocity 
spread of the components within a system $\Delta v_{syst}$ (km s$^{-1}$).}
\end{figure}

\clearpage
\begin{figure}
\figurenum{\scriptsize 11}
\epsscale{1.0}
%%\plotone{f11.eps}
\plotone{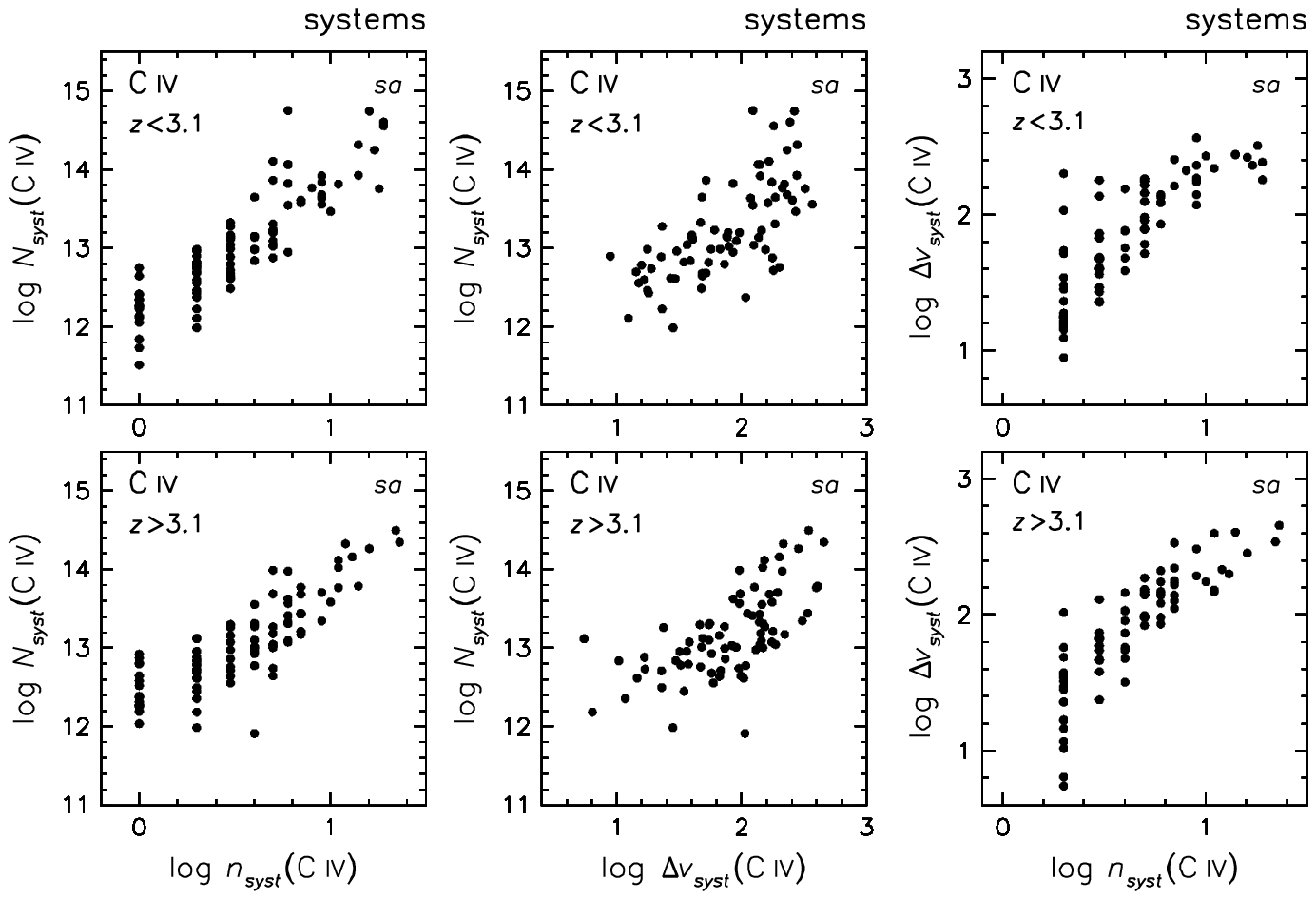}
\caption[f11.eps]{\scriptsize Relationships for all \CIV\ systems of
sample {\it sa} at redshifts $z < 3.1$ and $z > 3.1$, showing: {\it left panels}, 
system summed \CIV\ column density $N_{syst}$ (cm$^{-2}$) and number of 
components in the system $n_{syst}$; {\it middle panels}, $N_{syst}$ and overall 
velocity spread of system components $\Delta v_{syst}$ (km s$^{-1}$); {\it right 
panels}, $\Delta v_{syst}$ and $n_{\it syst}$.}
\end{figure}

\clearpage
\begin{figure}
\figurenum{\scriptsize 12}
\epsscale{1.0}
\plotone{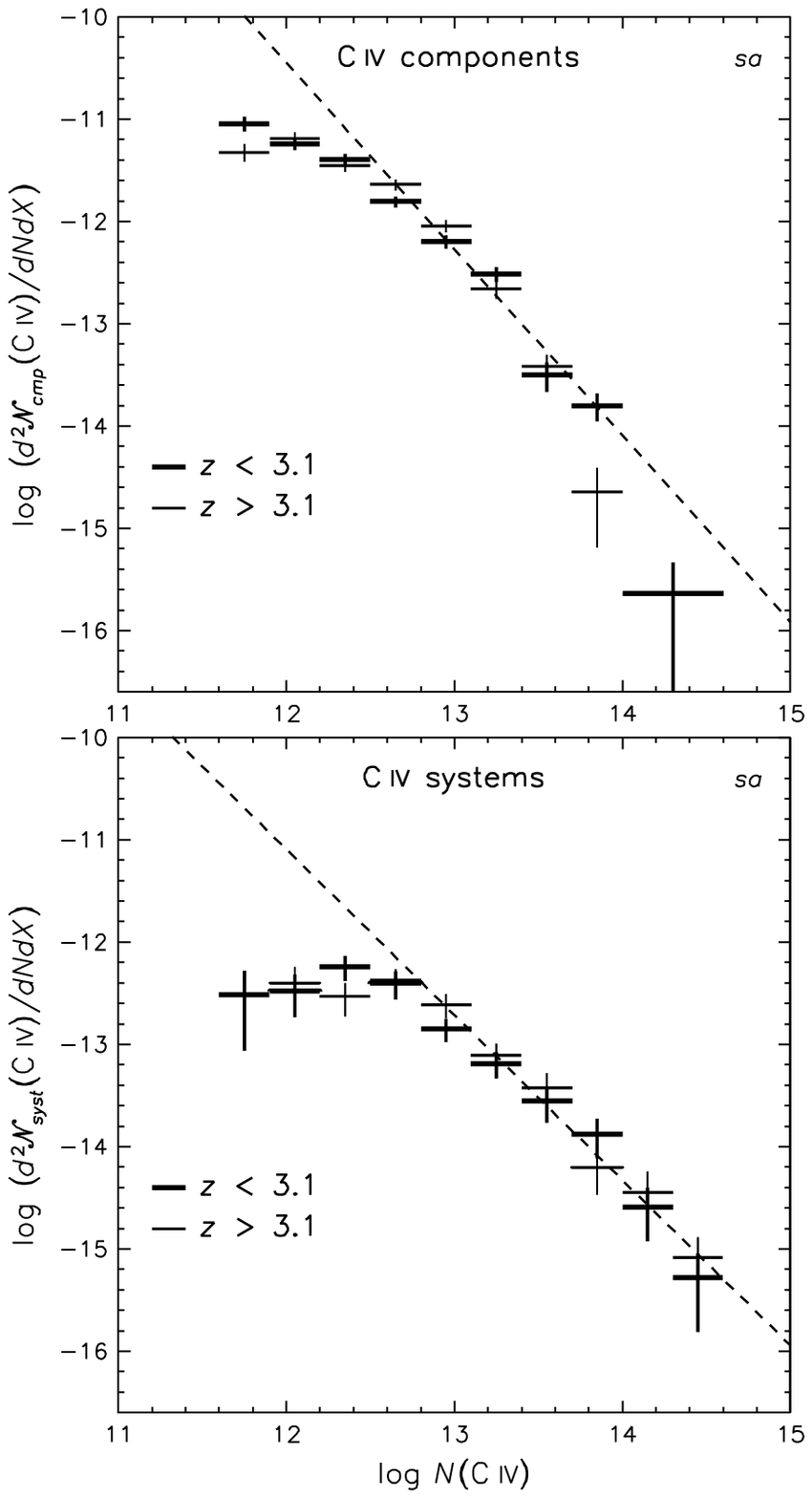}
\caption[f12.eps]{\scriptsize Differential column density distribution of 
\CIV\ components ({\it upper}) and systems ({\it lower}) for sample {\it sa} 
at redshifts $z < 3.1$ and $z > 3.1$. The bin size (shown by horizontal bars) is 
$10^{0.3}N$ (cm$^{-2}$) where $N$ is the column density; errors (vertical bars) are 
$\pm1\sigma$ values based on the number of absorbers ${\cal N}$ in each bin. The 
dashed lines are approximate power-law fits as described in the text with index 
$\beta = 1.84$ (components) and $\beta = 1.6$ (systems).}
\end{figure}

\clearpage
\begin{figure}
\figurenum{\scriptsize 13}
\epsscale{1.0}
%\epsscale{0.97}
\plotone{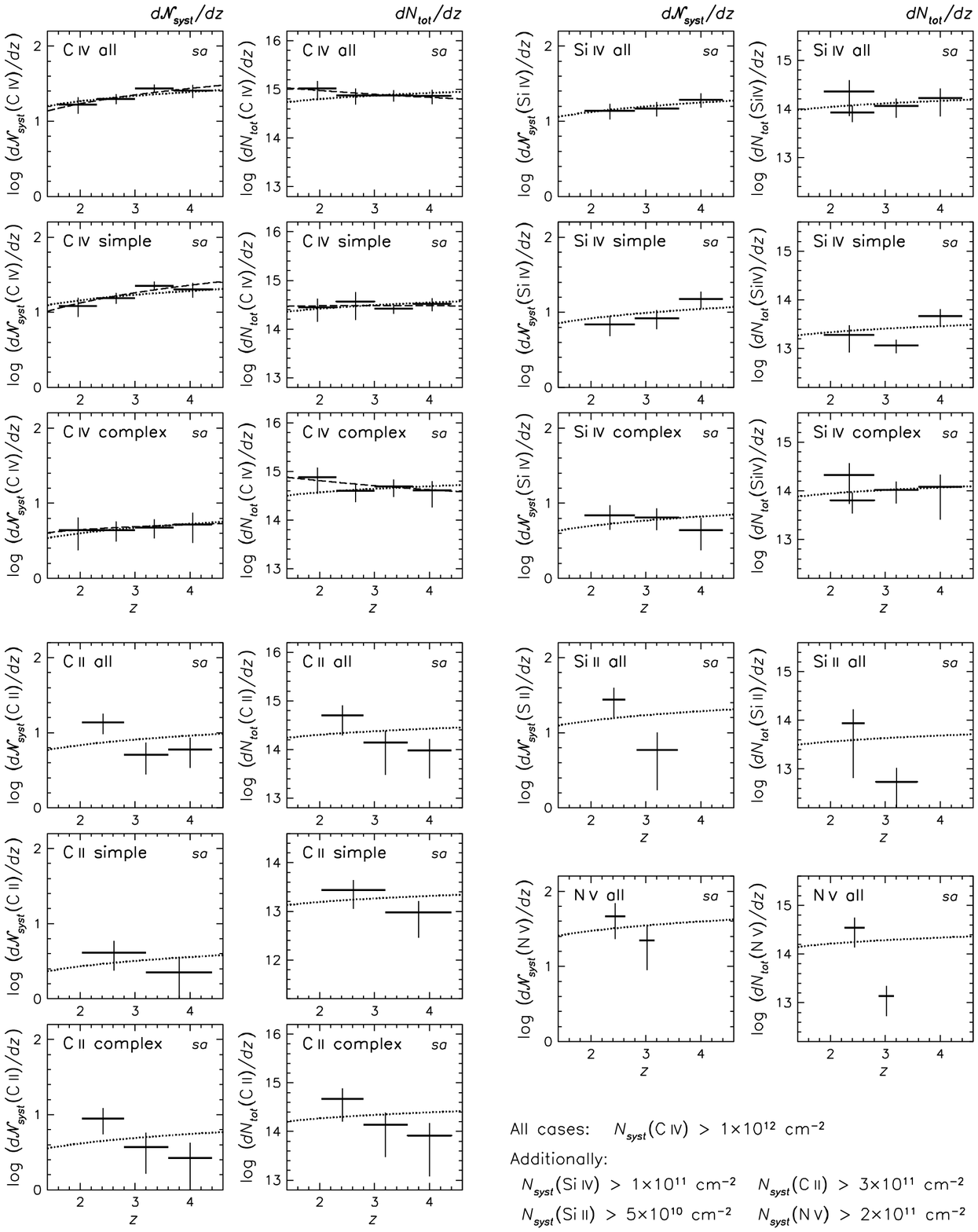}
\caption[f13.eps]{\scriptsize {\it Upper left panels}: Redshift evolution of 
\CIV\ total number of systems per unit redshift interval, $d{\cal N}_{syst}/dz$, 
and total column density (cm$^{-2}$) per unit redshift interval, $dN_{tot}/dz$, 
accounted over the redshift range shown by each bar, for systems in sample {\it sa} 
having $N_{syst}$(\CIV) $> 1 \times 10^{12}$ cm$^{-2}$, showing {\it all} 
systems, {\it simple} systems ($\leqslant 6$ \CIV\ components) and 
{\it complex} systems ($\geqslant 7$ \CIV\ components); for errors see text.
{\it Upper right and lower left panels}: Corresponding data for the same systems also 
having {\it detected} components in \SiIV\ and \CII\ above the indicated 
thresholds. {\it Lower right panels}: Similarly for {\it all} systems having 
{\it detected} components in \SiII\ and \NV. Note the various changes 
in vertical scale for $dN_{tot}/dz$. The dotted curves indicate non-evolving 
quantities (see text); dashed curves shown only for \CIV\ represent fits to the 
data.}
\end{figure}

\clearpage
\begin{figure}
\figurenum{\scriptsize 14}
\epsscale{1.0}
%%\plotone{f14.eps}
\plotone{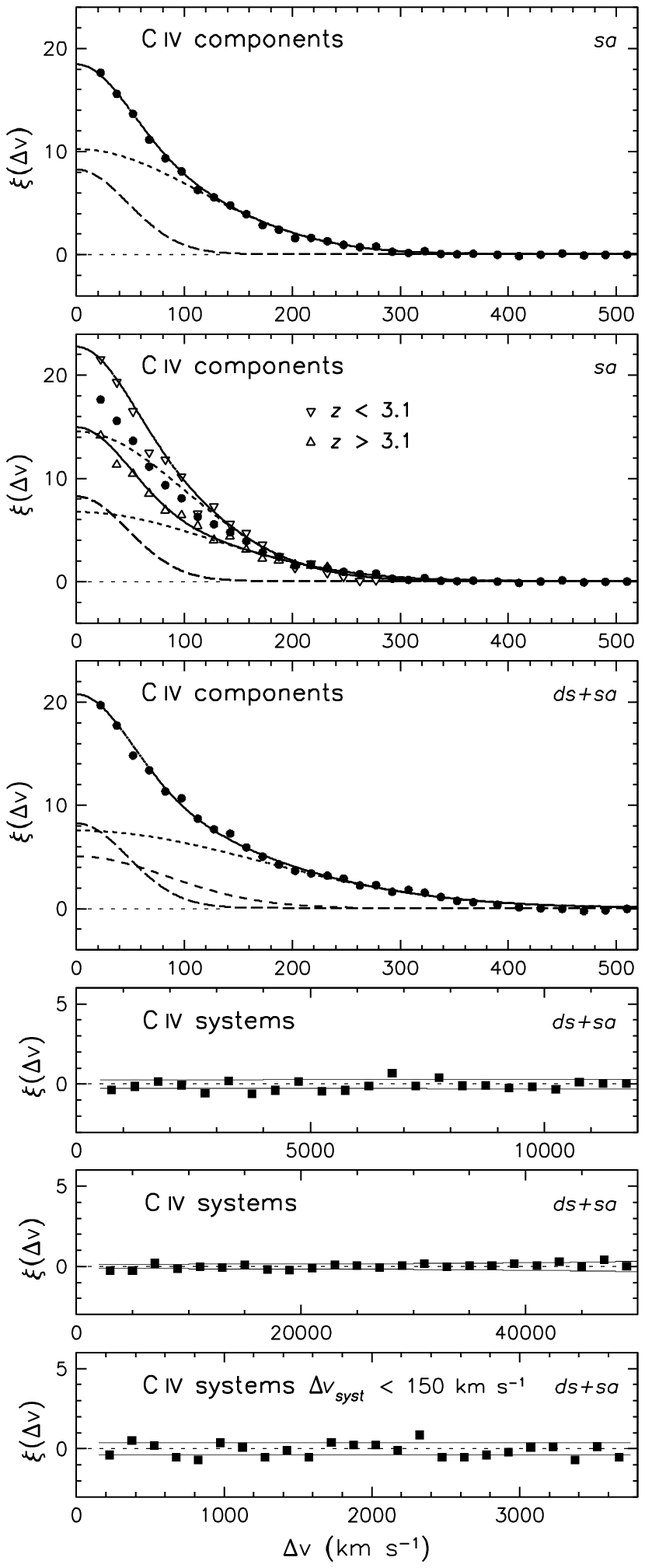}
\caption[f14.eps]{\scriptsize Two-point correlation functions 
$\xi(\Delta v)$ versus velocity separations $\Delta v$ for \CIV\ absorber 
redshifts spanning $1.6 \lesssim z \lesssim 4.4$. {\it Top panel:} Points give 
$\xi(\Delta v)$ for the individual components of sample {\it sa}, binned over 
15 km s$^{-1}$ for $\Delta v \leqslant 370$ km s$^{-1}$ and 20 km s$^{-1}$ for 
$\Delta v \geqslant 370$; $\pm1\sigma$ errors in the random distribution are smaller 
than the symbol size. A two-component Gaussian fit and the separate components of the 
fit are shown with parameters as given in the text. {\it Second panel:} Same as 
{\it top panel}, also showing results for subsets with $z < 3.1$ and $z > 3.1$. 
{\it Third panel:} Same as {\it top panel}, now adding all components of the 7 systems 
with significant Lyman~$\alpha$ damping wings and separated by 
$\gtrsim$ 3000 km s$^{-1}$ from the emission redshift, giving sample {\it ds$+$sa}; a 
three-component Gaussian fit is shown. {\it Fourth panel:} Result for the system 
redshifts of sample {\it ds$+$sa}, binned over 500 km s$^{-1}$ and extending to 
$\Delta v = 12000$ km s$^{-1}$; $\pm1\sigma$ errors in the random distribution are 
shown by bounding thin lines. {\it Fifth panel:} Same as {\it fourth panel} but binned 
over 2000 km s$^{-1}$ and extending to $\Delta v = 50000$ km s$^{-1}$. 
{\it Sixth panel:} Same as {\it fourth panel} but including only systems of velocity 
spread $\Delta v_{syst} < 150$ km s$^{-1}$, binned over 150 km s$^{-1}$ and extending 
to $\Delta v = 3800$ km s$^{-1}$.}
\end{figure}

\clearpage
\begin{figure}
\figurenum{\scriptsize 15}
\epsscale{1.0}
\plotone{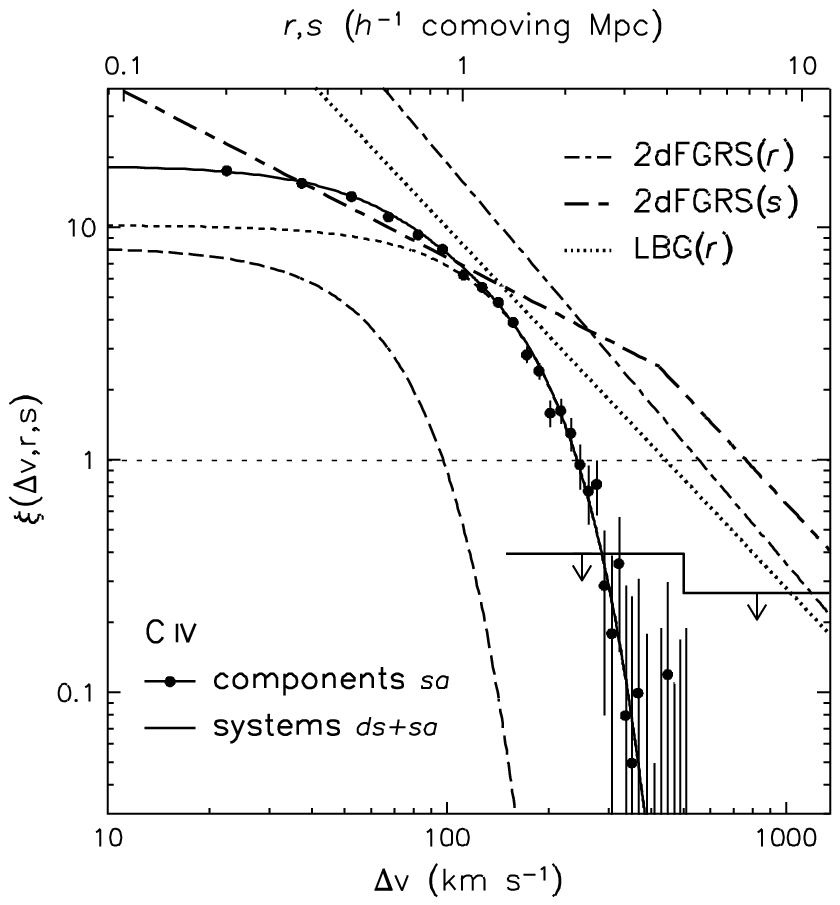}
\caption[f15.eps]{\scriptsize Comparison of \CIV\ absorber and galaxy 
two-point correlation functions in logarithmic form. The \CIV\ sample {\it sa} 
data for individual components are shown with the two-component Gaussian fit as in 
Figure 14; $\pm1\sigma$ errors in the random distributions are given with the data
points. Sample {\it ds+sa} data for systems as described in the text are shown here 
as upper limits using $+1\sigma$ errors from Figure 14. Fits to data from the 
2dF Galaxy Redshift Survey in real-space ($r$) and redshift-space ($s$) and for a 
sample of Lyman-break galaxies (LBG), all as described in the text, use the upper 
axis.}
\end{figure}

\clearpage
\begin{figure}
\figurenum{\scriptsize 16}
\epsscale{1.0}
%%\plotone{f16.eps}
\plotone{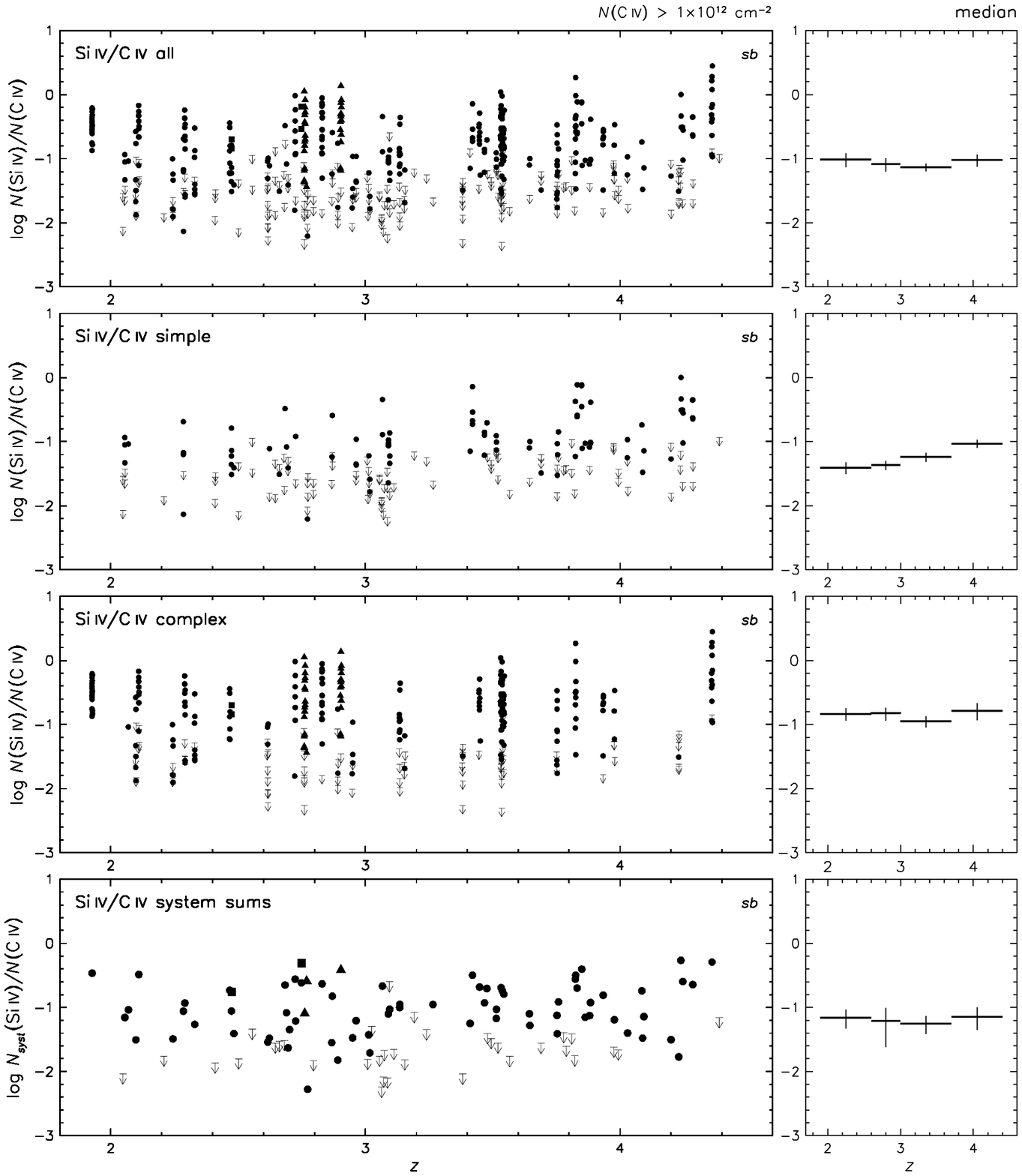}
\caption[f16.eps]{\scriptsize {\it Left panels}: Redshift evolution of 
\SiIV/\CIV\ column density ratios. Upper limits are $1\sigma$ values.
The {\it upper three panels} show individual component values having
$N$(\CIV) $> 1\times10^{12}$ cm$^{-2}$ from sample {\it sb} for 
{\it all} systems, {\it simple} systems ($\leqslant 6$ \CIV\ components) and 
{\it complex} systems ($\geqslant 7$ \CIV\ components). {\it Filled circles} 
show values obtained outside the Lyman forest; reliable values from lines in the 
Lyman forest are identified by {\it filled squares}; selected components clear of 
regions of high $N$(\HI) in systems containing mild Lyman~$\alpha$ damping 
wings are shown by {\it filled triangles}. The {\it bottom panel} gives values
obtained from {\it summed} column densities for all available \emph{systems} in sample 
{\it sb} having $N_{syst}$(\CIV) $> 1\times10^{12}$ cm$^{-2}$. {\it Right 
panels}: Redshift evolution of corresponding median values, obtained over the extent 
of each horizontal bar, indicated with $1\sigma$ error bars.}
\end{figure}

\clearpage
\begin{figure}
\figurenum{\scriptsize 17}
\epsscale{1.0}
%%\plotone{f17.eps}
\plotone{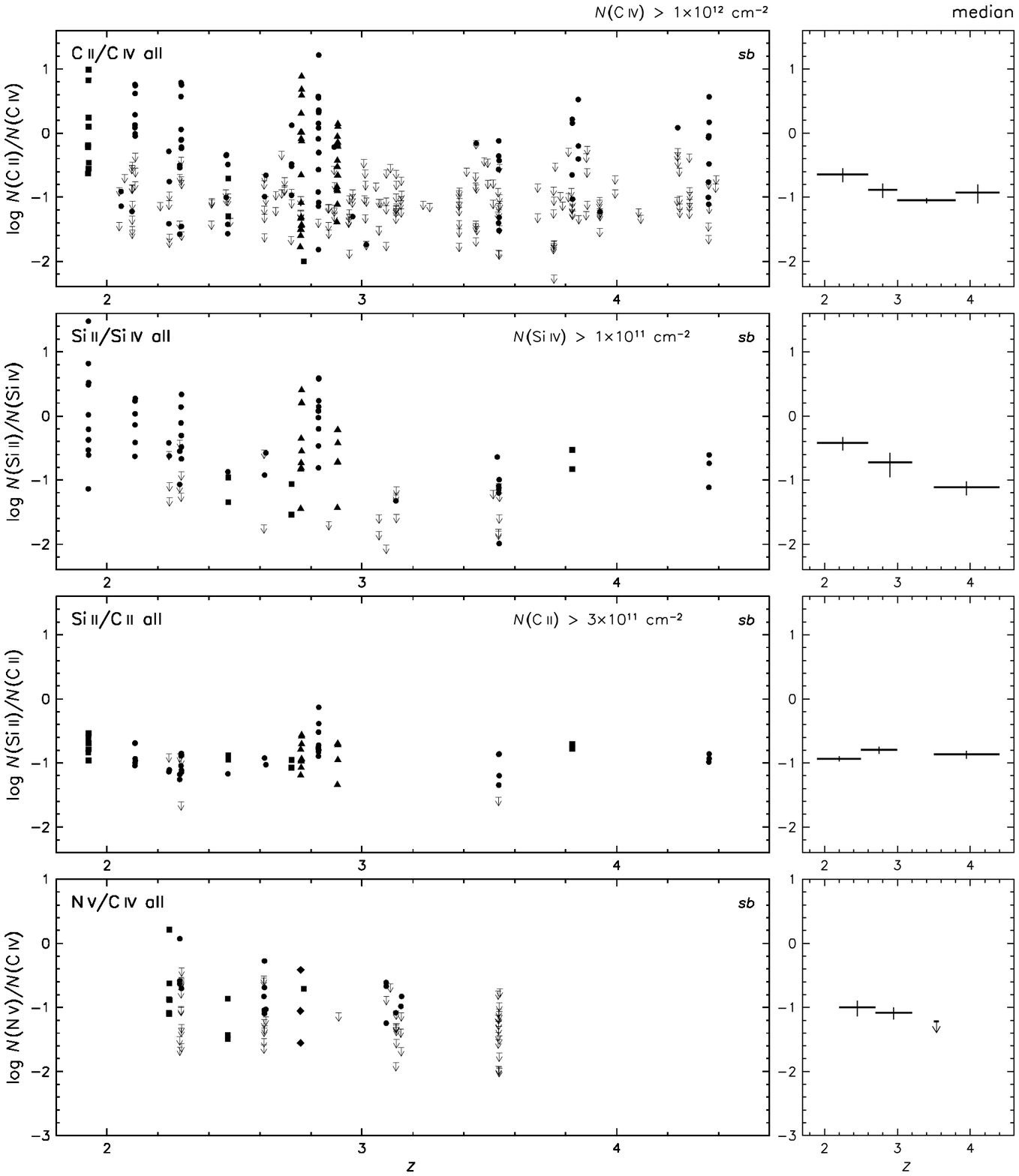}
\caption[f17.eps]{\scriptsize Same as for the {\it top} panels in 
Figure 16, here for the component column density ratios \CII/\CIV, \SiII/\SiIV\ 
(having {\it detected} \SiIV\ with $N$(\SiIV) $> 1 \times 10^{11}$ cm$^{-2}$), 
\SiII/\CII\ (having {\it detected} \CII\ with $N$(\CII) $> 1 \times 10^{11}$ cm$^{-2}$) 
and \NV/\CIV\ ({\it filled diamonds} indicate values from components that are both in 
the forest and selected from mildly damped systems). In all cases components have 
$N$(\CIV) $> 1\times10^{12}$ cm$^{-2}$.} 
\end{figure}

\clearpage
\begin{figure}
\figurenum{\scriptsize 18}
\epsscale{1.0}
%%\plotone{f18.eps}
\plotone{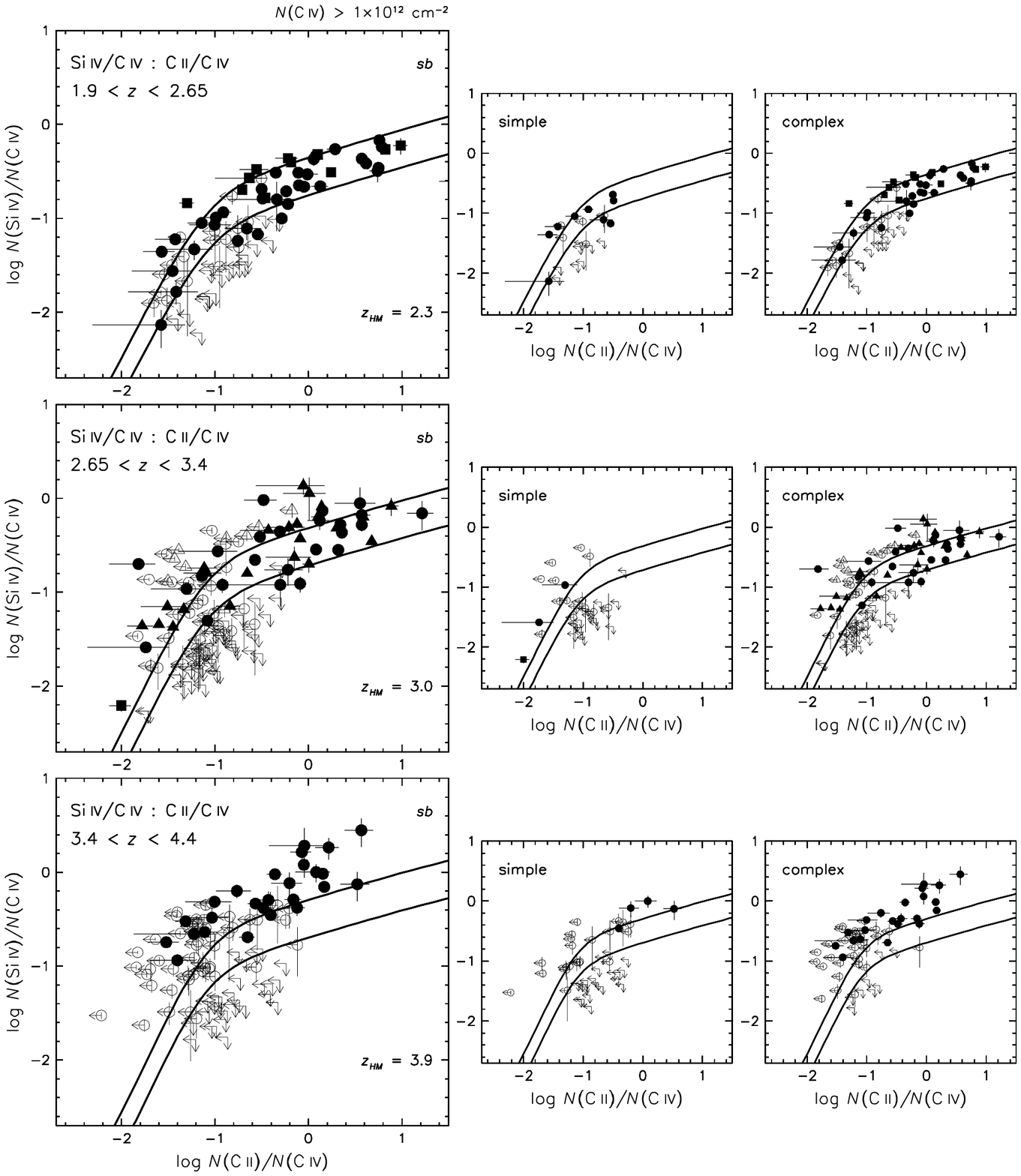}
\caption[f18.eps]{\scriptsize {\it Left panels}: Column density ratios 
\SiIV/\CIV~:~\CII/\CIV\ in three redshift intervals for {\it all} components having 
$N$(\CIV) $> 1\times10^{12}$ cm$^{-2}$ in sample {\it sb}. {\it Filled symbols} are 
defined in Figure 16. Error bars give $\pm1\sigma$ uncertainties; upper limit arrows 
point from $+1\sigma$ values. Cases where one of the two ionic ratios is an upper limit 
are indicated by an {\it open} symbol. The {\it thick lines} give model predictions of 
the Cloudy code for absorbers optically thin in the \HI\ Lyman continuum 
($N$(\HI) $= 1\times10^{15}$ cm$^{-2}$), low metallicity (0.003 $\times$ solar) and 
Si/C relative abundance values of solar and 2.5 $\times$ solar (see text), computed at 
appropriate redshifts $z_{HM}$ for Haardt \& Madau latest available versions of the 
metagalactic ionizing radiation background with the QSO contribution alone (model 
Q---see text for a full description); 
$J_{\nu_0} = 3.5 \times 10^{-22}$ erg s$^{-1}$ cm$^{-2}$ Hz$^{-1}$ sr$^{-1}$ at 
$z_{HM} = 2.3$, $2.5 \times 10^{-22}$ at $z_{HM} = 3.0$ and $1.6 \times 10^{-22}$ at 
$z_{HM} = 3.9$. The cosmic microwave background is included. 
{\it Right panels}: Subsets of {\it simple} systems ($\leqslant 6$ \CIV\ components) 
and {\it complex} systems ($\geqslant 7$ \CIV\ components).}  
\end{figure}

\clearpage
\begin{figure}
\figurenum{\scriptsize 19}
\epsscale{1.0}
%%\plotone{f19.eps}
\plotone{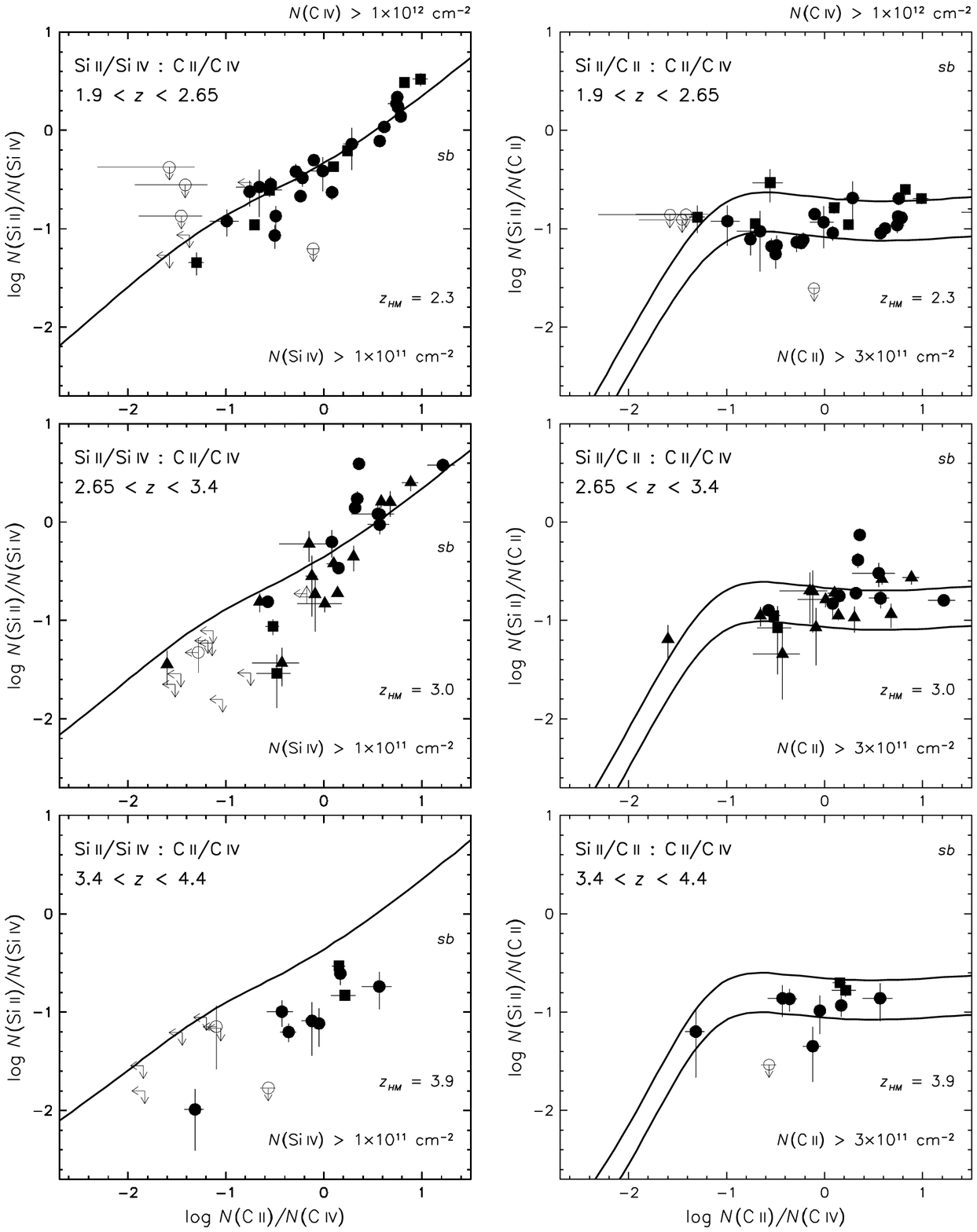}
\caption[f19.eps]{\scriptsize {\it Left panels}: Same as for {\it left} 
panels in Figure 18, here for \SiII/\SiIV~:~\CII/\CIV\ and additionally using only 
{\it detected} \SiIV\ components having $N$(\SiIV) $> 1\times10^{11}$ cm$^{-2}$. 
{\it Right panels}: Same for \SiII/\CII~:~\CII/\CIV, but using only {\it detected} 
\CII\ components having $N$(\CII) $> 3\times10^{11}$ cm$^{-2}$. Both Si/C relative 
abundance values used in Figure 18 are shown.}
\end{figure}

\clearpage
\begin{figure}
\figurenum{\scriptsize 20}
\epsscale{1.0}
\plotone{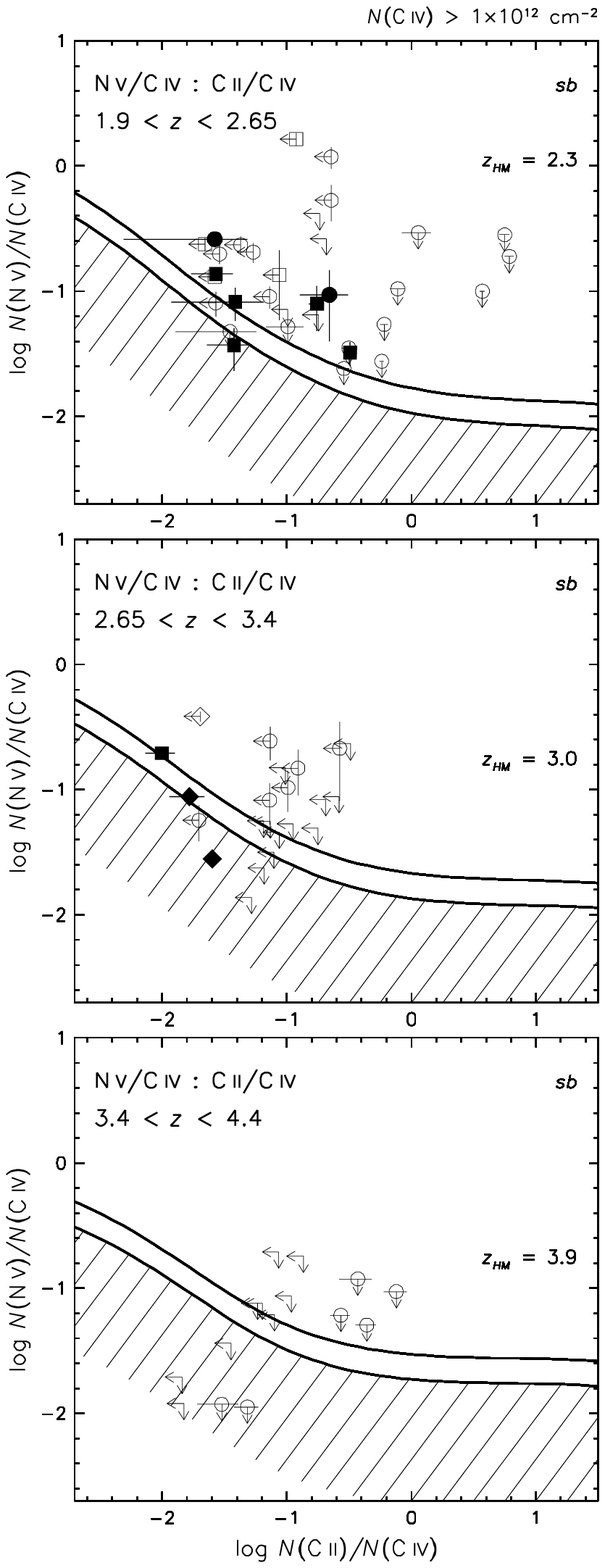}
\caption[f20.eps]{\scriptsize Same as {\it left} panels in Figure 18, here 
for \NV/\CIV~:~\CII/\CIV\ and N/C relative abundance values of solar and 
0.63 $\times$ solar with shading for the latter indicating the possible range extending 
lower by 1 dex (see text).} 
\end{figure}

\clearpage
\begin{figure}
\figurenum{\scriptsize 21}
\epsscale{1.0}
%%\plotone{f21.eps}
\plotone{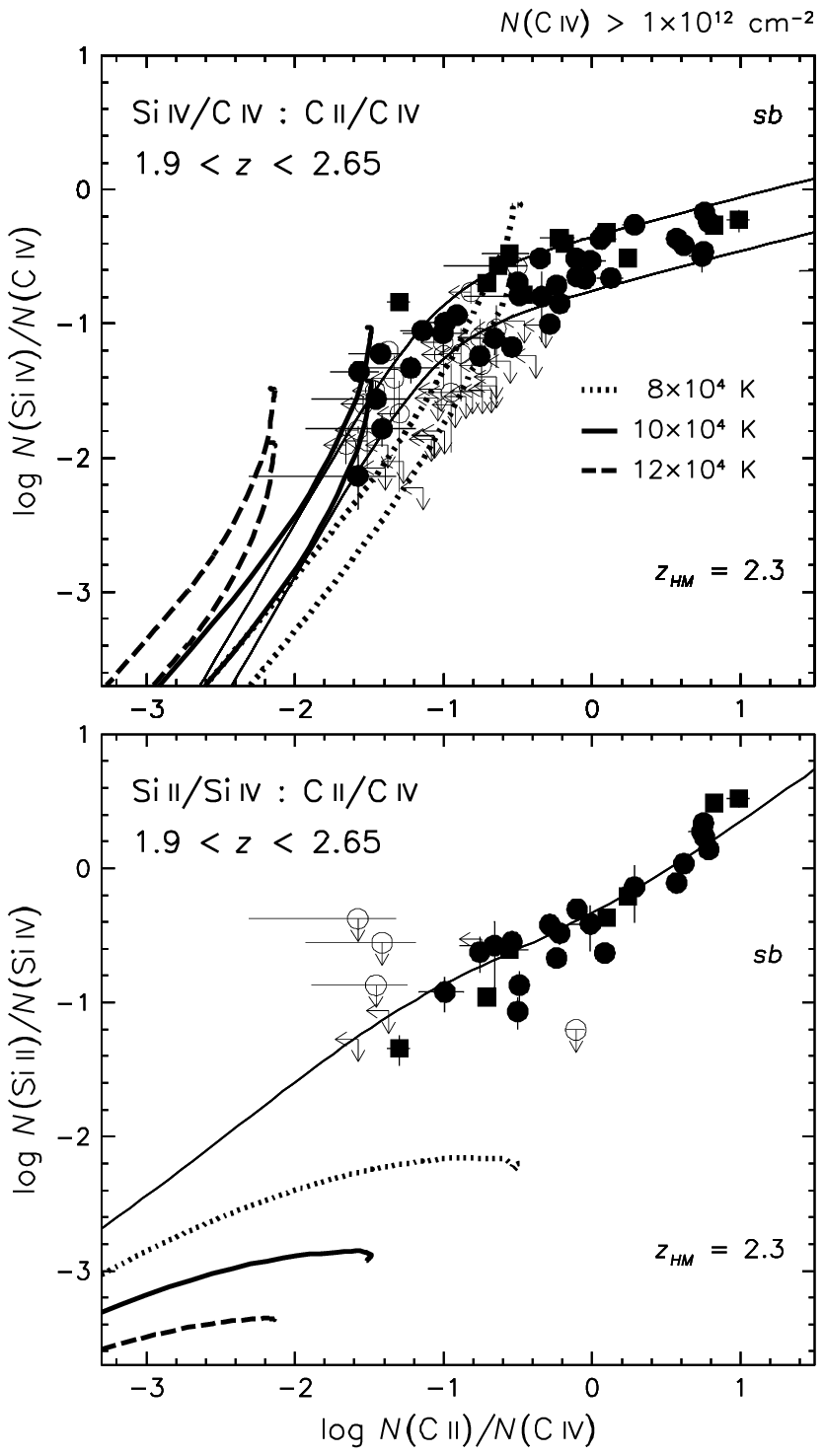}
\caption[f21.eps]{\scriptsize \SiIV/\CIV~:~\CII/\CIV\ and \SiII/\SiIV~:~\CII/\CIV\ as 
in the top panels of Figures 18 and 19, with extended axes. The curves show Cloudy 
models representing collisional ionization at fixed temperatures near $10^5$ K 
(indicated in the {\it upper} panel) in the presence of the latest available 
Haardt \& Madau pure QSO version of the metagalactic ionizing radiation background 
(model Q---see text) at $z_{HM} = 2.3$; for reference, models in photoionization 
equilibrium as in Figures 18 and 19 are shown here in {\it continuous lines}. Both 
values of Si/C relative abundance are included as in the previous figures. The 
collisional ionization curves terminate in the diagram where the process becomes 
independent of density.} 
\end{figure}

\clearpage
\begin{figure}
\figurenum{\scriptsize 22}
\epsscale{1.0}
\plotone{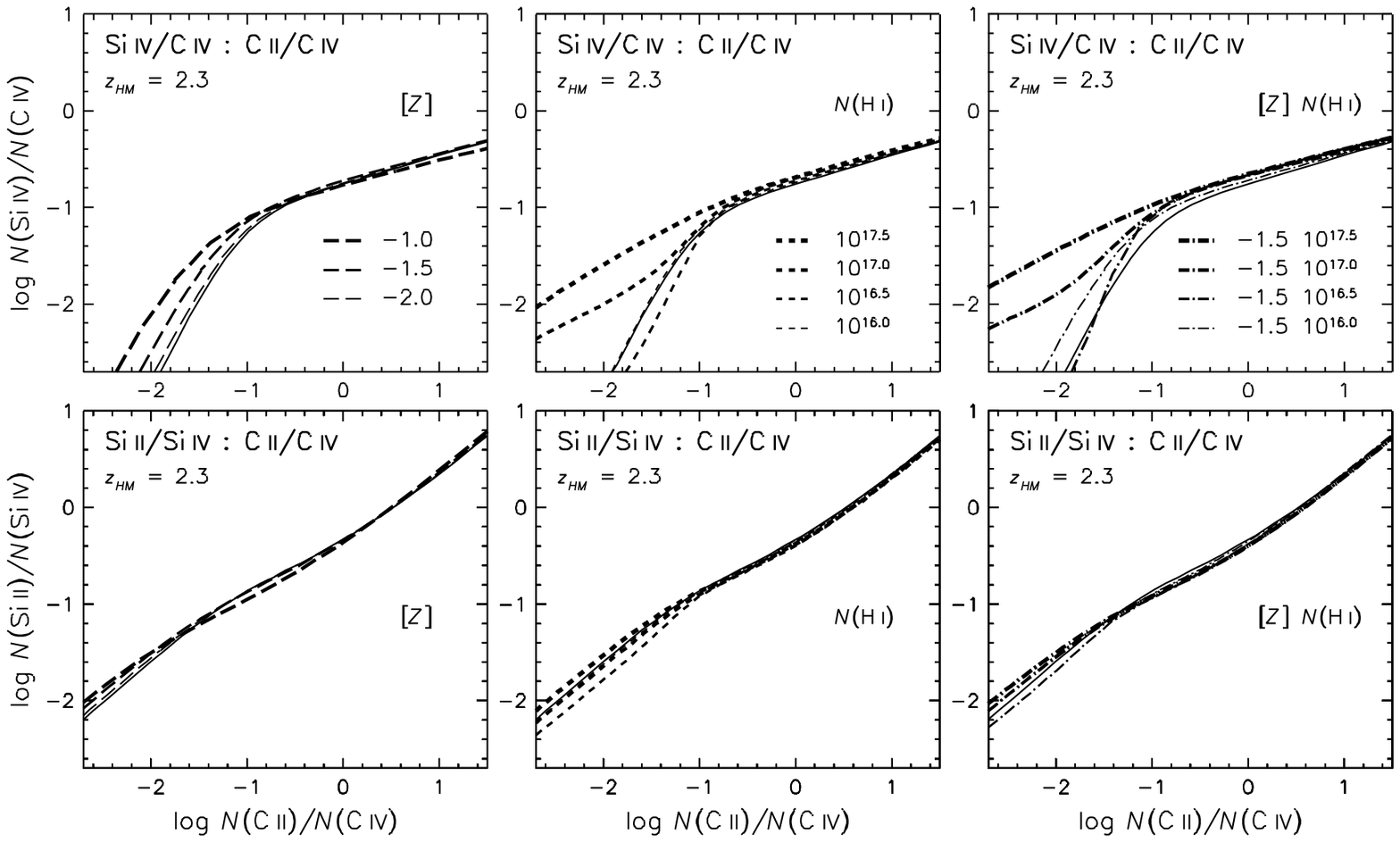}
\caption[f22.eps]{\scriptsize \SiIV/\CIV~:~\CII/\CIV\ and \SiII/\SiIV~:~\CII/\CIV\ 
Cloudy models in photoionization equilibrium using the latest available Haardt \& Madau 
pure QSO version of metagalactic ionizing radiation background (model Q---see text) at 
$z_{HM} = 2.3$ with absorber parameters arbitrarily differing in metallicity, 
[$Z$] $= -2.0$ to $-1.0$, and hydrogen column density, 
$N$(\HI) $= 10^{16}$ to $10^{17.5}$ cm$^{-2}$, as indicated; the nominal case with 
[$Z$] $= -2.5$ and $N$(\HI) $=10^{15.0}$ cm$^{-2}$ used in Figures 18 and 19 is shown 
here in {\it continuous lines}. For clarity only the solar Si/C relative abundance 
value is included.} 
\end{figure}

\clearpage
\begin{figure}
\figurenum{\scriptsize 23}
\epsscale{1.0}
\plotone{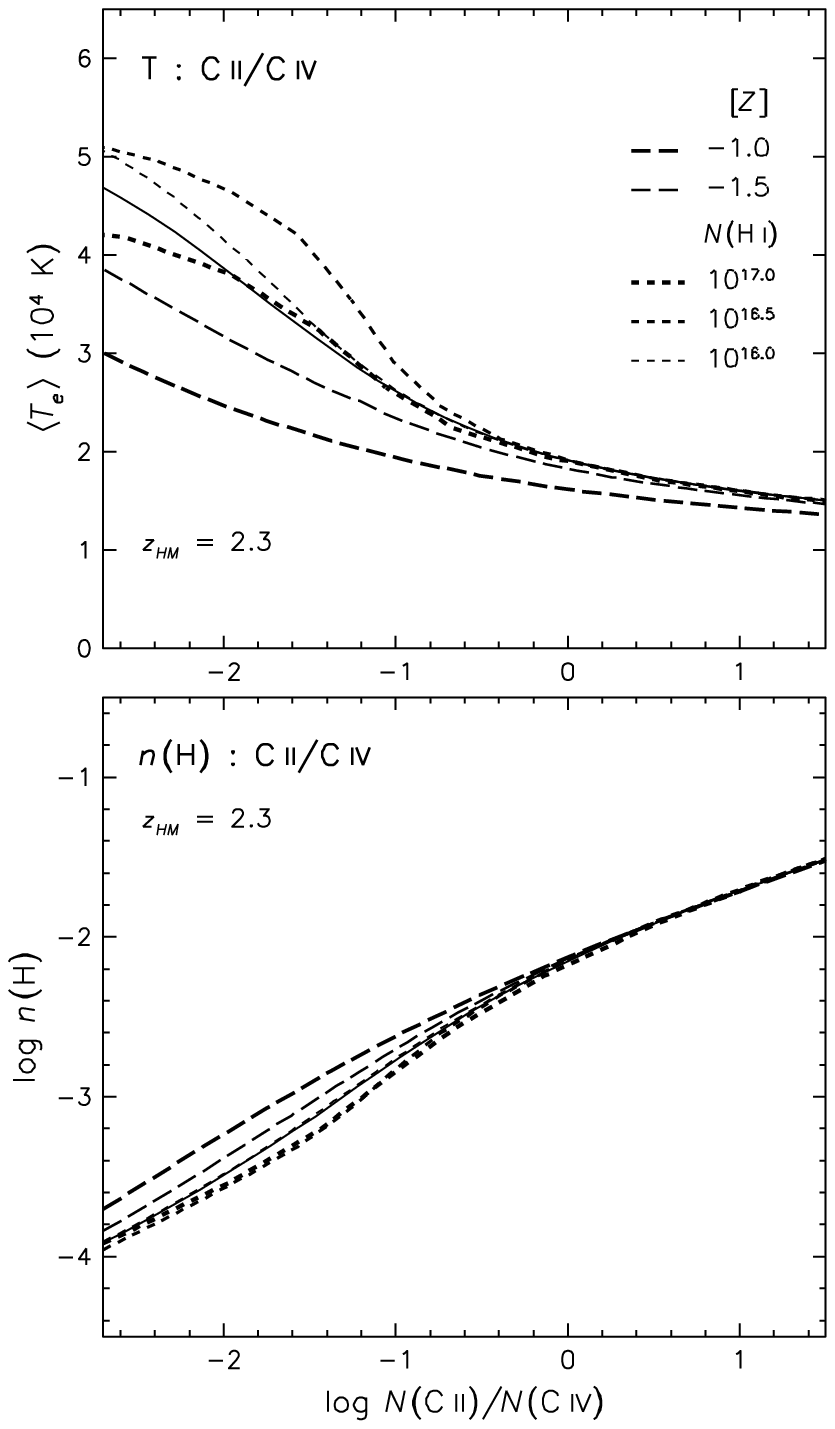}
\caption[f23.eps]{\scriptsize Photoionization equilibrium mean column temperature, 
$\langle T_e \rangle$ (K), and total hydrogen volume density, $n$(H) (cm$^{-3}$), 
versus \CII/\CIV, for the Cloudy-modelled case using the latest available 
Haardt \& Madau pure QSO version of metagalactic ionizing radiation background (model 
Q---see text) at $z_{HM} = 2.3$ with absorber parameters arbitrarily differing in 
metallicity [$Z$] and hydrogen column density $N$(\HI) as indicated, compared with the 
nominal case with [$Z$] $= -2.5$ and $N$(\HI) $=10^{15.0}$ cm$^{-2}$ shown in 
{\it continuous lines}.}
\end{figure}

\clearpage
\begin{figure}
\figurenum{\scriptsize 24}
\epsscale{1.0}
\plotone{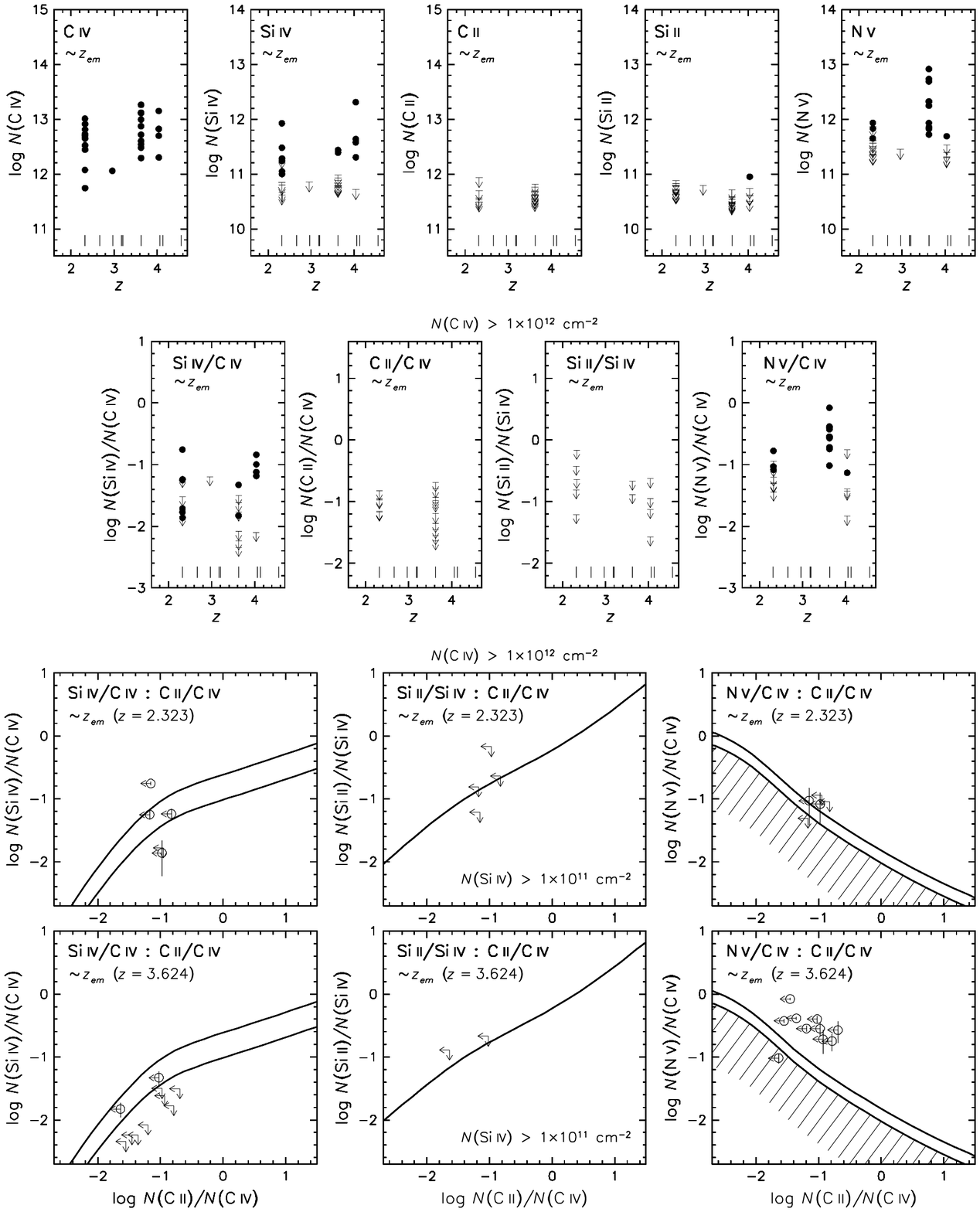}
\caption[f24.eps]{\scriptsize Systems with redshift close to the emission
redshift {\it z}$_{em}$ of their sightline QSO, (indicated by {\it vertical ticks}).
{\it Top panels}: \CIV, \SiIV, \CII, \SiII\ and \NV\ component column densities 
(cm$^{-2}$). {\it Middle panels}: Component ($N$(\CIV) $> 1\times10^{12}$ cm$^{-2}$) 
column density ratios \SiIV/\CIV, \CII/\CIV, \SiII/\SiIV\ 
($N$(\SiIV) $> 1\times10^{11}$ cm$^{-2}$) and \NV/\CIV. {\it Bottom set of 
panels}: Column density ratio combinations for components 
($N$(\CIV) $> 1\times10^{12}$ cm$^{-2}$) of the systems at $z = 2.323$ in Q1626$+$6433 
and $z = 3.624$ in Q1422$+$2309C compared with Cloudy results as in Figures 18--20, 
but using radiation with a power-law spectrum of index $-1.8$ having
$J_{\nu_0} = 3.5 \times 10^{-21}$ erg s$^{-1}$ cm$^{-2}$ Hz$^{-1}$ sr$^{-1}$; the 
cosmic microwave background is included.}
\end{figure}

\clearpage
\begin{figure}
\figurenum{\scriptsize 25}
\epsscale{1.0}
\plotone{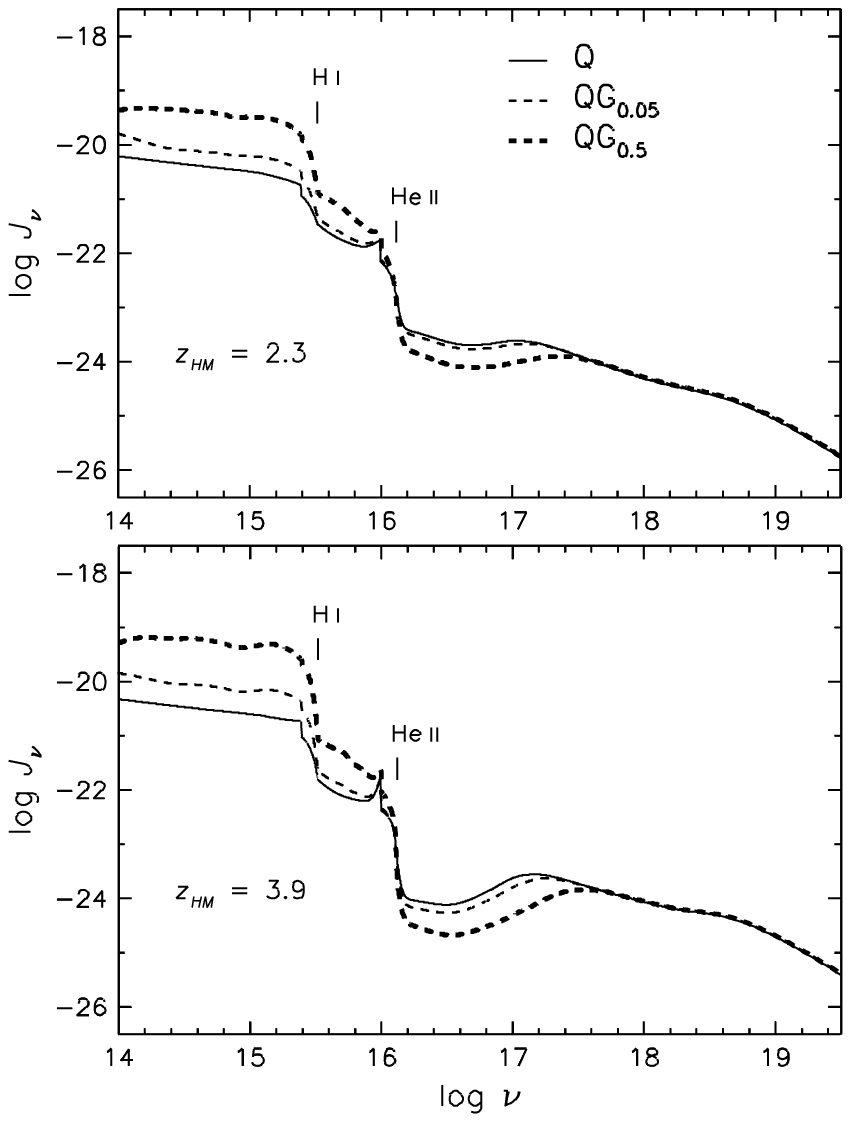}
\caption[f25.eps]{\scriptsize Rest spectral energy distributions plotted as 
mean intensity $J_{\nu}$ (erg cm$^{-2}$ s$^{-1}$ Hz$^{-1}$ sr$^{-1}$) over frequency 
$\nu$ (Hz), obtained from new Haardt \& Madau metagalactic ionizing radiation 
background models having contributions both from QSOs and galaxies for the two 
values $f_{\rm esc} = 0.05$ and 0.5 (see text), termed models QG$_{0.05}$ and 
QG$_{0.5}$, at redshifts $z_{HM} = 2.3$ and 3.9. The pure QSO case as used in 
Figures 18--20, model Q, is shown for comparison. The values of $J_{\nu_0}$ in the 
models containing galaxy contributions are respectively: $4.3 \times 10^{-22}$ and 
$1.2 \times 10^{-21}$ erg s$^{-1}$ cm$^{-2}$ Hz$^{-1}$ sr$^{-1}$ at $z_{HM} = 2.3$; 
$2.2 \times 10^{-22}$ and 
$8.4 \times 10^{-22}$ erg s$^{-1}$ cm$^{-2}$ Hz$^{-1}$ sr$^{-1}$ at $z_{HM} = 3.9$. The 
positions of the ionization thresholds for \HI\ and \HeII, at 1 and 4 
Rydberg, are indicated.}
\end{figure}

\clearpage
\begin{figure}
\figurenum{\scriptsize 26}
\epsscale{1.0}
\plotone{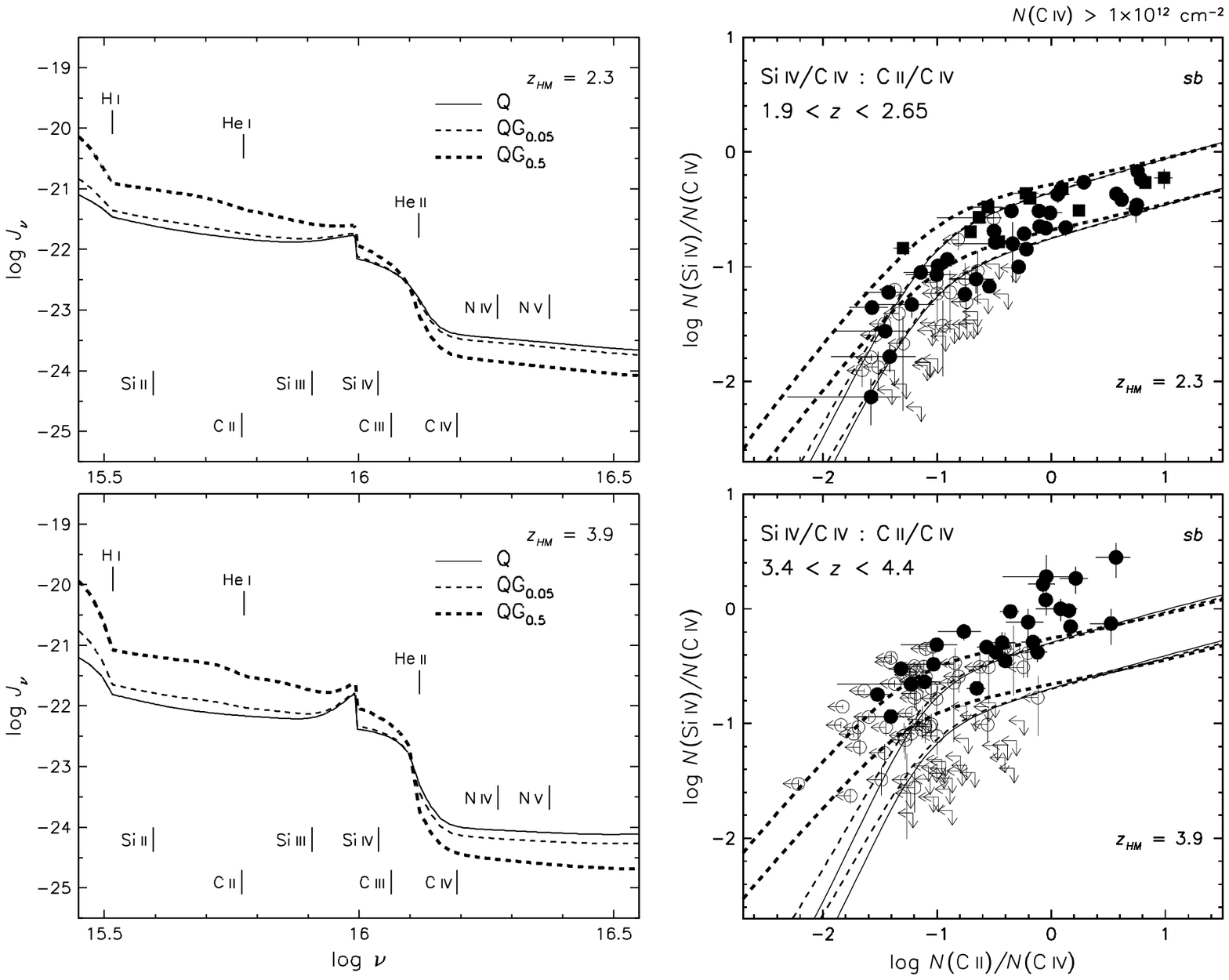}
\caption[f26.eps]{\scriptsize {\it Left panels}: Rest spectral energy 
distributions for the new Haardt \& Madau metagalactic ionizing radiation models 
QG$_{0.05}$ and QG$_{0.5}$ defined in Figure 25 but here concentrating on the region 
containing the ionization thresholds for H, He, C, Si and N significant for this paper. 
{\it Right panels}: Observed column density ratios \SiIV/\CIV~:~\CII/\CIV\ in two 
redshift intervals taken from Figure 18 compared with Cloudy results using models 
QG$_{0.05}$ and QG$_{0.5}$ (coding as indicated in the {\it left} panels) with absorber 
parameters as defined in Figure 18. The cosmic microwave background at the two 
redshifts is included to account for Compton cooling.}
\end{figure}

\clearpage
\begin{figure}
\figurenum{\scriptsize 27}
\epsscale{1.0}
\plotone{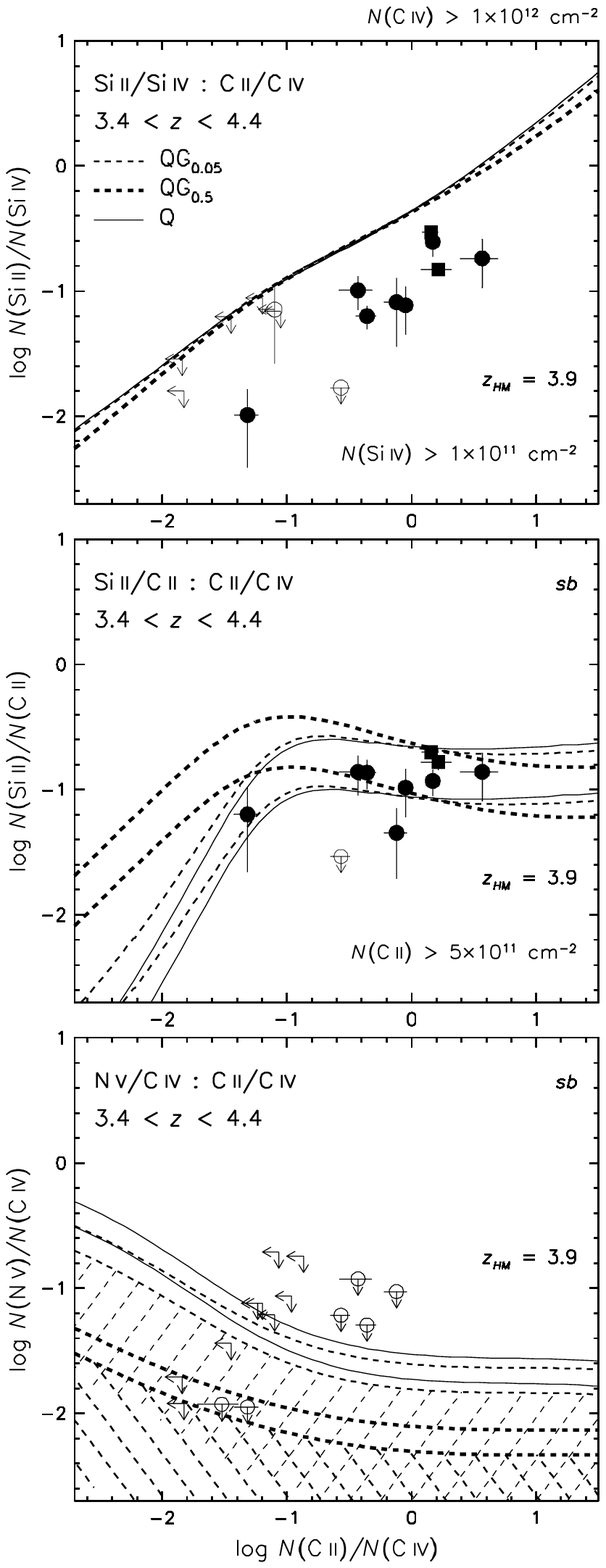}
\caption[f27.eps]{\scriptsize Same as for {\it lower right} 
panel in Figure 26, here for \SiII/\SiIV~:~\CII/\CIV, \SiII/\CII~:~\CII/\CIV\ and 
\NV/\CIV~:~\CII/\CIV, with the data values taken from Figures 19 and 20. The N/C 
relative abundance ranges as given in Figure 20 are also shown here but to avoid 
confusion the shading indicating possible lower values by up to 1 dex is omitted for 
model Q.}
\end{figure}

\clearpage
\begin{figure}
\figurenum{\scriptsize 28}
\epsscale{1.0}
\plotone{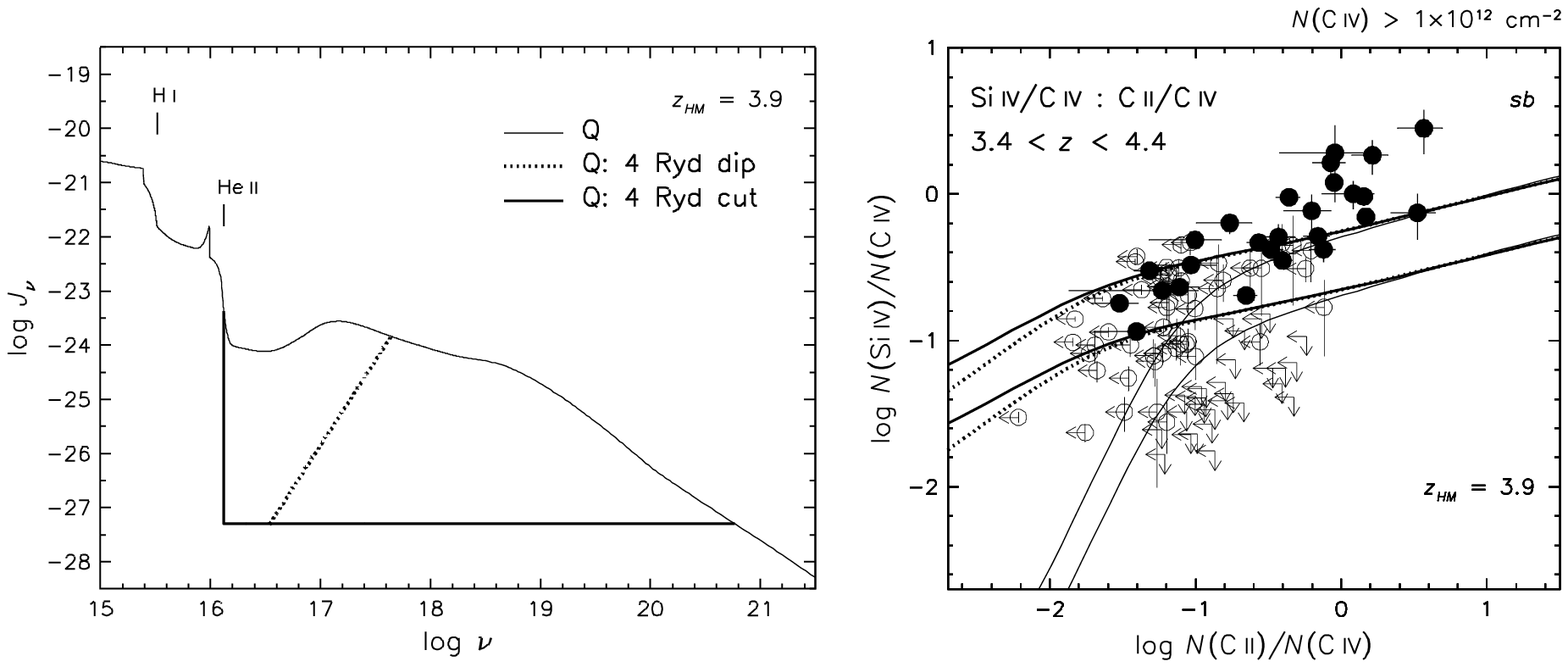}
\caption[f28.eps]{\scriptsize {\it Left panel}: Spectral energy 
distributions plotted similarly to those in Figure 25, for the QSO metagalactic 
ionizing radiation background model Q at $z_{HM} = 3.9$ with post-computation 
modifications in the \HeII\ continuum contrived for two cases, a horizontal cut and a 
deep depression, both initiated with a drop of 4 dex at the \HeII\ ionization edge. 
{\it Right panel}: Column density ratios \SiIV/\CIV~:~\CII/\CIV\ in our highest 
redshift interval compared with Cloudy results, presented as in Figure 18 but here 
using model Q with the modifications in the {\it left} panel and shown with the same 
coding; the unmodified case is shown for comparison. The cosmic microwave background is 
included.} 
\end{figure}

\clearpage
\begin{figure}
\figurenum{\scriptsize 29}
\epsscale{1.0}
\plotone{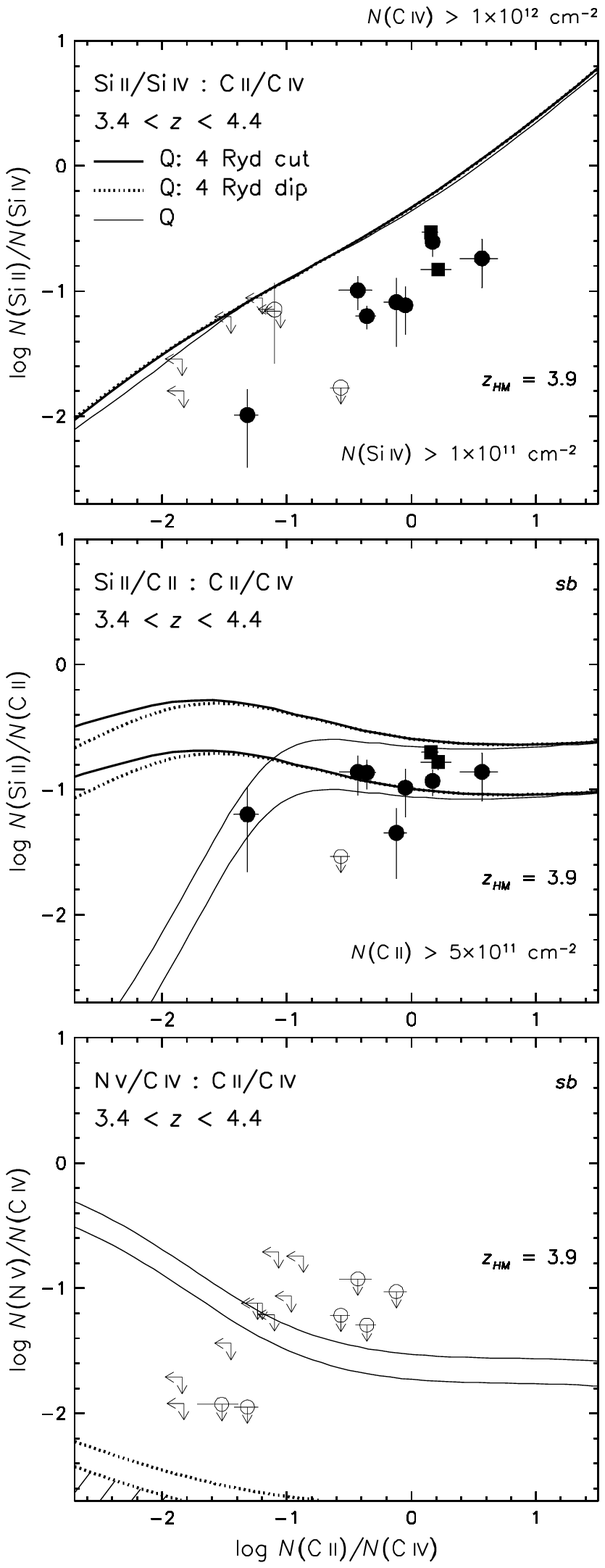}
\caption[f29.eps]{\scriptsize Same as for {\it right} panel 
in Figure 28, here for \SiII/\SiIV~:~\CII/\CIV, \SiII/\CII~:~\CII/\CIV\ and 
\NV/\CIV~:~\CII/\CIV (see comment in Figure 27 regarding N/C relative abundance, and 
shading), with the observed ratios taken from Figures 19 and 20.} 
\end{figure}

\clearpage
\begin{figure}
\figurenum{\scriptsize 30}
\epsscale{1.0}
\plotone{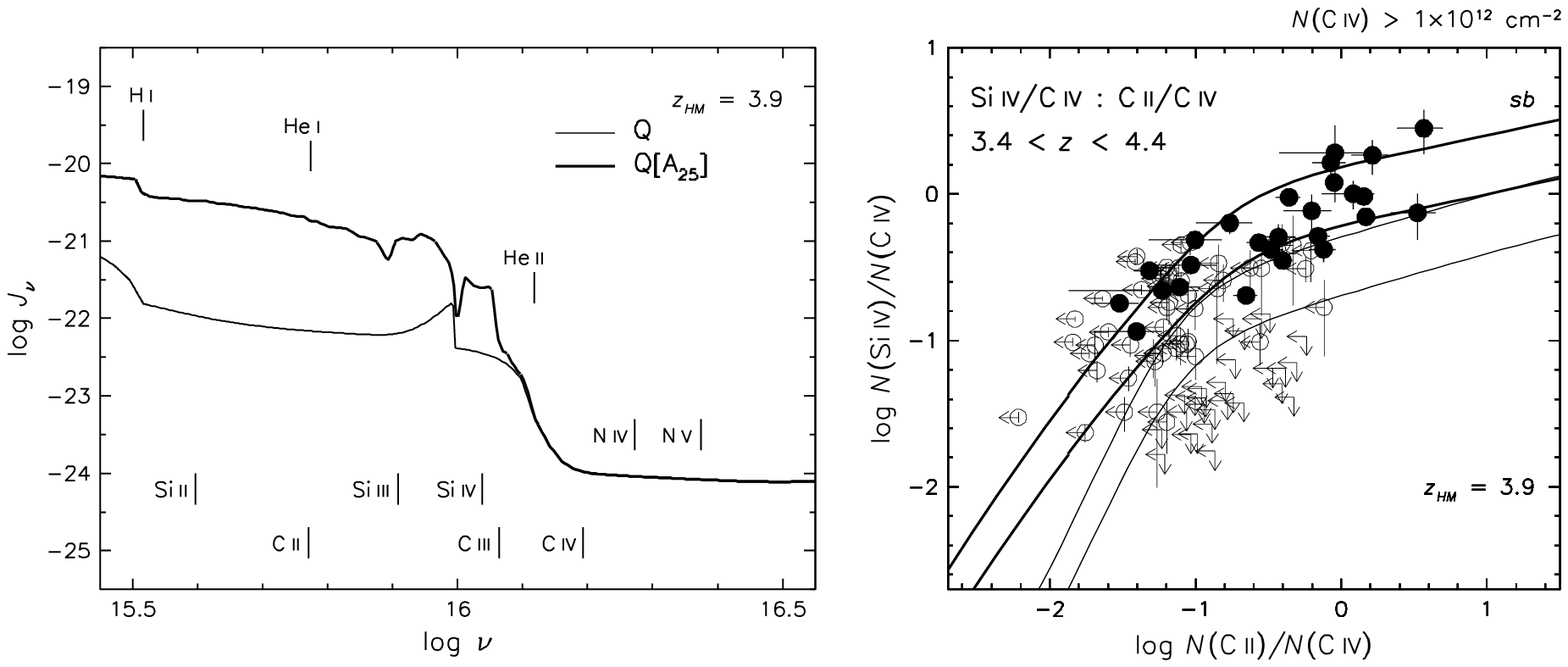}
\caption[f30.eps]{\scriptsize {\it Left panel}: Spectral energy 
distribution as presented in Figure 26, here using the new Haardt \& Madau 
metagalactic ionizing radiation background model containing only QSO sources, 
model Q, combined with a contribution from {\it local} sources represented by a 
specific 45,000 K stellar spectral model (see text) scaled at the \HI\ Lyman 
limit by $f_{\rm loc} = 25q$ ({\it i.e.} 25 times the mean intensity of the model Q 
background); this model is termed Q[A$_{25}$]. The model Q case is shown for 
comparison. {\it Right panel}: Column density ratios \SiIV/\CIV~:~\CII/\CIV\ in our 
highest redshift interval compared with Cloudy results as presented in Figure 18, here 
using model Q[A$_{25}$] with model Q for comparison (coding as indicated in the 
{\it left panel}). The cosmic microwave background is included.}
\end{figure}

\clearpage
\begin{figure}
\figurenum{\scriptsize 31}
\epsscale{1.0}  
\plotone{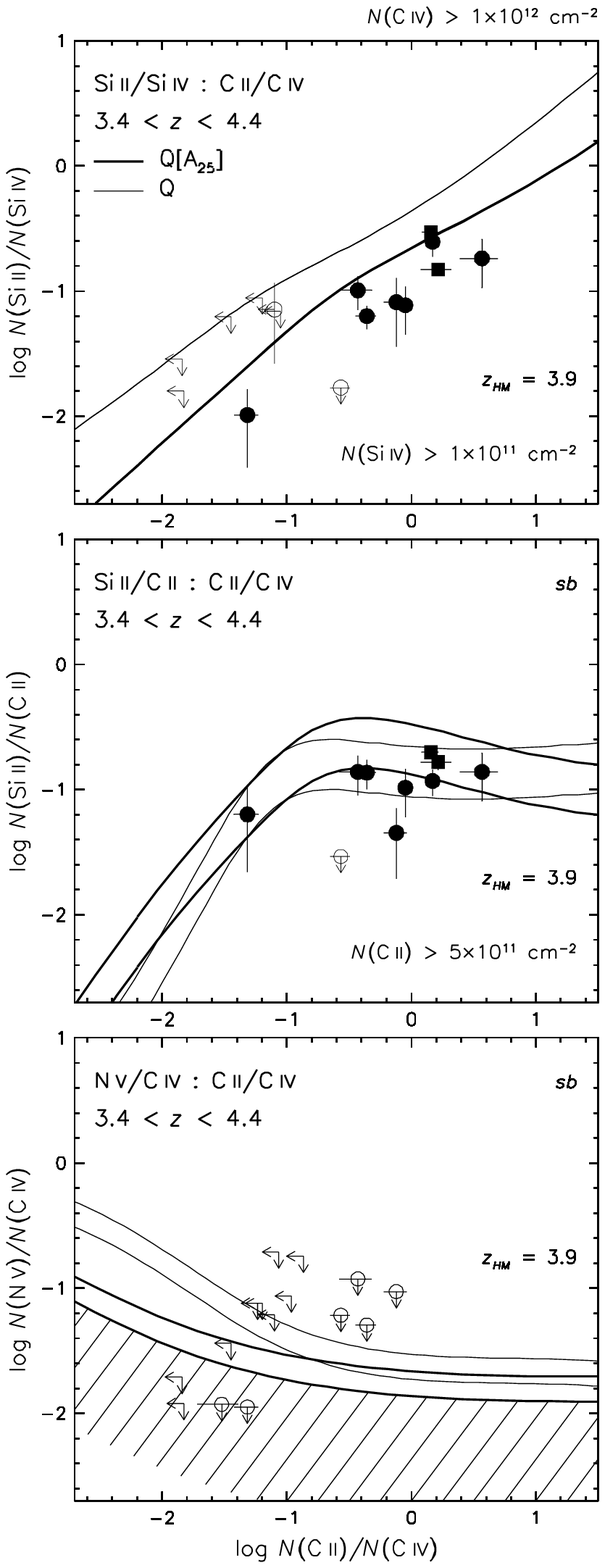}
\caption[f31.eps]{\scriptsize Same as for {\it right} 
panel in Figure 30, here for \SiII/\SiIV~:~\CII/\CIV, \SiII/\CII~:~\CII/\CIV\ and 
\NV/\CIV~:~\CII/\CIV\ (see comment in Figure 27 regarding N/C relative abundance, and 
shading), with the observed ratios taken from Figures 19 and 20.} 
\end{figure}

\clearpage
\begin{figure}
\figurenum{\scriptsize 32}
\epsscale{1.0}
\plotone{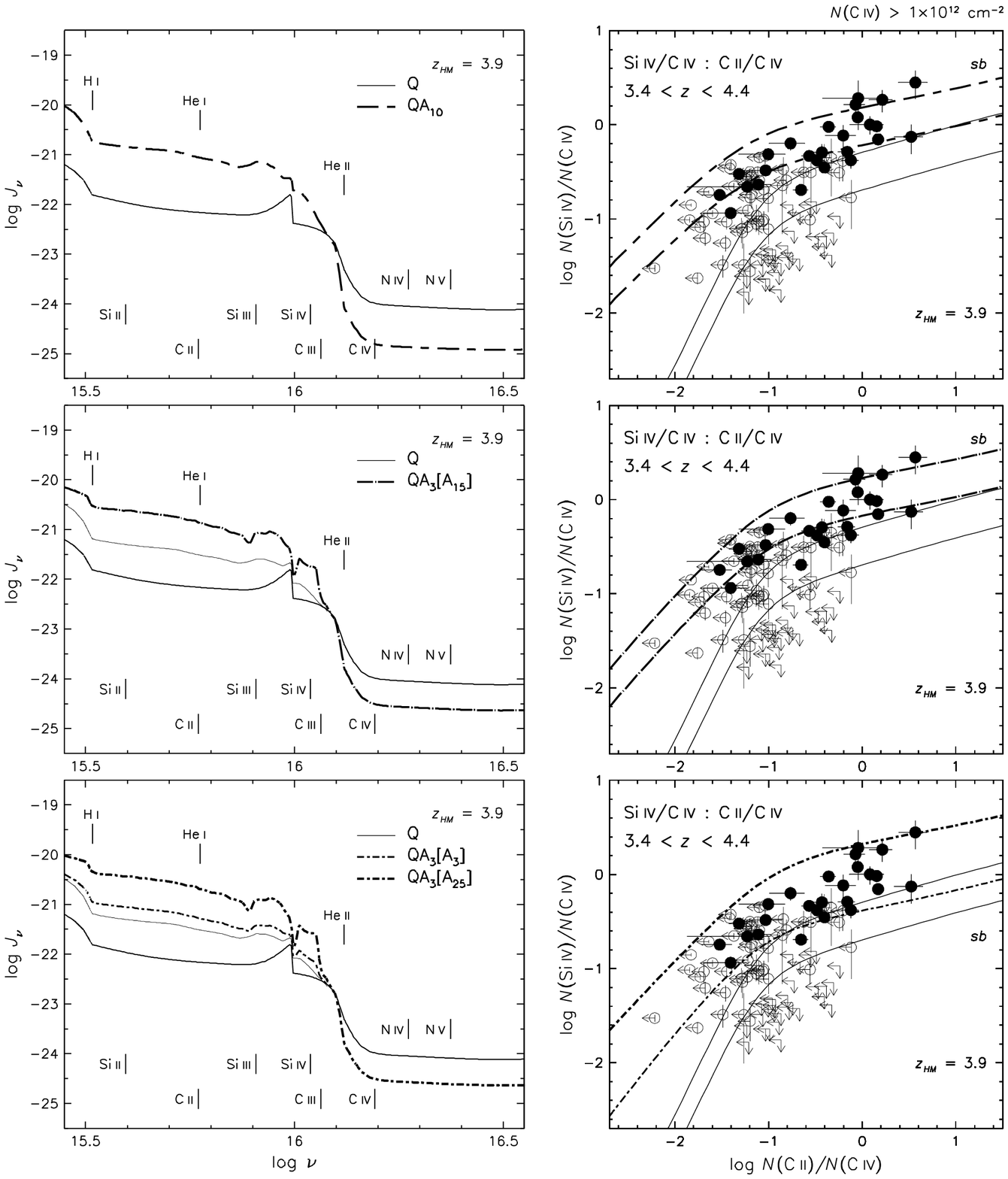}
\caption[f32.eps]{\scriptsize {\it Top left panel}: Spectral energy 
distribution as presented in Figure 26, here with a contrived Haardt \& Madau
QSO $+$ galaxy metagalactic background using ``galaxies'' made up from the 45,000 K 
stellar spectral model employed in Figure 30 {\it included} in the cosmological 
radiative transfer computation with a scaling in volume emissivity at the \HI\ Lyman 
limit relative to the QSOs by factor $f_{\rm met} = 10$ (see text), termed model 
QA$_{10}$; 
$J_{\nu_0} = 1.8 \times 10^{-21}$ erg s$^{-1}$ cm$^{-2}$ Hz$^{-1}$ sr$^{-1}$. Model 
Q is shown for comparison. {\it Middle left panel}: Similar to {\it top} panel with 
$f_{\rm met} = 3$ (shown by the thin continuous line; 
$J_{\nu_0} = 6.6 \times 10^{-22}$ erg s$^{-1}$ cm$^{-2}$ Hz$^{-1}$ sr$^{-1}$), 
combined with the 45,000 K stellar model representing {\it local} sources as in 
Figure 30 scaled by ${\it f}_{\rm loc} =$ 15q, termed model QA$_{3}$[A$_{15}$];
$J_{\nu_0} = 3.0 \times 10^{-21}$ erg s$^{-1}$ cm$^{-2}$ Hz$^{-1}$ sr$^{-1}$.
{\it Bottom left panel}: Similar to {\it middle} panel, showing bounds representing 
cosmic variance in the local source contribution with $f_{\rm loc} = 3q$ and $25q$, 
termed models QA$_{3}$[A$_{3}$] and QA$_{3}$[A$_{25}$]; 
$J_{\nu_0} = 1.1 \times 10^{-21}$ and 
$4.6 \times 10^{-21}$ erg s$^{-1}$ cm$^{-2}$ Hz$^{-1}$ sr$^{-1}$, respectively.
{\it Right panels}: Column density ratios \SiIV/\CIV~:~\CII/\CIV\ in our highest 
redshift interval compared with Cloudy results as in Figure 18, obtained using the 
models in the {\it left} panels with coding as indicated; for the curves in the 
{\it bottom} panel, QA$_{3}$[A$_{3}$] is coupled with absorber Si/C solar relative 
abundance and QA$_{3}$[A$_{25}$] with 2.5 $\times$ solar. The cosmic microwave 
background is included.}
\end{figure}

\clearpage
\begin{figure}
\figurenum{\scriptsize 33}
\epsscale{1.0}
\plotone{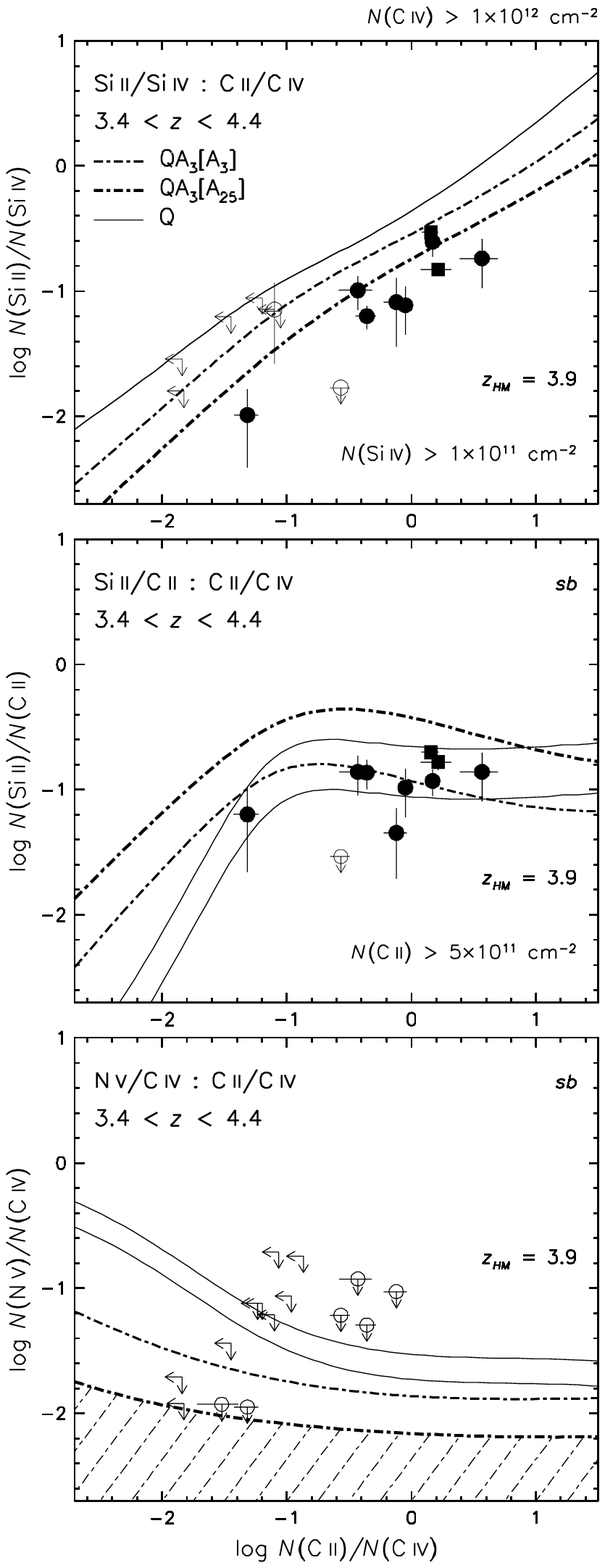}
\caption[f33.eps]{\scriptsize Same as for {\it bottom right} 
panel in Figure 32, for \SiII/\SiIV~:~\CII/\CIV, \SiII/\CII~:~\CII/\CIV\ and 
\NV/\CIV~:~\CII/\CIV\ (see comment in Figure 27 regarding shading), with the observed 
ratios taken from Figures 19 and 20. QA$_{3Q}$+A$_{3q}$ is coupled with absorber Si/C 
and N/C solar relative abundance and QA$_{3Q}$+A$_{25q}$ with 2.5 $\times$ and 
0.63 $\times$ solar, respectively.}
\end{figure}

\clearpage
\begin{figure}
\figurenum{\scriptsize 34}
\epsscale{1.0}
\plotone{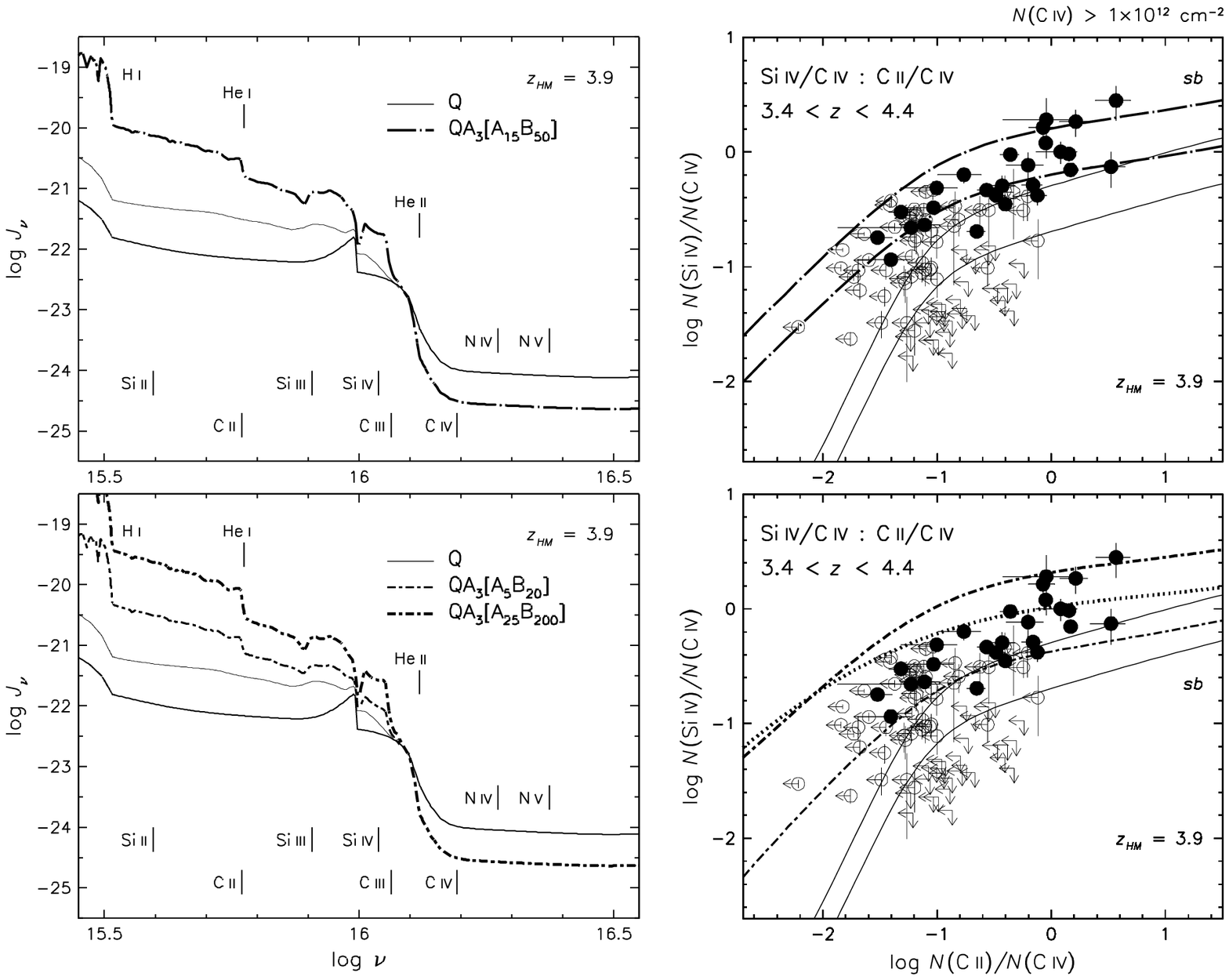}
\caption[f34.eps]{\scriptsize {\it Upper left panel}: Spectral energy 
distribution as presented in {\it middle left} panel of Figure 32, here with an added 
local stellar component introducing a significant \HeI\ ionization edge (see text) 
scaled by $f_{\rm loc} = 50q$, termed model QA$_{3}$[A$_{15}$B$_{50}$]; 
$J_{\nu_0} = 1.1 \times 10^{-20}$ erg s$^{-1}$ cm$^{-2}$ Hz$^{-1}$ sr$^{-1}$.
The faint trace is the metagalactic contribution within this case as given in 
Figure 32. {\it Lower left panel}: Similar to {\it upper} panel, showing bounds 
representing cosmic variance in the local source contributions, termed models 
QA$_{3}$[A$_{5}$B$_{20}$] and QA$_{3}$[A$_{25}$B$_{200}$];
$J_{\nu_0} = 5.3 \times 10^{-21}$ and 
$3.5 \times 10^{-20}$ erg s$^{-1}$ cm$^{-2}$ Hz$^{-1}$ sr$^{-1}$, respectively.
{\it Right panels}: Column density ratios \SiIV/\CIV~:~\CII/\CIV\ in our highest 
redshift interval compared with Cloudy results as in Figure 18, obtained using the 
models in the {\it left} panels with coding as indicated; for the curves in the 
{\it lower} panel, model QA$_{3}$[A$_{5}$B$_{20}$] is coupled with absorber Si/C 
solar relative abundance and QA$_{3}$[A$_{5}$B$_{20}$] with 2.5 $\times$ solar. 
The additional heavy dotted line in the {\it lower} panel is a case using model 
QA$_{3}$[A$_{25}$B$_{200}$] with absorbers of \emph{solar} Si/C relative abundance 
and metallicity [$Z$] $= -1.5$. The cosmic microwave background is included in all
computations.}
\end{figure}

\clearpage
\begin{figure}
\figurenum{\scriptsize 35}
\epsscale{1.0}
\plotone{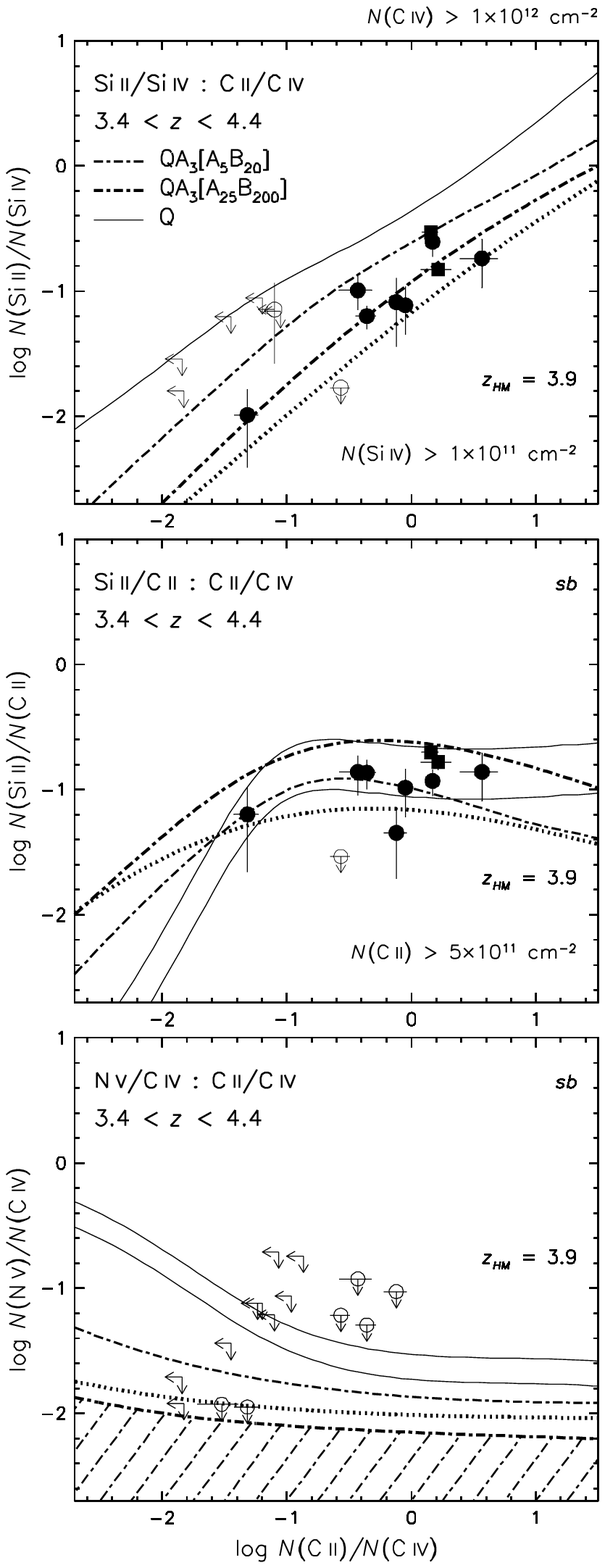}
\caption[f35.eps]{\scriptsize Same as for {\it lower 
right} panel in Figure 34, for \SiII/\SiIV~:~\CII/\CIV, \SiII/\CII~:~\CII/\CIV\ and 
\NV/\CIV~:~\CII/\CIV\ (see comment in Figure 27 regarding shading, applying also to 
the dotted line), with the observed ratios taken from Figures 19 and 20. Model 
QA$_{3}$[A$_{5}$B$_{20}$] is coupled with absorber Si/C and N/C solar relative 
abundance and QA$_{3}$[A$_{25}$B$_{200}$] with 2.5 $\times$ and 0.63 $\times$ 
solar, respectively.} 
\end{figure}

\clearpage
\begin{figure}
\figurenum{\scriptsize 36}
\epsscale{1.0}
\plotone{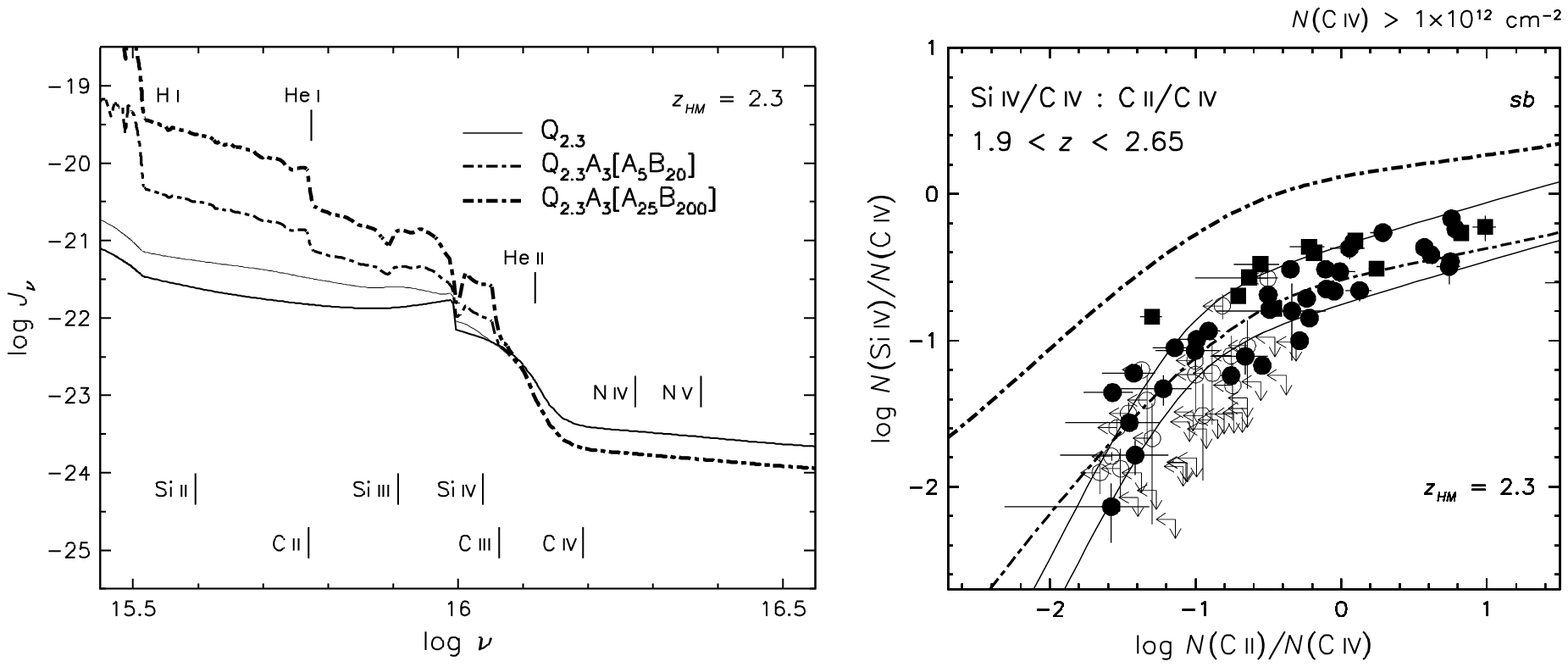}
\caption[f36.eps]{\scriptsize Similar to {\it lower} panels in Figure 
34, substituting the QSO source flux at $z_{HM} = 2.3$ but otherwise using the same 
stellar source quantities, termed models Q$_{2.3}$A$_{3}$[A$_{5}$B$_{20}$] and 
Q$_{2.3}$A$_{3}$[A$_{25}$B$_{200}$], and comparing the Cloudy results with the 
data in our lowest redshift interval presented in Figure 18. The appropriate cosmic 
microwave background is included.}
\end{figure}

\end{document}